\documentclass[aps,prd,preprint,groupedaddress,nofootinbib]{revtex4}
\usepackage{graphicx}
\usepackage{amsmath,amssymb}
\usepackage{bm}
\usepackage[caption=false]{subfig}
\usepackage{color}
\usepackage{floatrow}
\usepackage{setspace}
\usepackage[x11names]{xcolor}
\usepackage[colorlinks,linkcolor=purple,citecolor=teal]{hyperref}

\def\Lag{\mathcal{L}}
\def\p{\partial}

\def\arg{\mathop{\mathrm{arg}}\nolimits}

\def\=:{=\hspace{-.7em}\raisebox{1.1ex}{.}\hspace{.1em}\raisebox{-0.2ex}{.}}

\newcommand{\beq}{\begin{eqnarray}}
\newcommand{\eeq}{\end{eqnarray}}
\newcommand{\non}{\nonumber\\}

\newcommand{\bphi}{\boldsymbol{\phi}}
\newcommand{\U}{{\rm U}}
\newcommand{\SU}{{\rm SU}}
\newcommand{\Og}{{\rm O}}
\newcommand{\SO}{{\rm SO}}

\begin{document}

\title{Baryonic handles:\\ Skyrmions as open vortex strings on a domain wall}

\author{Sven Bjarke Gudnason}
\email{gudnason(at)keio.jp}
\affiliation{Department of Physics, and Research and Education Center for Natural 
Sciences, Keio University, Hiyoshi 4-1-1, Yokohama, Kanagawa 223-8521, Japan
}
\author{Muneto Nitta}
\email{nitta(at)phys-h.keio.ac.jp}
\affiliation{Department of Physics, and Research and Education Center for Natural 
Sciences, Keio University, Hiyoshi 4-1-1, Yokohama, Kanagawa 223-8521, Japan
}
\date{\today}
\begin{abstract}
We consider the BEC-Skyrme model which is based on a Skyrme-type model
with a potential motivated by Bose-Einstein condensates (BECs) and, in
particular, we study the Skyrmions in proximity of a domain wall that
inhabits the theory. The theory turns out to have a rich flora of
Skyrmion solutions that manifest themselves as twisted vortex rings or
vortons in the bulk and vortex handles attached to the domain wall. 
The latter are linked open vortex strings. We further study the
interaction between the domain wall and the Skyrmion and between the
vortex handles themselves as well as between the vortex handle and the
vortex ring in the bulk. We find that the domain wall
provides a large binding energy for the solitons and it is
energetically preferred to stay as close to the domain wall as
possible; other configurations sticking into the bulk are metastable. 
We find that the most stable 2-Skyrmion is
a torus-shaped braided string junction ending on the domain wall, 
which is produced by a collision of two vortex handles on the wall, 
but there is also a metastable configuration which is a doubly twisted
vortex handle produced by a collision of a vortex handle on
the wall and a vortex ring from the bulk.
\end{abstract}
\maketitle

\tableofcontents

\section{Introduction}

Skyrmions are topological textures that are the minimizers in energy
of a map from 3-space to SU(2) isospin space
\cite{Skyrme:1961vq,Skyrme:1962vh} and 
provide a low-energy effective description of baryons in large-$N$ QCD
\cite{Witten:1983tw,Witten:1983tx}. 
In condensed matter physics, Skyrmions usually refer to 2-dimensional
solitons living in a magnetization vector of a suitable host material 
\cite{Bogdanov:1989,Muhlbauer:2009,Yu:2010,Cortes-Ortuno:2017}. 
Nonetheless, considerable strides are being made in condensed matter 
physics to realize a 3-dimensional Skyrmion in a two-component
Bose-Einstein condensate (BEC)
\cite{Ruostekoski:2001fc,Battye:2001ec,Khawaja:2001,Khawaja:2001zz,Savage:2003hh,Ruostekoski:2004pj,Wuster:2005,Oshikawa:2006,Tokuno:2009,Kawakami:2012zw,Nitta:2012hy}
(see Ref.~\cite{Kasamatsu:2005} for a nice review).
A potential which is motivated by two-component BECs was considered in
Refs.~\cite{Gudnason:2014gla,Gudnason:2014hsa,Gudnason:2014jga} and it
was shown that the Skyrmion is deformed into a twisted vortex ring (or
vorton) due to the explicit breaking of the SU(2) symmetry that is
normally possessed by Skyrmions (i.e.~isospin symmetry). 

A convenient parametrization of the Skyrme field (chiral Lagrangian
field) is to write the O(4) vector field as a two component complex
scalar field, $\phi_{1,2}$, with the constraint
$|\phi_1|^2+|\phi_2|^2=1$. 
The BEC-inspired potential takes the form $V=M^2|\phi_1|^2|\phi_2|^2$
and thus there are two different vacua: either $\phi_1$ or $\phi_2$
must vanish in the vacuum.
In the vacuum where one component vanishes, 
the vacuum manifold $S^1$ is parametrized by the phase of the other
component. 
For instance, in the vacuum where $\phi_2$ vanishes, $|\phi_1|=1$ 
and vice versa. Hence, there exists a global vortex 
having logarithmically divergent energy 
\cite{Gudnason:2014gla,Gudnason:2014hsa,Gudnason:2014jga}.
In the vortex core, the other component is confined; 
for instance, in the core of a vortex having a winding in the $\phi_1$
component, the $\phi_2$ component is confined which yields a U(1)
modulus. 
This is a global analog of the superconducting cosmic string
\cite{Witten:1984eb}.
These properties are also shared by a model with the potential
$V=M^2|\phi_1|^2$ \cite{Gudnason:2016yix}.
A Skyrmion in the BEC-Skyrme model takes the form of a twisted vortex
ring or a vorton, i.e.~a vortex ring along which the phase of the
confined component winds from $0$ to $2\pi$. If it is twisted $B$
times (i.e.~it winds from $0$ to $2\pi B$), then it carries baryon
number $B$. 
This was actually first found in two-component BECs 
\cite{Ruostekoski:2001fc,Battye:2001ec,Khawaja:2001,Khawaja:2001zz,Savage:2003hh,Ruostekoski:2004pj,Wuster:2005,Oshikawa:2006,Tokuno:2009,Kawakami:2012zw,Nitta:2012hy},
and is shared by this model as well as the above mentioned model with
potential $V=M^2|\phi_1|^2$ \cite{Gudnason:2016yix}.

The BEC-Skyrme model also admits a domain wall interpolating between
the two vacua. 
Vortices can terminate on the domain wall, just like in two-component
BECs
\cite{Kasamatsu:2010aq,Kasamatsu:2013lda,Kasamatsu:2013qia,Nitta:2012hy}
(see also Ref.~\cite{Mateo:2016bai}). 
This configuration resembles a D-brane in string theory, 
thereby called a D-brane soliton 
\cite{Gauntlett:2000de,Isozumi:2004vg,Shifman:2002jm,Gudnason:2014uha}.
Vortex endpoints are also called boojums, since they resemble those in
Helium-3 superfluids \cite{Mermin,Volovik}. 
One question is what happens if the two ends of a vortex terminate on
the same domain wall.  
A Skyrme model with the potential term 
$V=M^2(\Re(\phi_1))^2$ also admits a domain wall, 
and Skyrmions are absorbed into the wall becoming 
lumps or baby-Skyrmions on the wall 
\cite{Nitta:2012wi,Nitta:2012rq,Gudnason:2014nba}.
Thus a further question is what happens to the Skyrmions in the
presence of a domain wall in the BEC-Skyrme model.
We will answer this question with the results of this paper. 

In this paper, we will study all the physics in the proximity of the
above-mentioned domain wall. If we start in the bulk and place a
Skyrmion, it is a twisted vortex ring, as mentioned above. If the
distance to the domain wall is not too large, an attractive force will
pull the Skyrmion into the domain wall and it will be absorbed. The
vortex ring is then connected to the vacuum on the other side of the
domain wall and it thus manifests itself as a handle sitting on the
domain wall, sticking into the bulk of the side it came from.
This description sounds asymmetric and, in fact, we shall discover in
this paper that the true energy minimizer symmetrizes itself to become 
symmetric under the exchange of $\phi_1$ and $\phi_2$;
it becomes a link of two vortex handles.
This is in stark contrast to the above-mentioned model with 
the potential $V=M^2(\Re(\phi_1))^2$,  
in which case a Skyrmion absorbed into the wall becomes 
a baby-Skyrmion and no complicated structure appears 
\cite{Nitta:2012wi,Nitta:2012rq,Gudnason:2014nba}.
The attraction between the vortex ring and the domain wall is also a
result of a large reduction of the static energy by absorption into  
the domain wall. We perform an explicit calculation and find that
there is a large binding energy for the vortex ring to gain at the
cost of transforming itself into a handle sitting on the domain wall.

We shall also study the interactions between two vortex handles on the
domain wall. Since, as we already spoiled, the handle configuration
will turn out to be symmetric, it does not matter on which side of the
domain wall the handle sits. We find that there is an attractive
channel and a repulsive channel between the two vortex handles. 
If we place the handles in the attractive channel, they will combine
themselves into a braided string junction of a toroidal shape -- a
2-Skyrmion absorbed into the domain wall.

We also investigate the interaction between the vortex handle and the
vortex ring in the bulk and find that they always attract each other
(or quickly rotate themselves into the attractive channel). Next the
vortex ring goes through a string reconnection mechanism to transform
the configuration into a doubly twisted vortex handle, which is 
the analog of a vortex ring that is twisted 2 times (hence baryon number
two) being absorbed into the domain wall.

Finally, we compare the energies of the two Skyrmion configurations
with baryon number two and find that the braided string junction has
less energy than the doubly twisted vortex handle. 
 
This paper is organized as follows. In Sec.~\ref{sec:model} we
will briefly review the BEC-Skyrme model and its symmetry and vacuum
properties. Sec.~\ref{sec:dw} reviews the domain wall. In
Sec.~\ref{sec:boojum}, we construct the boojum for the first time in
the BEC-Skyrme model; it is a semi-infinite string attached to the
domain wall. In Sec.~\ref{sec:openstringloop} the vortex handle is
constructed. In Sec.~\ref{sec:ring} the interaction between the domain
wall and the vortex ring is studied. In Sec.~\ref{sec:doublehandle} we
add a twist to the modulus of the vortex handle, producing a 
2-Skyrmion. In Sec.~\ref{sec:int} we study the interactions between
two vortex handles and find a new 2-Skyrmion: the braided string
junction of toroidal shape. Sec.~\ref{sec:ringint} considers the
interaction between the vortex handle on a wall and the vortex ring in
the bulk, which reproduces the doubly twisted vortex handle found in
Sec.~\ref{sec:doublehandle}. In Sec.~\ref{sec:energy} the energies of
the two 2-Skyrmions are compared. Sec.~\ref{sec:higher} considers the
construction of higher-charged configurations. Finally, we conclude
with a discussion in Sec.~\ref{sec:discussion}.

\section{The BEC-Skyrme model}\label{sec:model}

The model that we will consider in this paper is the generalized
Skyrme model, consisting of the kinetic term, the Skyrme term
\cite{Skyrme:1961vq,Skyrme:1962vh}, the 
BPS-Skyrme term \cite{Adam:2010fg,Adam:2010ds}
\begin{align}
  \Lag &= \Lag_2 + c_4\Lag_4 + c_6\Lag_6 - V,\label{eq:Lag}\\
  \Lag_2 &= -\frac12\p_\mu\bphi^\dag\p^\mu\bphi,\\
  \Lag_4 &= \frac18(\p_\mu\bphi^\dag\p_\nu\bphi)(\p^{[\mu}\bphi^\dag\p^{\nu]}\bphi)
    + \frac18(\p_\mu\bphi^\dag\sigma^2\p_\nu\bphi)(\p^{[\mu}\bphi^\dag\sigma^2\p^{\nu]}\bphi) \non
    &= -\frac14(\p_\mu\bphi^\dag\p^\mu\bphi)^2
    +\frac{1}{16}(\p_\mu\bphi^\dag\p_\nu\bphi + \p_\nu\bphi^\dag\p_\mu\bphi)^2,\\
  \Lag_6 &= \frac14(\epsilon^{\mu\nu\rho\sigma}\bphi^\dag\p_\nu\bphi\p_\rho\bphi^\dag\p_\sigma\bphi)^2,
\end{align}
and finally the BEC-inspired potential
\cite{Gudnason:2014gla,Gudnason:2014hsa,Gudnason:2014jga} 
\beq
V = \frac18M^2\left[1 - (\bphi^\dag\sigma^3\bphi)^2\right]
  = \frac12M^2|\phi_1|^2|\phi_2|^2,
\label{eq:V}
\eeq
where $\sigma^a$ are the Pauli matrices. 
The vector $\bphi\equiv(\phi_1(x),\phi_2(x))^{\rm T}$ is a complex
2-vector field, the spacetime indices $\mu,\nu,\rho,\sigma$ run over 0
through 3, the flat Minkowski metric is taken to be of the mostly
positive signature, and finally, the nonlinear sigma model constraint
is imposed as $\bphi^\dag\bphi=|\phi_1|^2+|\phi_2|^2=1$. 
The relation of the complex vector field $\bphi$, or equivalently the
two complex fields $\phi_{1,2}$, to the usual chiral Lagrangian field
used in the Skyrme model is given by 
\beq
U =
\begin{pmatrix}
\bphi & -i\sigma^2 \bar{\bphi}
\end{pmatrix}
=
\begin{pmatrix}
  \phi_1 & -\bar\phi_2\\
  \phi_2 & \bar\phi_1
\end{pmatrix},
\eeq
and thus the nonlinear sigma model constraint reads
$\det U=|\phi_1|^2+|\phi_2|^2=1$.

The target space (i.e.~the vacuum manifold) for $M=0$ is
$\mathcal{M}\simeq \Og(4)/\Og(3) \simeq \SU(2)\simeq S^3$,
and thus the 
one-point compactified space $\mathbb{R}^3\cup\{\infty\}\simeq S^3$
supports topological solitons (Skyrmions) characterized by 
\beq
\pi_3(\mathcal{M})=\mathbb{Z}.
\eeq
The topological degree of the map $\bphi$ is $B\in\pi_3(S^3)$ and can
be calculated as
\beq
B = \frac{1}{4\pi^2}\int d^3x\; \epsilon^{i j k}
  \bphi^\dag\p_i\bphi\p_j\bphi^\dag\p_k\bphi.
\eeq
Once we turn on a nonvanishing potential $M>0$, the Skyrmions survive,
but the vacuum of the theory and the physics of the solitons change.

The two vacua of the model with nonvanishing potential $V$ in
\eqref{eq:V}, are 
\begin{align}
\odot&:\quad\bphi = (e^{i\alpha},0)^{\rm T},\non
\otimes&:\quad\bphi = (0,e^{i\beta})^{\rm T},
\label{eq:vacua}
\end{align}
which by the nonlinear sigma-model constraint yield the other
component to be at its maximum.

The symmetry of the model with the potential $V$ in Eq.~\eqref{eq:V}
is explicitly broken from $\Og(4)$ down to
\beq
G = \U(1) \times \Og(2)
\simeq \U(1)_0\times [\U(1)_3\rtimes(\mathbb{Z}_2)_{1,2}],
\eeq
where the group is defined by the symmetries
\begin{align}
\U(1)_0&:\quad \bphi\to e^{i\alpha}\bphi,\\
\U(1)_3&:\quad \bphi\to e^{i\beta\sigma^3}\bphi,\\
(\mathbb{Z}_2)_{1,2}&:\quad e^{i\pi\sigma^{1,2}/2}\bphi,
\end{align}
and $\U(1)_3$ acts on $\mathbb{Z}_2$ in such a way that they define a
semi-direct product denoted by $\rtimes$.
The unbroken symmetry groups in the vacua \eqref{eq:vacua}, are thus
\begin{align}
H_\odot = \U(1)_{0-3} &:\quad \bphi\to e^{i\alpha}e^{-i\alpha\sigma^3}\bphi,\\
H_\otimes = \U(1)_{0+3} &:\quad \bphi\to e^{i\alpha}e^{+i\alpha\sigma^3}\bphi,
\end{align}
The target space (vacuum manifold) is thus given by the coset group
\beq
\mathcal{M} \simeq G/H
= \frac{\U(1)_0\times [\U(1)_3\rtimes (\mathbb{Z}_2)_{1,2}]}{\U(1)_{0\pm 3}}
\simeq \SO(2)_{0\mp 3}\rtimes (\mathbb{Z}_2)_{1,2} = \Og(2),
\eeq
and the nontrivial homotopy groups of this manifold read
\beq
\pi_0(\mathcal{M}) = \mathbb{Z}_2,\qquad
\pi_1(\mathcal{M}) = \mathbb{Z}.
\eeq
The theory thus supports both domain walls and vortices in addition to
the Skyrmions.

Although we have included both the Skyrme term, $\Lag_4$, and the
BPS-Skyrme term, $\Lag_6$, in the model, we will only use either of
the terms as follows
\beq
2+4 \ {\rm model}: &\qquad& c_4=1, \ \ c_6=0,\non
2+6 \ {\rm model}: &\qquad& c_4=0, \ \ c_6=1.
\eeq
It turns out that the two models give qualitatively the same results,
so we will only show some of the results for both models in the next
section. 

In Ref.~\cite{Gudnason:2014jga} we studied the domain wall, the vortices
and the Skyrmions in one vacuum (the $H_\otimes$ vacuum).

In this paper we study the theory in the presence of the domain wall
of Ref.~\cite{Gudnason:2014jga} with Skyrmions as closed or open
vortex strings in proximity of the domain wall.

\section{Vortices and the domain wall}

We will now consider that the 3-dimensional space has a domain wall
separating two phases with vacua $\otimes$ and $\odot$, respectively. 
Without loss of generality, we will consider the vortices 
in the $\otimes$ phase; the results apply to the $\odot$
phase by interchanging the complex scalar fields,
$\phi_1\leftrightarrow\phi_2$.

\subsection{The domain wall}\label{sec:dw}

As we will place everything in this paper in the presence of the
domain wall, we will make a short review of the domain wall solution
\cite{Gudnason:2014gla,Gudnason:2014hsa,Gudnason:2014jga} here.
Since the domain wall is a codimension-1 soliton, only the potential
\eqref{eq:V} and the kinetic term, $\Lag_2$, contribute to its
energy
\beq
2E = |\p_z\phi_1|^2
  + |\p_z\phi_2|^2
  + M^2|\phi_1|^2|\phi_2|^2.
\eeq
The solution is thus
\beq
\bphi = \frac{1}{\sqrt{1 + e^{\pm 2M(z - z_0)}}}
\begin{pmatrix}
  e^{i\chi}\\
  e^{\pm M(z - z_0) + i\vartheta}
\end{pmatrix},
\eeq
with $\chi$ and $\vartheta$ being constant phase parameters. 
We will choose the upper sign throughout this paper and hence the
$\otimes$ vacuum is always at $z>0$ (up) and the $\odot$ vacuum is
at $z<0$ (down).
Furthermore, we will set $z_0=0$ from now on.
This is the translational modulus of the domain wall and we can always
adjust our coordinate system such that the domain wall is at $z=0$.

\subsection{The boojum or D-brane soliton}\label{sec:boojum}

We will now consider attaching a single (infinitely long) open
vortex string to the domain wall from the side of the $\otimes$
phase. 
The energy of such a system is thus divergent for three different
reasons: the domain wall carries an infinite energy, the semi-infinite
vortex string has infinite energy and finally, the fact that it is a
global vortex implies that the energy in the direction transverse to
the string in the $\otimes$ phase diverges logarithmically due to the
winding contribution to the kinetic term. 
This solution serves mostly as an illustration.

It will be convenient to parametrize the fields as follows
\begin{align}
  \bphi =
  \begin{pmatrix}
    e^{i\chi}\cos f \\
    e^{i\vartheta}\sin f
  \end{pmatrix}.
  \label{eq:phi_parametrization}
\end{align}
The vortex is a solution that winds in $\phi_2$ (since we are in the 
$\otimes$ phase) which thus means that $\vartheta$ is a winding
phase. The ``profile function'' of the (global) vortex is $\sin f$ 
and due to the winding phase, it must vanish at the vortex center; we
thus choose $f=0$ there.
Asymptotically, $f$ tends to its vacuum value in the $\otimes$ phase,
which is $f=\tfrac{\pi}{2}$.

An initial condition for the numerical calculation of the boojum can 
thus be constructed by combining the domain wall solution and the 
vortex Ansatz
\begin{align}
\bphi &=
\begin{pmatrix}
  \cos f & 
  e^{i\vartheta}\sin f
\end{pmatrix},\\
\sin f &= \left\{
\begin{array}{ll}
  \frac{e^{M z}}{1+e^{2M z}}, & z\leq 0,\\
  \frac{e^{M z}}{1+e^{2M z}} \mathcal{F}(\rho), & z>0,
\end{array}\right.\\
\vartheta &= \left\{
\begin{array}{ll}
  0, & z\leq 0,\\
  \theta, & z>0,
\end{array}\right.
\end{align}
where $\mathcal{F}(\rho)$ is a suitable guess for the vortex profile
function obeying $\mathcal{F}(0)=0$ and $\mathcal{F}(\infty)=1$. 

\begin{figure}[!htp]
  \begin{center}
    \mbox{\subfloat[]{\includegraphics[width=0.49\linewidth]{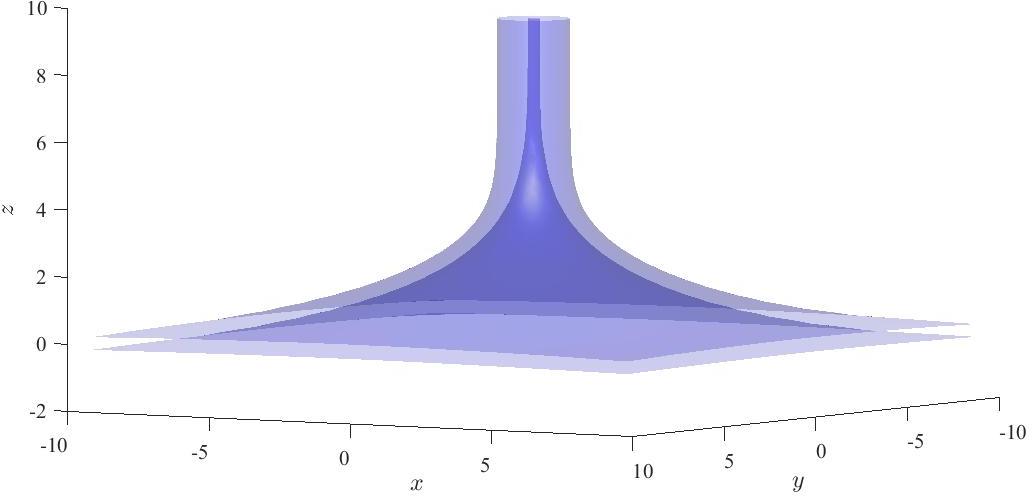}}
      \subfloat[]{\includegraphics[width=0.49\linewidth]{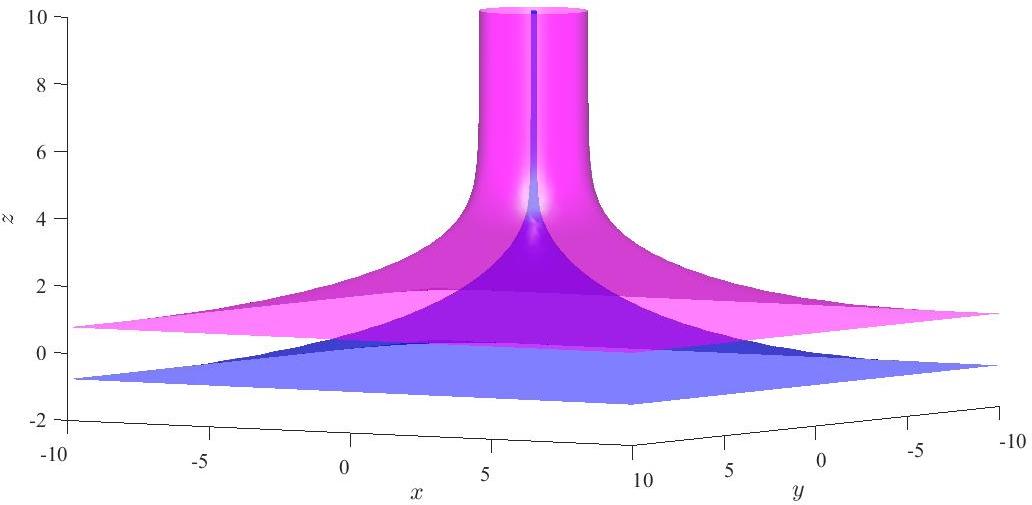}}}
    \caption{The vortex string ending on the domain wall from the
      $\otimes$ phase in the 2+4 model: (a) energy isosurface(s), the
      maximum of the energy is between the two surfaces.
      (b) isosurfaces of $0.1|\phi_1|$ (magenta) and $0.1|\phi_2|$ (blue),
      showing the surfaces close to the vacua.
      The typical logarithmic bending of the domain wall
      characteristic of a boojum is clearly visible.
      The parameters $c_4=1$ and $M=3$ were used for this
      calculation. } 
    \label{fig:boojum}
  \end{center}
\end{figure}

Fig.~\ref{fig:boojum} shows the numerical solution for the boojum.
A well-known characteristic of the boojum junction of the vortex
string and the domain wall is that the domain wall bends
logarithmically due to the junction.

\section{Skyrmions as a vortex handle on a domain wall}

Without loss of generality, we will again consider Skyrmions  
in the $\otimes$ phase; the results apply to the $\odot$
phase by interchanging the complex scalar fields,
$\phi_1\leftrightarrow\phi_2$.
We will consider the Skyrmions to be closed vortex strings (of
$\phi_2$) in the $\otimes$ phase and open vortex strings attached to 
the domain wall, from the side of the domain wall with the $\otimes$
phase.
This soliton picture closely resembles the physics in string theory. 

\subsection{The vortex handle -- open string loop}\label{sec:openstringloop}

We will now consider a single Skyrmion ($B=1$) as an open vortex
string loop emanating from the domain wall (at $z=0$) into the
$\otimes$ phase and returning back to the domain wall.

The vortex a priori does not bear baryon charge ($\pi_3$ charge
density).
We will now explain how it comes about.
Using the parametrization \eqref{eq:phi_parametrization} of $\bphi$,
the vortex position implies $f=0$ in the $(x,y)$-plane at $z=0$ and
the phase $\vartheta$ winds locally around the vortex zero.
The string extends into the $\otimes$ phase and eventually returns to
the domain wall providing another zero in $f$ in a position different
from the starting point.
Since the phase winds around the vortex all the way into the bulk
phase and on the way back, the vortex that comes back to the domain
wall looks to the domain wall like an antivortex. 
The solution so far can be made with 2 components of the $\bphi$ field
equal and hence it is clear that the baryon charge vanishes.
The way this soliton becomes a Skyrmion is by turning on a twisting of
its U(1) modulus, which lives in the string world volume.
More precisely, we will let the phase field $\chi$ wind from $0$ to
$\pi$ on the way out and away from the domain wall, and from $\pi$ to
$2\pi$ on the way back to the domain wall.
This extra ``winding'' will make the string handle cover the 3-sphere
(the target space) and hence comprise a unit baryon charge. 

The above description is quite idealistic, however, the vortex string 
will have a tension which will tend to shorten the string as much as
possible.
For this reason, the unit charge Skyrmion or the 1-Skyrmion as a
single vortex string handle attached to the domain wall, will be quite
short.

There is a simple way to extend the above construction to a
$B$-Skyrmion, i.e.~by twisting the phase function $\chi$ not 1 time,
but $B$ times. More precisely, we can twist the function $\chi$ from
$0$ to $\pi B$ on the way into the bulk and from $\pi B$ to $2\pi B$
on the way back to the domain wall.
This can produce a vortex handle with baryon charge or topological
degree $B$.
We will see shortly that the configuration becomes more complicated
for higher twists, than what we described here.

\begin{figure}[!htp]
  \begin{center}
    \mbox{\sidesubfloat[]{\includegraphics[width=0.65\linewidth]{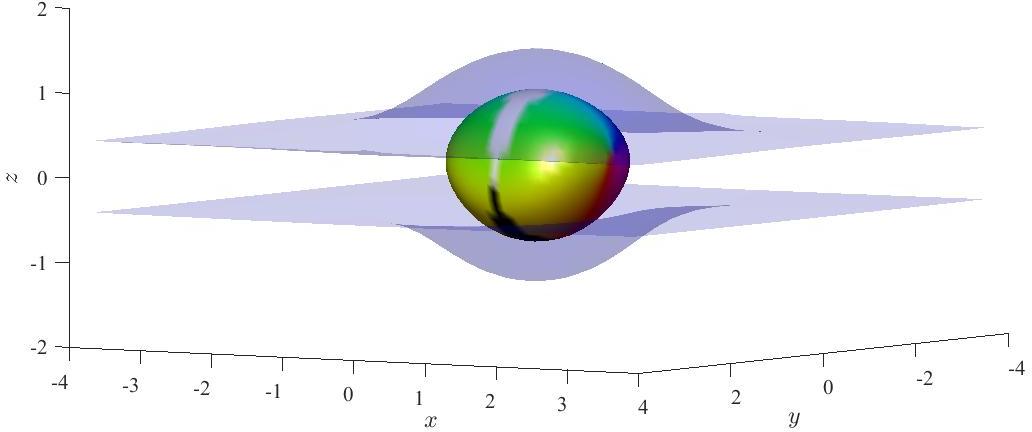}}}
    \mbox{\sidesubfloat[]{\includegraphics[width=0.65\linewidth]{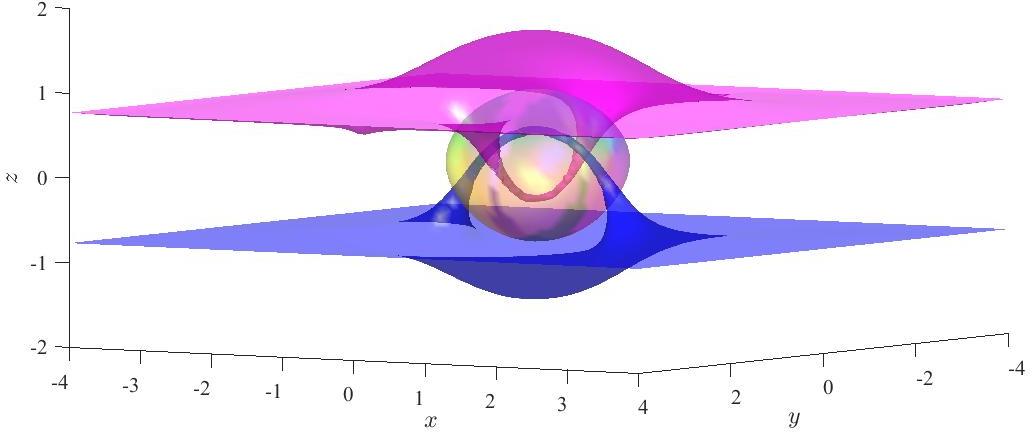}}}
    \mbox{\sidesubfloat[]{\includegraphics[width=0.4\linewidth]{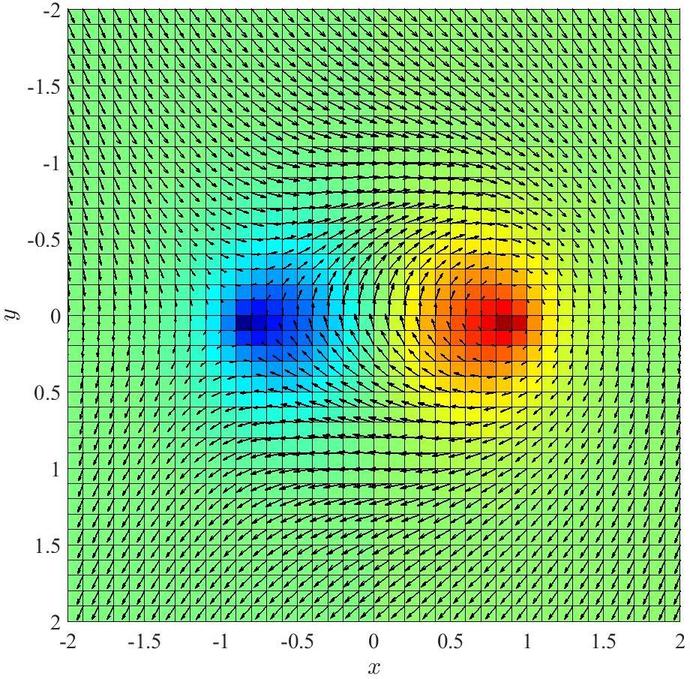}}\ \ 
      \sidesubfloat[]{\includegraphics[width=0.4\linewidth]{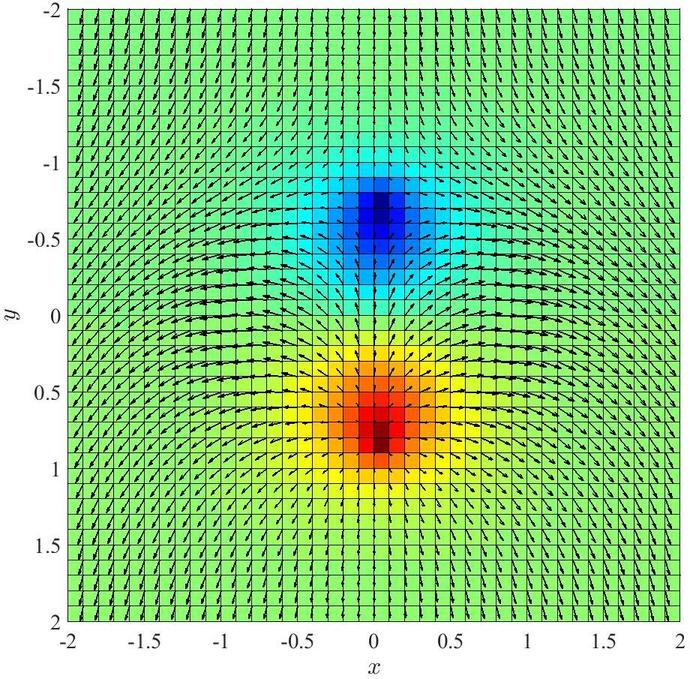}}}
    \caption{Skyrmion as a vortex handle with a single twist on the
      domain wall in the 2+4 model. (a) the energy density is shown
      with blue transparent isosurfaces illustrating the domain wall
      and the baryon charge density is shown with an isosurface with
      the color scheme described in the text. (b) the blue isosurface
      (bottom) represents the zeros of $\phi_2$ and the magenta
      isosurface (top) is the zeros of $\phi_1$. The baryon charge is
      added transparently. (c) the vortex (red) and antivortex (blue)
      pair of $\phi_2$. (d) the vortex (red) and antivortex (blue)
      pair of $\phi_1$. In this figure, we have taken $M=3$.  }
    \label{fig:24B1}
  \end{center}
\end{figure}

\begin{figure}[!htp]
  \begin{center}
    \mbox{\sidesubfloat[]{\includegraphics[width=0.65\linewidth]{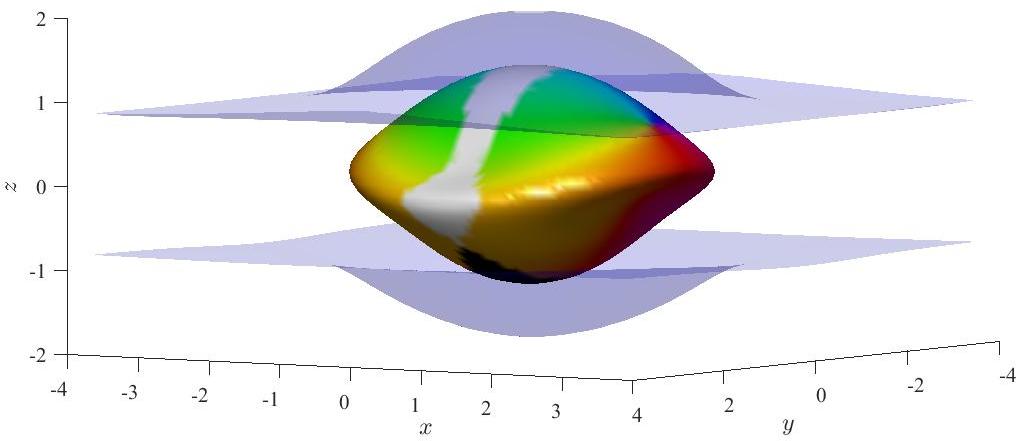}}}
    \mbox{\sidesubfloat[]{\includegraphics[width=0.65\linewidth]{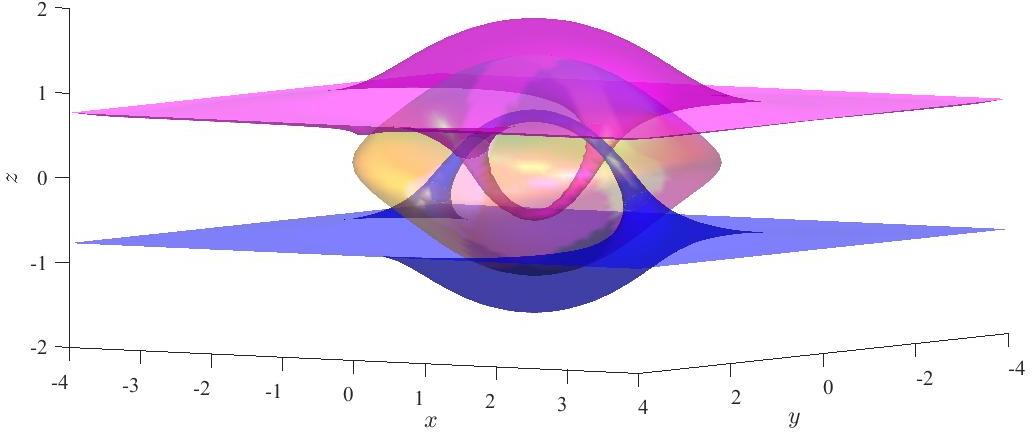}}}
    \mbox{\sidesubfloat[]{\includegraphics[width=0.4\linewidth]{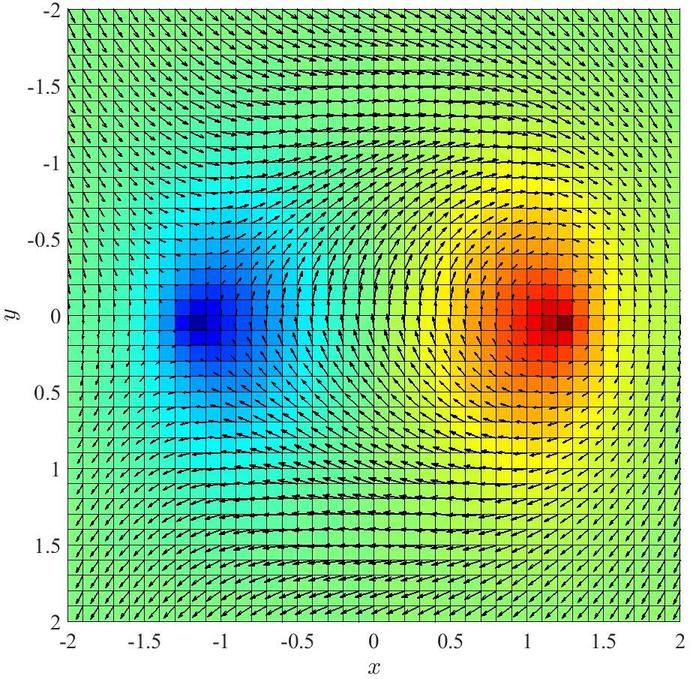}}\ \ 
      \sidesubfloat[]{\includegraphics[width=0.4\linewidth]{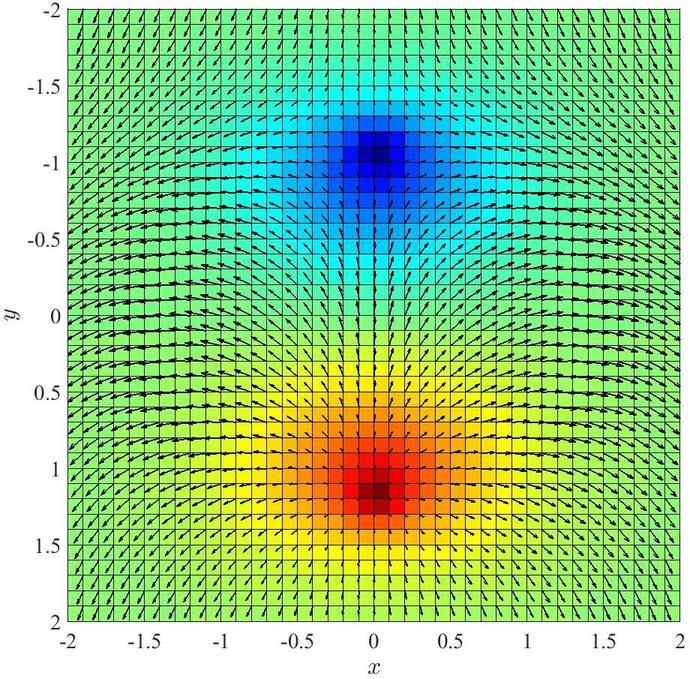}}}
    \caption{Skyrmion as a vortex handle with a single twist on the
      domain wall in the 2+6 model. (a) the energy density is shown
      with blue transparent isosurfaces illustrating the domain wall
      and the baryon charge density is shown with an isosurface with
      the color scheme described in the text. (b) the blue isosurface
      (bottom) represents the zeros of $\phi_2$ and the magenta
      isosurface (top) is the zeros of $\phi_1$. The baryon charge is
      added transparently. (c) the vortex (red) and antivortex (blue)
      pair of $\phi_2$. (d) the vortex (red) and antivortex (blue)
      pair of $\phi_1$. In this figure, we have taken $M=3$. }
    \label{fig:26B1}
  \end{center}
\end{figure}

We are now ready to present the numerical solution of the single
vortex string handle attached to the domain wall.
The numerical solution is shown in Fig.~\ref{fig:24B1}.
Fig.~\ref{fig:24B1}(a) shows the energy isosurfaces in transparent
blue and the baryon charge isosurfaces with an unconventional
color scheme that we will describe shortly.
Fig.~\ref{fig:24B1}(b) shows the vacua, or more precisely the values
$0.1|\phi_1|$ in magenta and $0.1|\phi_2|$ in blue.
The domain wall approaches the vacuum exponentially, so the true
vacuum value is only reached asymptotically.
The vortex, on the other hand, approaches the $\otimes$ vacuum
linearly, so the true vortex zero is quite close to the tube-like
surface surrounding it.
Figs.~\ref{fig:24B1}(c) and (d) show the vortices of $\phi_2$ and
$\phi_1$, respectively, in the $(x,y)$-plane at $z=0$ (in the middle
of the domain wall). 
The arrows are unit vectors pointing in the direction given by
$\arg(x+iy)=\arg(\phi_2)$. The length of the arrows is normalized to
the unit cell of the lattice showing the configuration.
Although it is possible to identify the vortex as opposed to the
antivortex from the arrows alone, we have overlaid a color scheme
which is based on the following empirical expression
\beq
Q = - \frac{\Re(\p_{[x}\phi_2)\Im(\p_{y]}\phi_2)}{2\pi(|\phi_2|^2+\epsilon)}.
\eeq
Large positive values of $Q$ are plotted with red and large negative 
values are plotted with blue; green is zero and other colors are
interpolation values between red and blue.
The brackets around the spatial indices indicate that the indices
are antisymmetrized and finally, $\epsilon$ is an ad-hoc small number
that is regularizing the quantity at the vortex cores.
$Q$ is used solely for the intent of clarifying which vortices are
vortices and which are antivortices. 

We will now explain the color scheme utilized for coloring in the
baryon charge isosurface that shows the Skyrmion configuration in
Fig.~\ref{fig:24B1}(a).
The color scheme is based on the parametrization
\eqref{eq:phi_parametrization} as
\beq
\left\{
\begin{array}{ll}
  0\leq \vartheta\leq \frac{\pi}{10}, &\qquad {\rm white}\\
  \frac{\pi}{10}<\vartheta<\frac{9\pi}{10}, &\qquad {\rm color}\\
  \frac{9\pi}{10}\leq \vartheta\leq \frac{11\pi}{10}, &\qquad {\rm black}\\
  \frac{11\pi}{10}<\vartheta<\frac{19\pi}{10}, &\qquad {\rm color}\\
  \frac{19\pi}{10}\leq \vartheta\leq 2\pi, &\qquad {\rm white}
\end{array}
\right.
\eeq
The color is defined as a map from $\chi$ to the color circle (the
hue), such that $\chi=0$ is red, $\chi=2\pi/3$ is green, $\chi=4\pi/3$
is blue, $\chi=\pi/3$ is yellow, $\chi=\pi$ is cyan and $\chi=5\pi/3$
is magenta.

There is a surprise, that is not obvious from the construction we
described above; it turns out that there is a dual string, meaning a
string in the $\phi_1$ field in addition to the string in the $\phi_2$
field that we pictured so far.
The nature of the Skyrmion, covering the entire 3-sphere (target
space), implies that the closed string in the ($\otimes$) bulk will
have a dual string (of $\phi_1$) piercing through the Skyrmion.
When the Skyrmion is attached to the domain wall, the vortex ring (of
$\phi_2$) becomes a handle, but the dual string piercing the Skyrmion
becomes a dual handle.
Therefore, the 1-Skyrmion is actually symmetric between the two
phases. Had we described everything in terms of strings in the $\odot$
phase, the vortex would be a string in the $\phi_1$ field and it also
becomes a handle attached to the domain wall, albeit from the other
side, see Fig.~\ref{fig:24B1}(b).
The vortex zero in $\phi_2$ is depicted by a blue isosurface and the
vortex zero in $\phi_1$ is shown with a magenta isosurface. 
The two vortices (blue and magenta) link each other once (if we
include their respective vacua).

Fig.~\ref{fig:26B1} shows the same 1-Skyrmion configuration as in
Fig.~\ref{fig:24B1}, but in the 2+6 model instead of in the 2+4 model.
The 2+4 model is given by the Lagrangian \eqref{eq:Lag} with $c_4=1$,
$c_6=0$ and the 2+6 model has $c_4=0$, $c_6=1$.

\subsection{Interactions between wall and a closed vortex string}\label{sec:ring}

In this section, we will start from a 1-Skyrmion that can stably exist
in the bulk \cite{Gudnason:2014jga} and put it near the domain wall.
The 1-Skyrmion in the $\otimes$ bulk exists as a vortex ring (closed
vortex string) in the $\phi_2$ field (blue).
If we were to place it in the $\odot$ phase instead, the 1-Skyrmion
would be a vortex ring in the $\phi_1$ field (magenta). 
If the 1-Skyrmion is placed far way from the domain wall, the force
between them is exponentially suppressed and it will take a long time
before an attraction will accelerate the 1-Skyrmion towards the domain
wall.
Therefore, we will place the 1-Skyrmion in the bulk in near proximity
to the domain wall and the attraction happens quite rapidly.

\begin{figure}[!htp]
  \begin{center}
    \mbox{\subfloat{\includegraphics[width=0.245\linewidth]{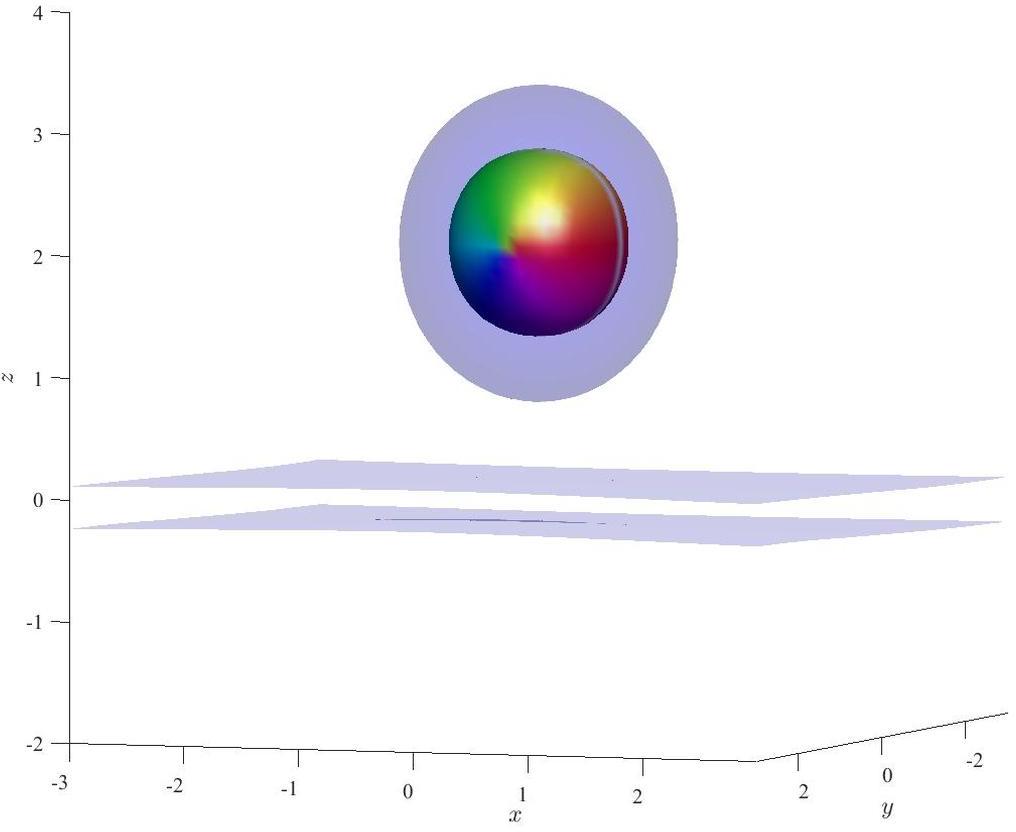}}
      \subfloat{\includegraphics[width=0.245\linewidth]{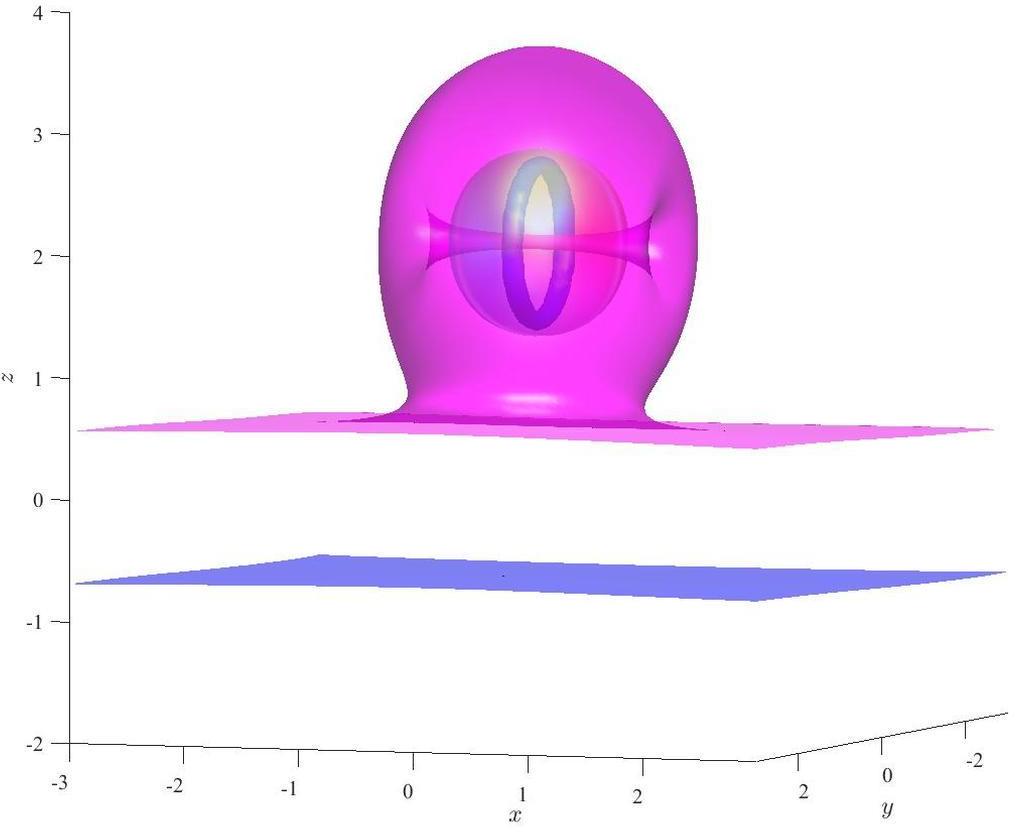}}
      \subfloat{\includegraphics[width=0.245\linewidth]{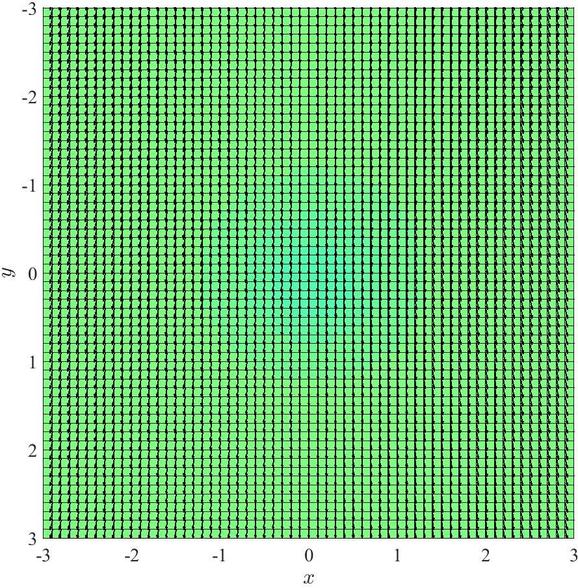}}
      \subfloat{\includegraphics[width=0.245\linewidth]{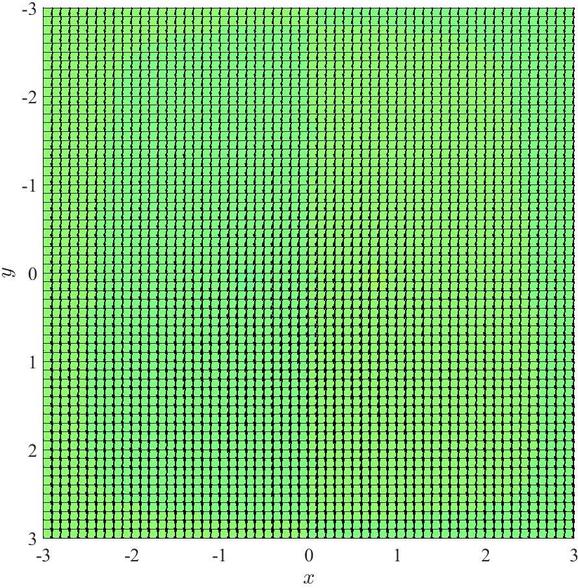}}}
    \mbox{\subfloat{\includegraphics[width=0.245\linewidth]{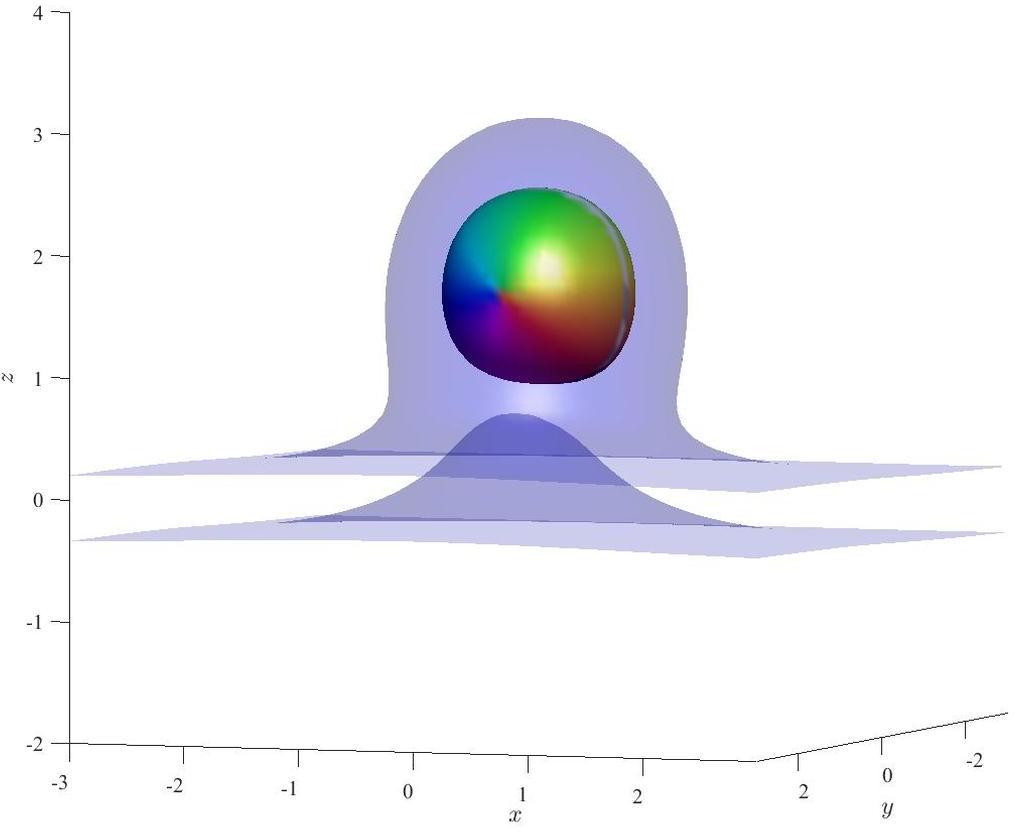}}
      \subfloat{\includegraphics[width=0.245\linewidth]{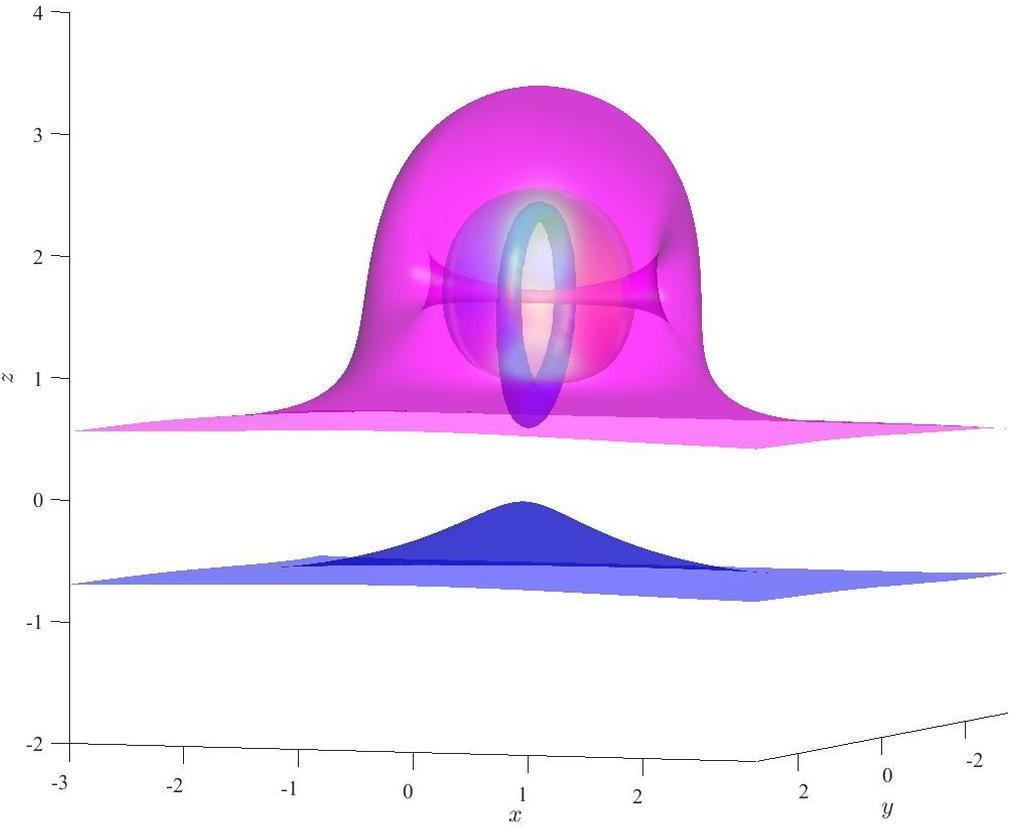}}
      \subfloat{\includegraphics[width=0.245\linewidth]{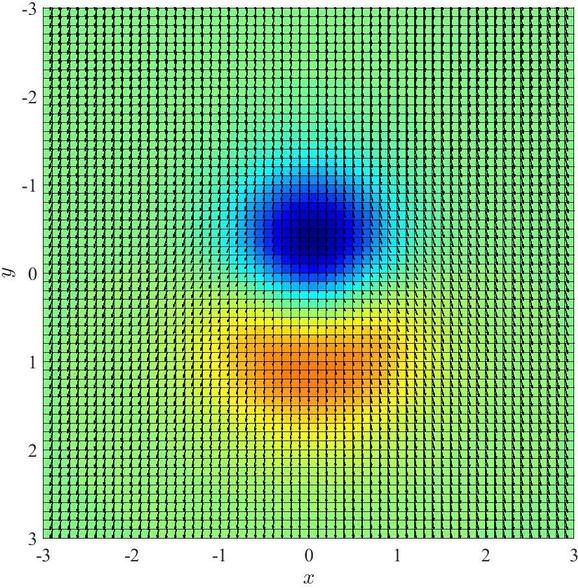}}
      \subfloat{\includegraphics[width=0.245\linewidth]{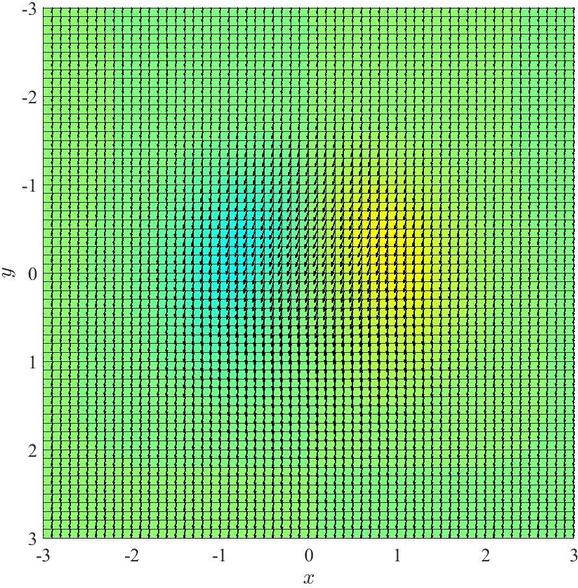}}}
    \mbox{\subfloat{\includegraphics[width=0.245\linewidth]{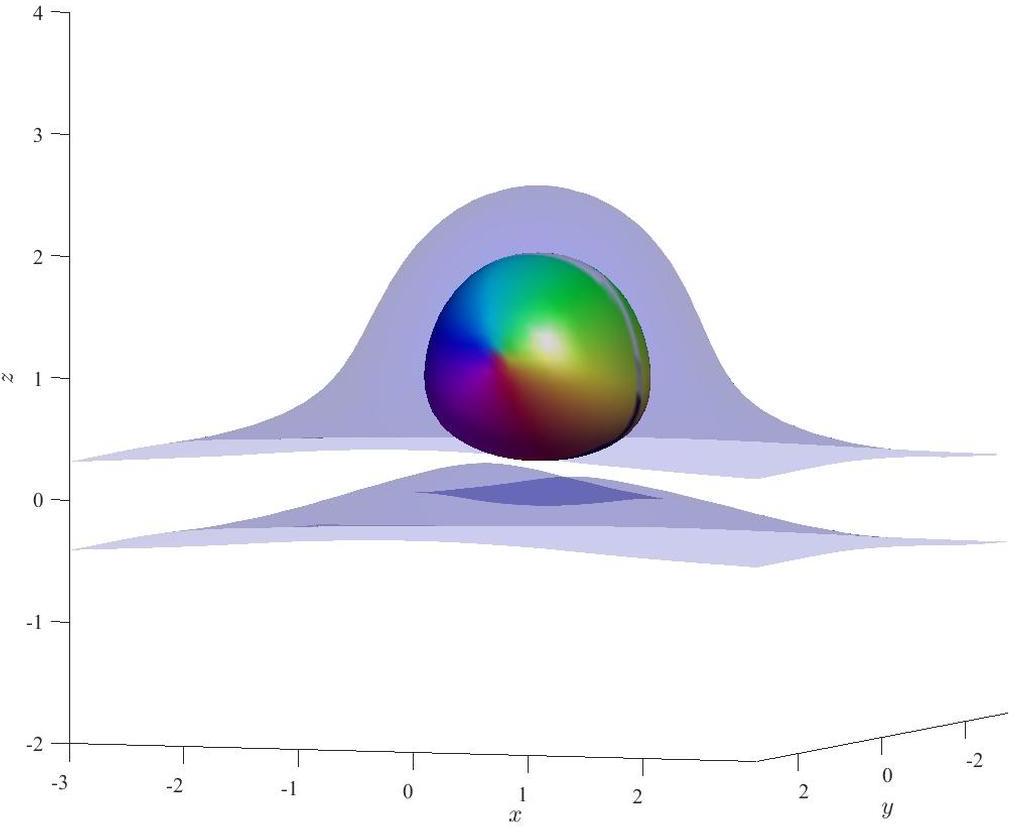}}
      \subfloat{\includegraphics[width=0.245\linewidth]{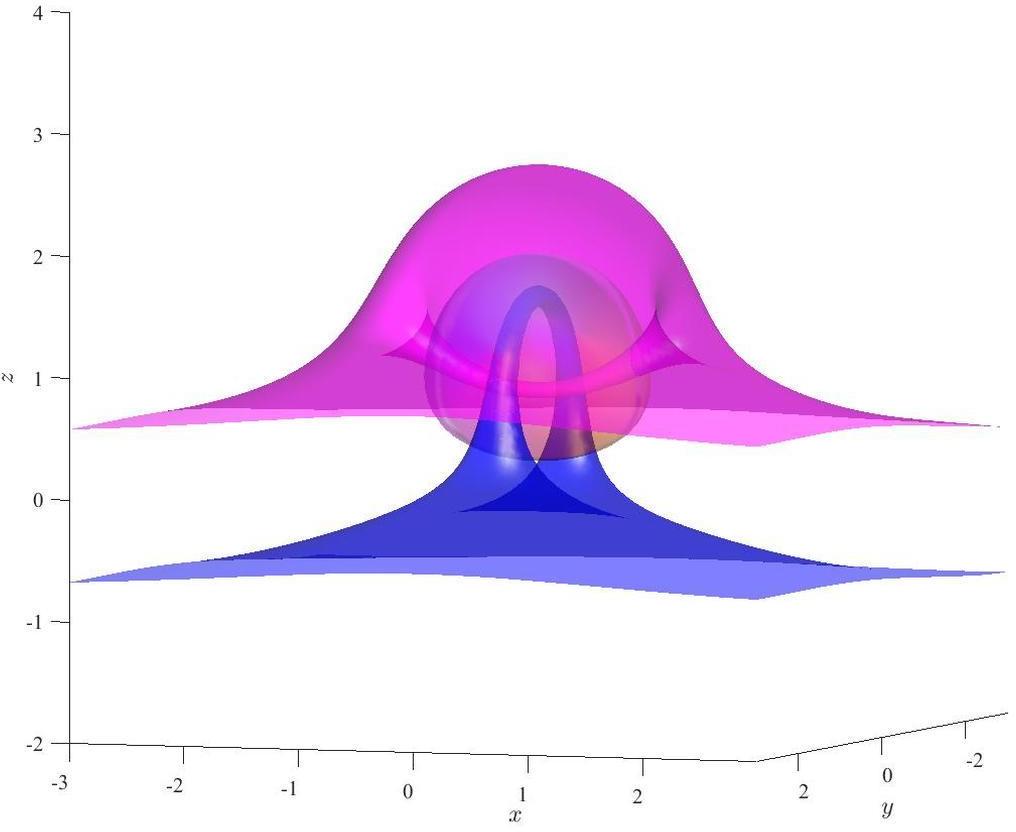}}
      \subfloat{\includegraphics[width=0.245\linewidth]{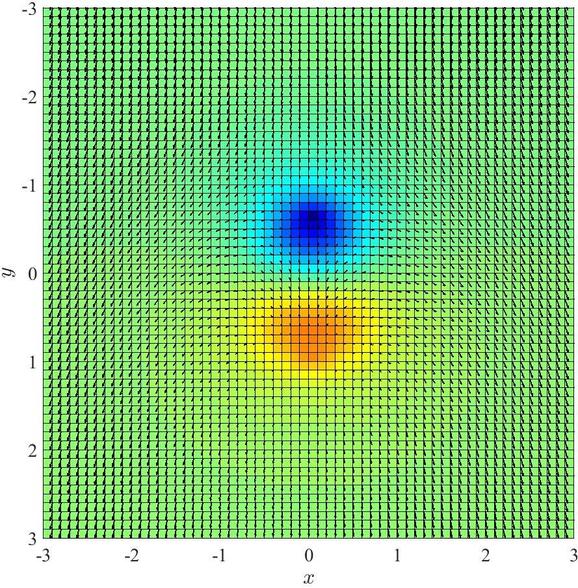}}
      \subfloat{\includegraphics[width=0.245\linewidth]{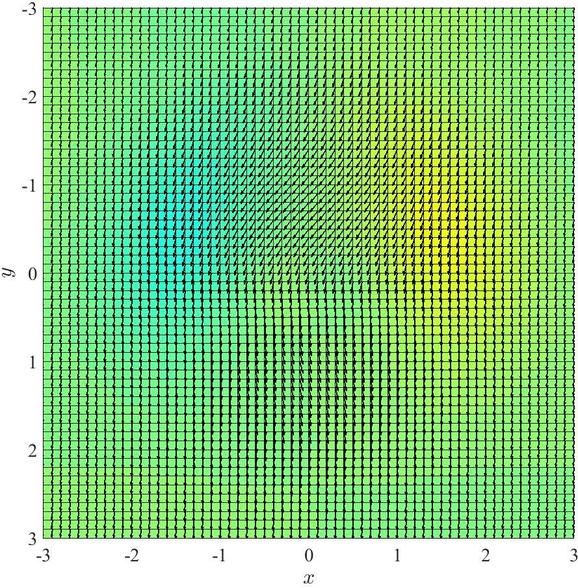}}}
    \mbox{\subfloat{\includegraphics[width=0.245\linewidth]{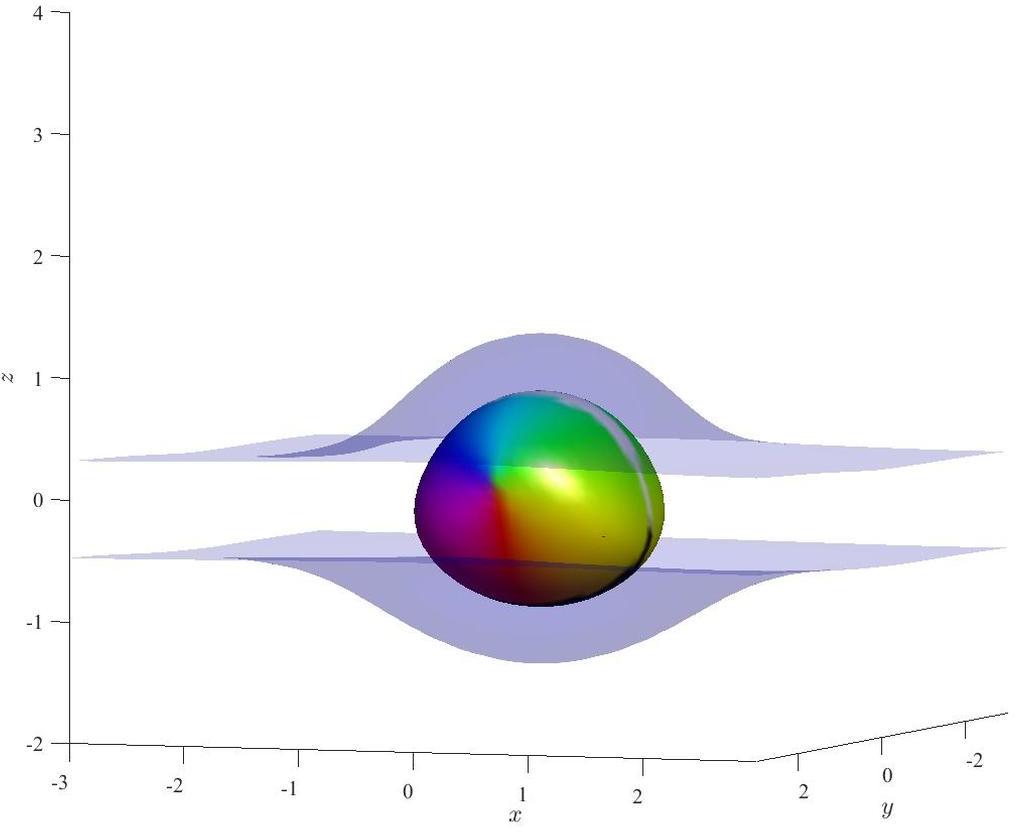}}
      \subfloat{\includegraphics[width=0.245\linewidth]{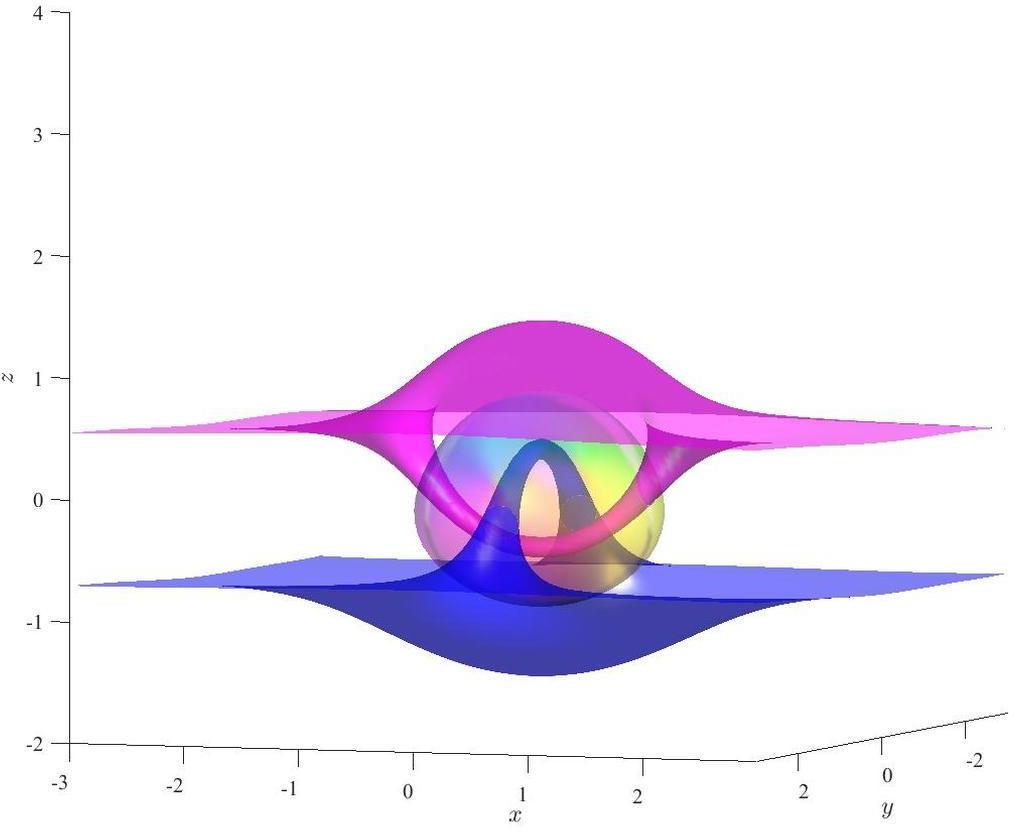}}
      \subfloat{\includegraphics[width=0.245\linewidth]{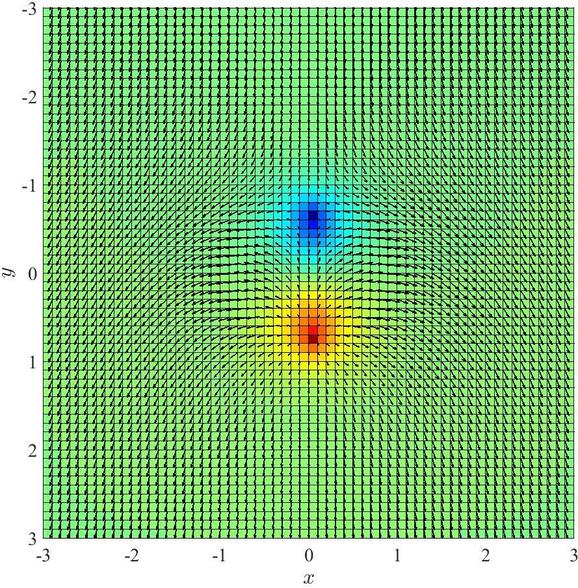}}
      \subfloat{\includegraphics[width=0.245\linewidth]{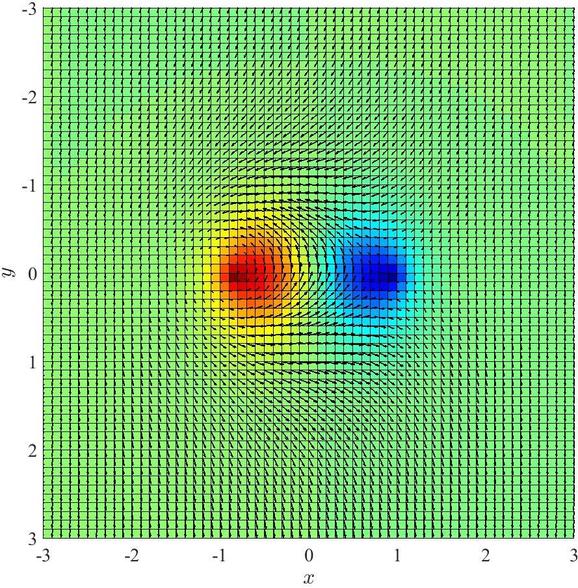}}}
    \caption{Skyrmion as a bulk vortex ring in the $\otimes$ phase
      interacting with the domain wall in the 2+4 model.
      The four columns display (1) the energy isosurfaces and baryon
      charge isosurface; (2) the zeros of $\phi_2$ (blue) and $\phi_1$
      (magenta); (3) the vortex (red) and antivortex (blue) pair of
      $\phi_2$; (4) the vortex (red) and antivortex (blue) pair of
      $\phi_1$. The rows correspond to (imaginary) relaxation time of
      the simulation. The rows are not equidistant in relaxation
      time. In this figure, we have taken $M=3$. }
    \label{fig:ring3c4}
  \end{center}
\end{figure}

The simulation is made not with a relativistic kinetic term, but with
the relaxation method, which is dissipative. If the dynamics was made
with a real relativistic kinetic term, the interaction would be
oscillating many times and only come to a final fixed point when all
the excess energy has been radiated away.
Instead we will evolve the dynamics of the simulation with a
first-order kinetic term, which is dissipative as mentioned
already. This means that the energy is not conserved and the
configuration will quickly approach the fixed point losing the excess
energy to the dissipative term in the evolution.

The numerical calculation is shown in Fig.~\ref{fig:ring3c4}.
We have placed the 1-Skyrmion near to the domain wall in the $\otimes$
phase with an orientation such that the plane of the vortex ring is
perpendicular to the domain wall. This happens to be the optimal
orientation for attraction between the 1-Skyrmion and the domain
wall.
The four columns in Fig.~\ref{fig:ring3c4} depict the energy/baryon
charge, the vacua, the (anti)vortices of $\phi_2$ on the domain wall
and the (anti)vortices of $\phi_1$ on the domain wall, respectively.
The rows are snapshots in imaginary time corresponding to the
evolution of the relaxation time.
The attraction between the 1-Skyrmion and the domain wall happens
quickly and the first thing that happens is that the vortex ring in
$\phi_2$ (blue) is attracted and partially absorbed into the vacuum on
the other side of the domain wall (the $\odot$ phase), see rows 2 and
3. This is the creation of the vortex handle, and because it is only
partially absorbed, the remaining part is a handle of the $\phi_2$
vortex sticking out into the $\otimes$ bulk (positive-$z$ direction).

From this simulation, we can also see the origin of the dual handle,
i.e.~the vortex handle that exists in the $\phi_1$ field. In the
$\otimes$ bulk it was just a string that pierced through the
1-Skyrmion making it into a torus as claimed in
Ref.~\cite{Gudnason:2014jga}.
This piercing string connects the vacua on both sides of the torus and
energy-wise we find it more intuitive to think about the 1-Skyrmion as
being a wrapped-up vortex ring in the $\phi_2$ field.
However, once the vortex ring in the $\phi_2$ field is absorbed into
the wall and becomes a vortex handle, the dual string (in the $\phi_1$
field) is drawn down into the domain wall and becomes a dual handle.
In fact the configuration is symmetric if we flip the $z\to-z$ and
make a 90-degree rotation with a flip as $x\to y$ and $y\to x$. The
second flip is necessary for keeping the baryon charge positive
(otherwise it would become an anti-Skyrmion).

Now that we have understood the 1-Skyrmion absorbed into the domain
wall from two different perspectives, let us consider the 2-Skyrmions
in the next section.

\subsection{The handle with double twist}\label{sec:doublehandle}

In this section, we will take the approach of adding a twist as
described briefly in section \ref{sec:openstringloop}. The string
in $\phi_2$ that emanates from the domain wall into the $\otimes$
phase is twisted in the $\chi$ field from 0 to $2\pi$ and on the way
back to the domain wall it further twists to $4\pi$, yielding a total
twist of $4\pi=2\pi B$; hence $B=2$.
As mentioned in section \ref{sec:openstringloop}, there is unavoidably
a dual string in $\phi_1$ in the Skyrmion and we will see that it is
related to the number of twists. More precisely, once the twist is
higher than one, the dual string is either multiple-wound or there are
several dual strings. 

\begin{figure}[!htp]
  \begin{center}
    \mbox{\sidesubfloat[]{\includegraphics[width=0.65\linewidth]{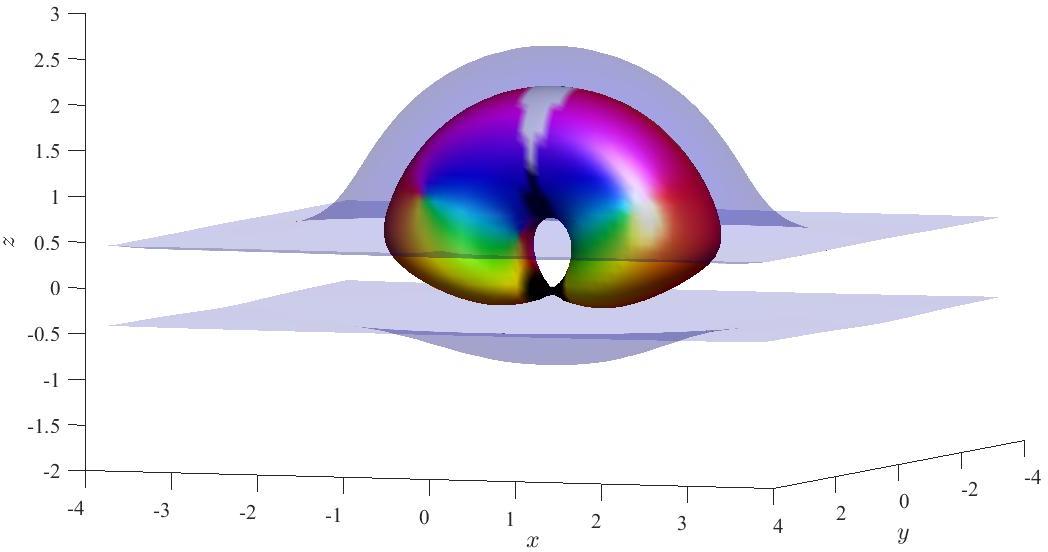}}}
    \mbox{\sidesubfloat[]{\includegraphics[width=0.65\linewidth]{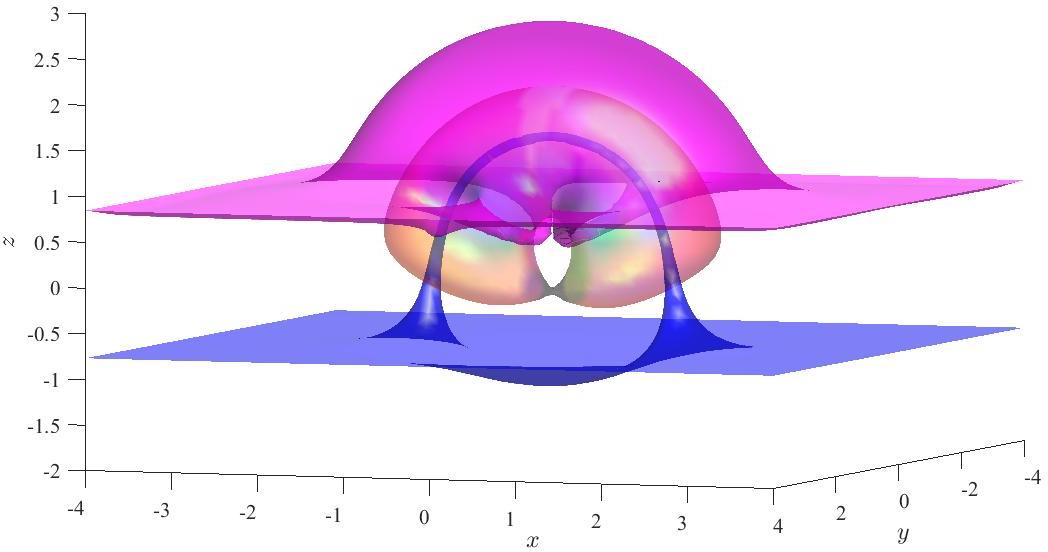}}}
    \mbox{\sidesubfloat[]{\includegraphics[width=0.4\linewidth]{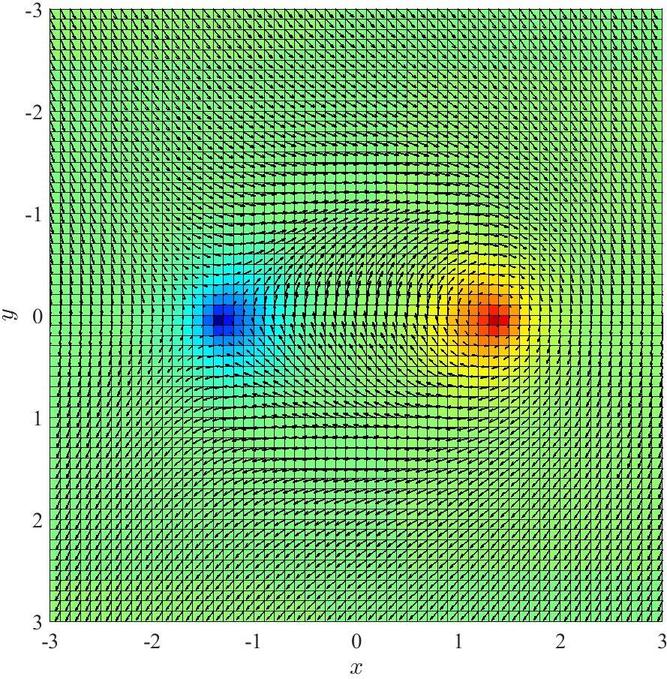}}\ \ 
      \sidesubfloat[]{\includegraphics[width=0.4\linewidth]{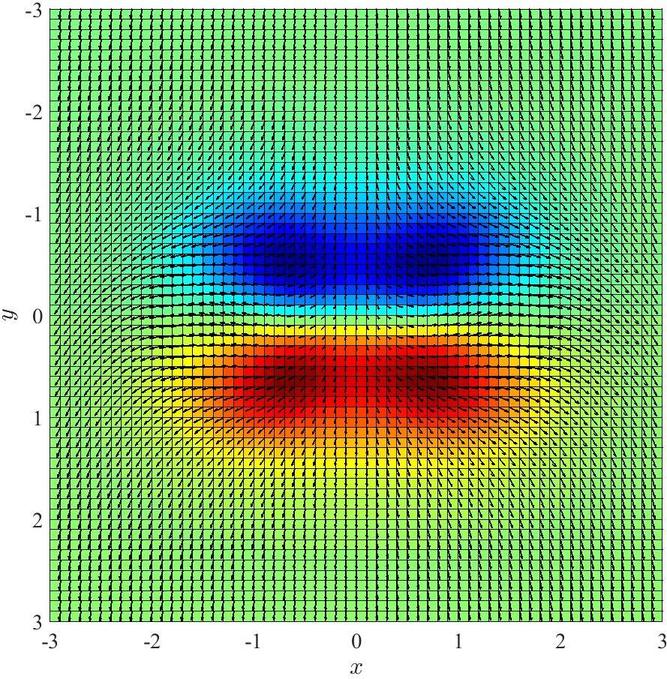}}}
    \caption{2-Skyrmion as a vortex handle with double twist on the
      domain wall in the 2+4 model. (a) the energy density is shown
      with blue transparent isosurfaces illustrating the domain wall
      and the baryon charge density is shown with an isosurface with
      the color scheme described in the text. (b) the blue isosurface
      (bottom) represents the zeros of $\phi_2$ and the magenta
      isosurface (top) is the zeros of $\phi_1$. The baryon charge is
      added transparently. (c) the vortex (red) and antivortex (blue)
      pair of $\phi_2$. (d) the vortex (red) and antivortex (blue)
      pairs of $\phi_1$. In this figure, we have taken $M=3$. }
    \label{fig:24B2}
  \end{center}
\end{figure}

\begin{figure}[!htp]
  \begin{center}
    \mbox{\sidesubfloat[]{\includegraphics[width=0.65\linewidth]{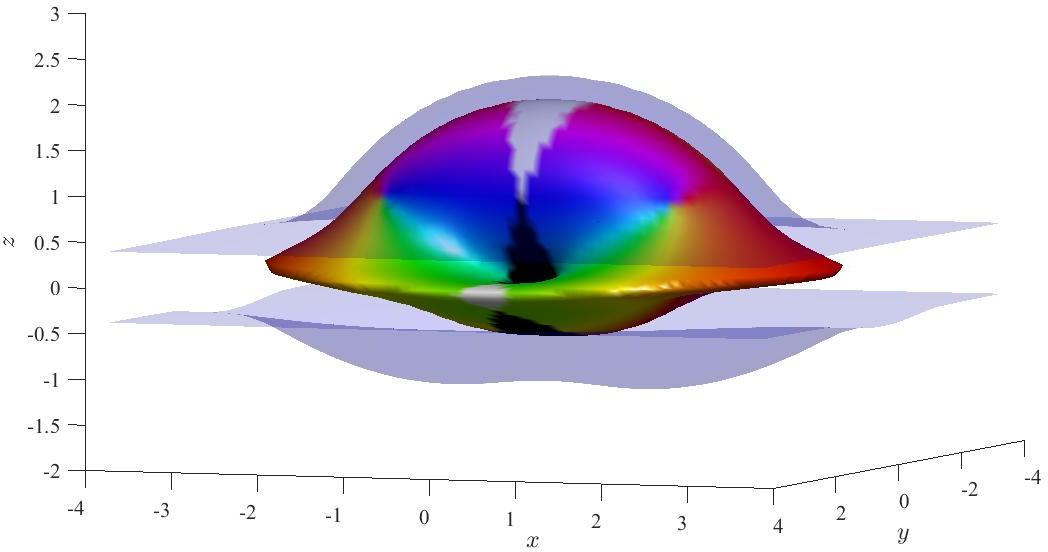}}}
    \mbox{\sidesubfloat[]{\includegraphics[width=0.65\linewidth]{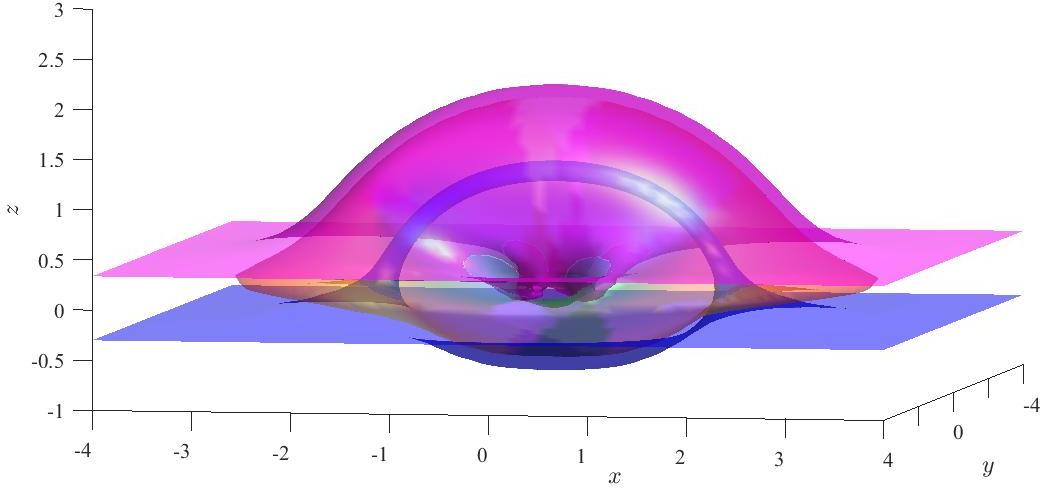}}}
    \mbox{\sidesubfloat[]{\includegraphics[width=0.4\linewidth]{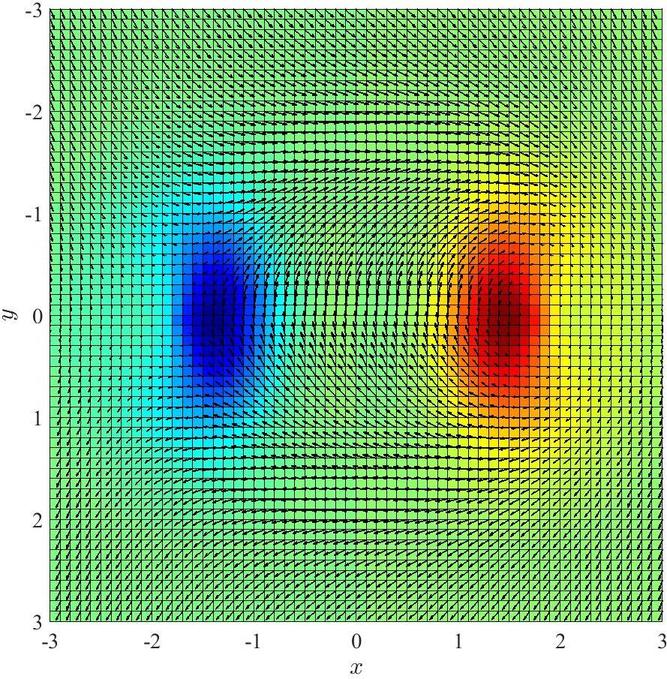}}\ \ 
      \sidesubfloat[]{\includegraphics[width=0.4\linewidth]{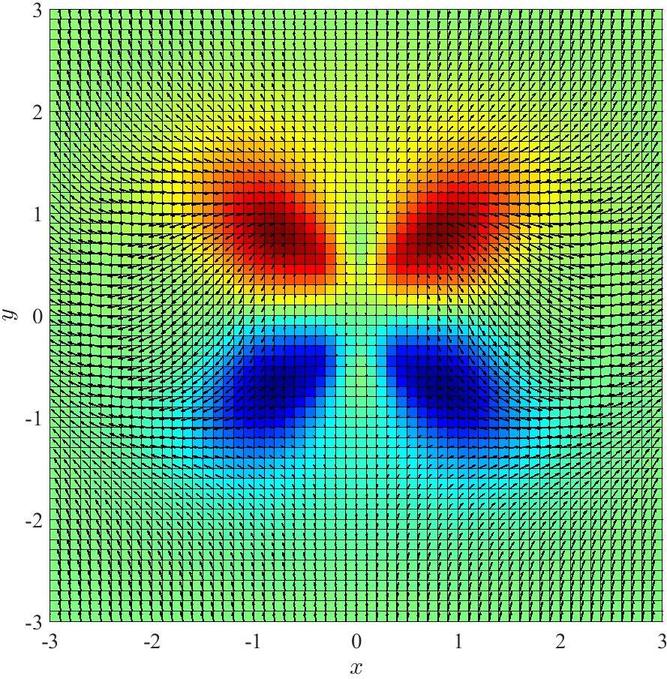}}}
    \caption{2-Skyrmion as a vortex handle with double twist on the
      domain wall in the 2+6 model. (a) the energy density is shown
      with blue transparent isosurfaces illustrating the domain wall
      and the baryon charge density is shown with an isosurface with
      the color scheme described in the text. (b) the blue isosurface
      (bottom) represents the zeros of $\phi_2$ and the magenta
      isosurface (top) is the zeros of $\phi_1$. The baryon charge is
      added transparently. (c) the vortex (red) and antivortex (blue)
      pair of $\phi_2$. (d) the vortex (red) and antivortex (blue)
      pairs of $\phi_1$. In this figure, we have taken $M=7$. }
    \label{fig:26B2}
  \end{center}
\end{figure}

The numerical calculations for the 2+4 model and the 2+6 model are
shown in Figs.~\ref{fig:24B2} and \ref{fig:26B2}, respectively.
All the subpanels are in the same format as presented in
Fig.~\ref{fig:24B1} of section \ref{sec:openstringloop}.
We can see that the handle, being the vortex string in $\phi_2$
(blue), is elongated and the dual string in $\phi_1$ (magenta) has
become two separate dual strings, instead of one.
The linking number between the blue and magenta strings (including the
vacuum) is two, see Figs.~\ref{fig:24B2}(b) and \ref{fig:26B2}(b).
It is clear that there are two vortex antivortex pairs in the
field $\phi_1$ in Figs.~\ref{fig:24B2}(d) and \ref{fig:26B2}(d).
We can also see that the doubly twisted handle is much bigger in the
2+6 model than in the 2+4 model with the normalization of the terms
used in the Lagrangian \eqref{eq:Lag}.
From the doubly twisted handle, it is clear why we would like to
associate the handle to the vortex in $\phi_2$ (blue); however, we
could just as well take the opposite point of view and say that it is
two separate handles in $\phi_1$ (magenta) which are closely bound
together because they are linked with the same vortex in
$\phi_2$ (blue).

\subsection{Interactions between two handles and the 
braided string junction terminating on a wall as a $B=2$ Skyrmion}\label{sec:int}

The way that we created a 2-Skyrmion on the domain wall (sticking a
bit out into the $\otimes$ bulk) in the previous section, was by
adding an extra twist to the 1-Skyrmion, topping the baryon number up
to two. This created an asymmetric configuration where the $\phi_2$
vortex (blue) is linked with two individual $\phi_1$ vortices
(magenta).

In this section we study the interactions of two individual
1-Skyrmions both already absorbed into the domain wall. Under normal
circumstances, this is also a way of creating multi-Skyrmions.
The recipe is to find the attractive channel between the two
1-Skyrmions and let them combine into the (probably) optimal
2-Skyrmion. 
It turns out that the theory is much more complex when we have a
domain wall, as we shall see shortly.

\begin{figure}[!htp]
  \begin{center}
    \mbox{\subfloat{\includegraphics[width=0.245\linewidth]{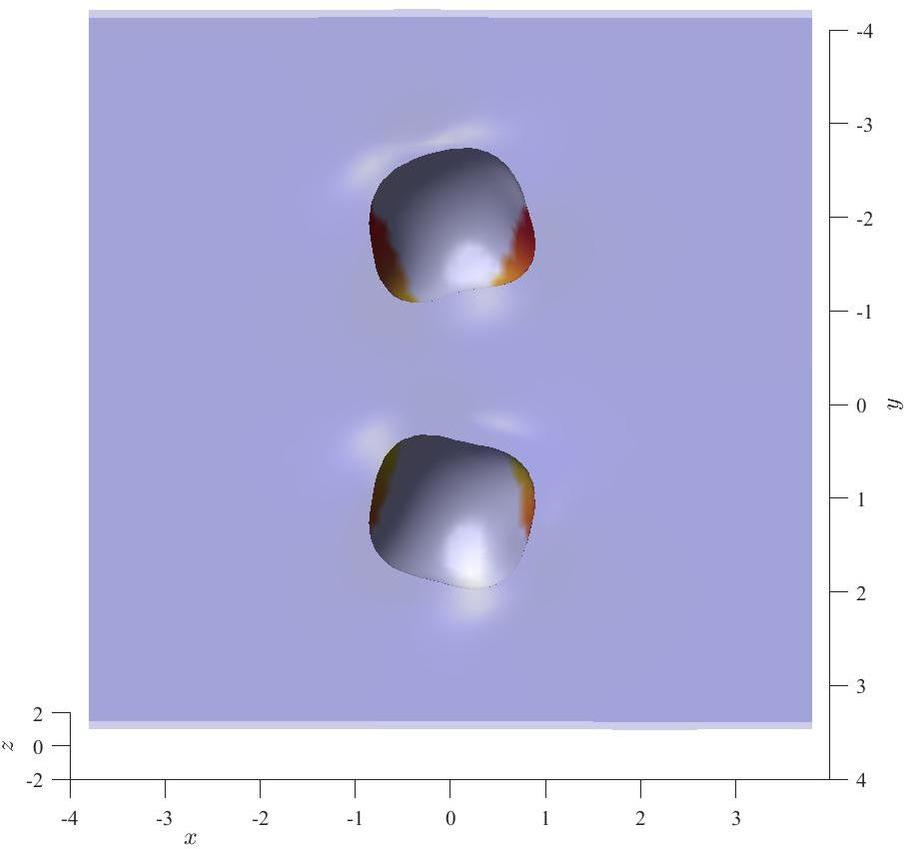}}
      \subfloat{\includegraphics[width=0.245\linewidth]{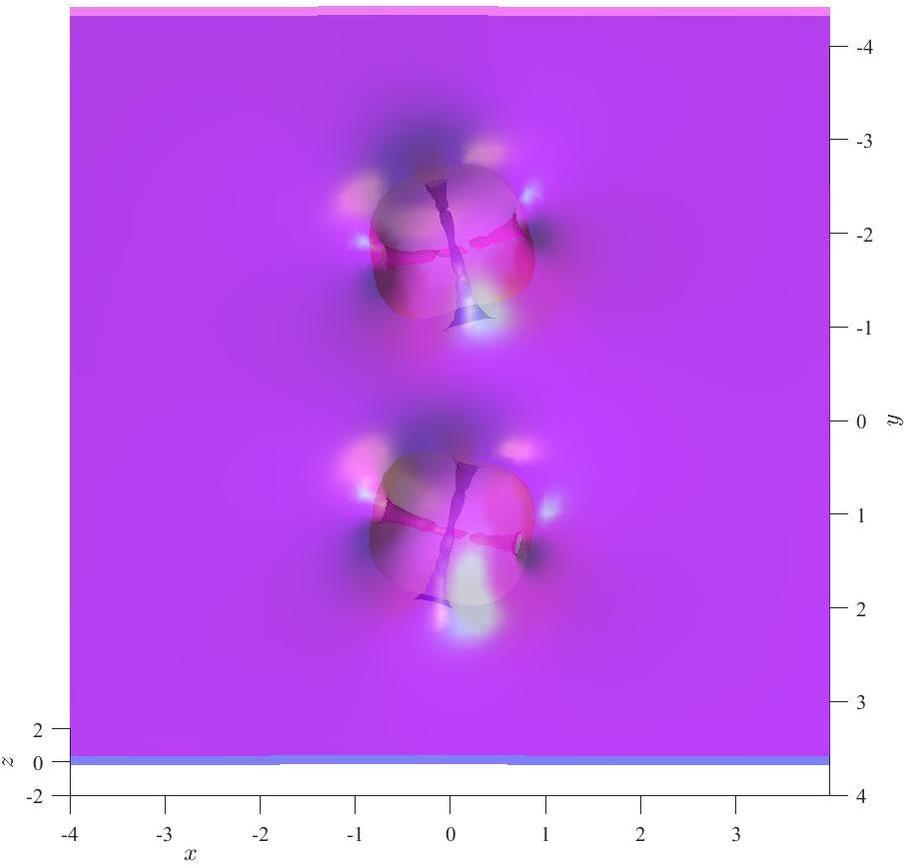}}
      \subfloat{\includegraphics[width=0.245\linewidth]{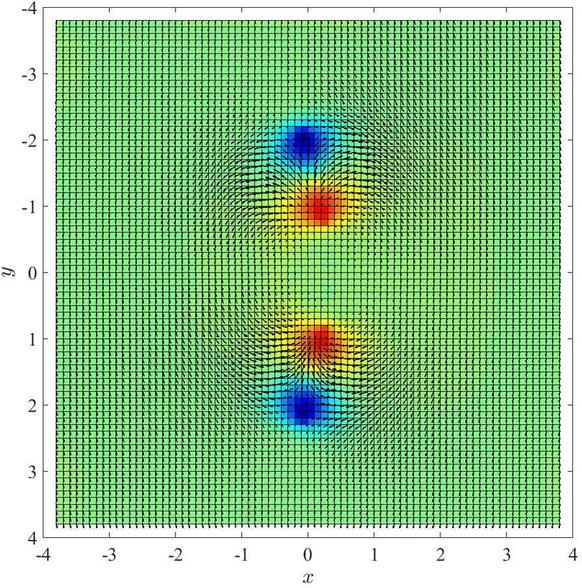}}
      \subfloat{\includegraphics[width=0.245\linewidth]{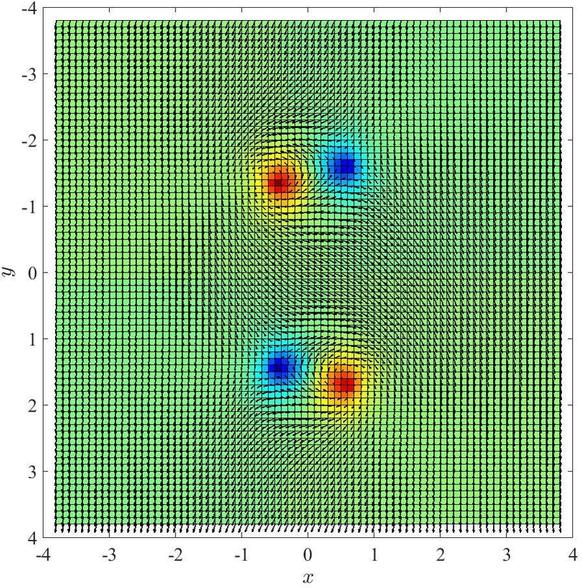}}}
    \mbox{\subfloat{\includegraphics[width=0.245\linewidth]{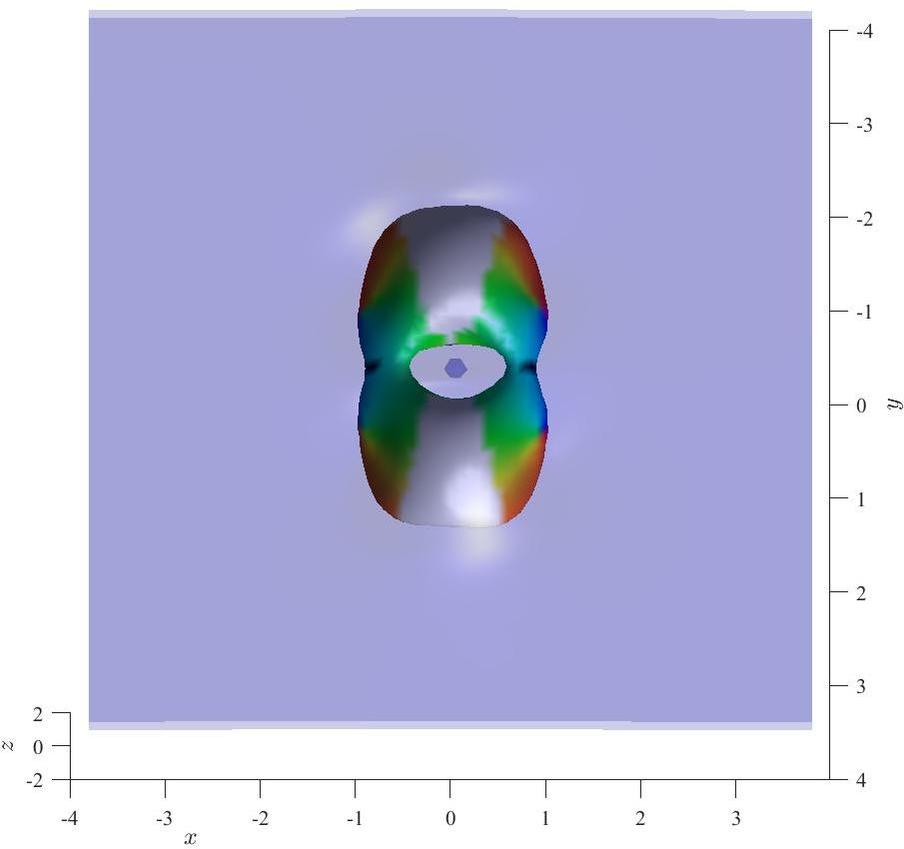}}
      \subfloat{\includegraphics[width=0.245\linewidth]{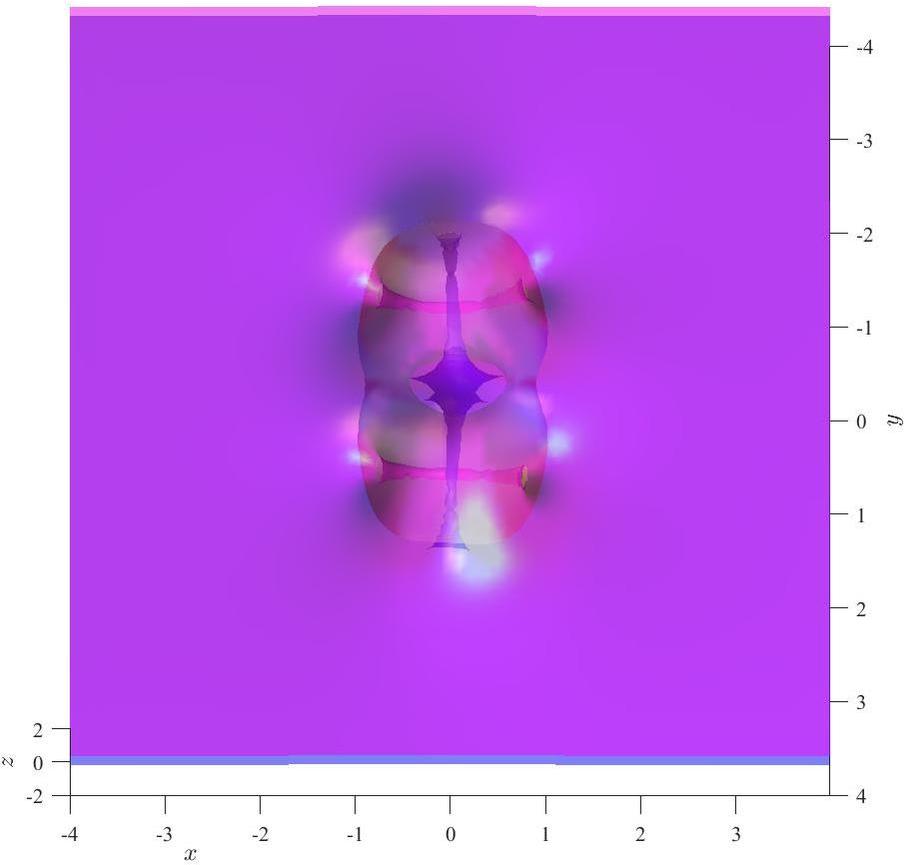}}
      \subfloat{\includegraphics[width=0.245\linewidth]{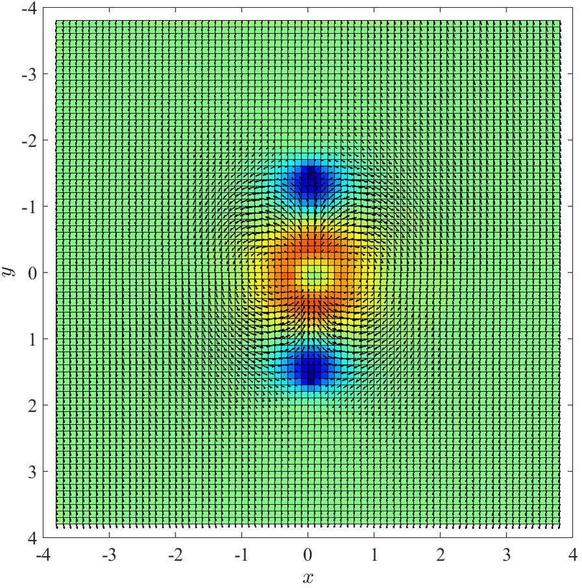}}
      \subfloat{\includegraphics[width=0.245\linewidth]{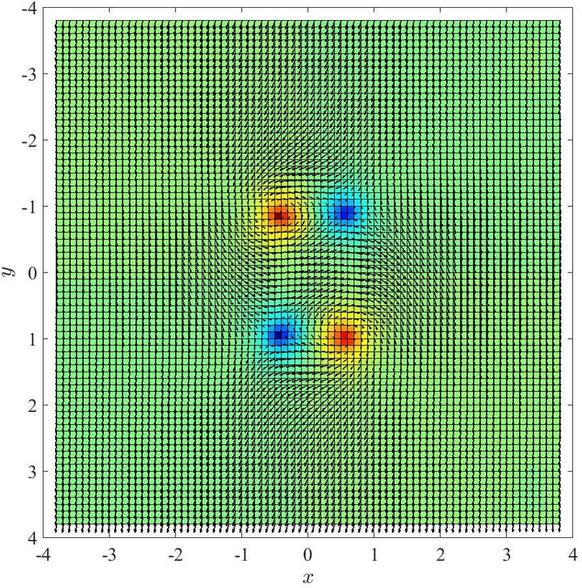}}}
    \mbox{\subfloat{\includegraphics[width=0.245\linewidth]{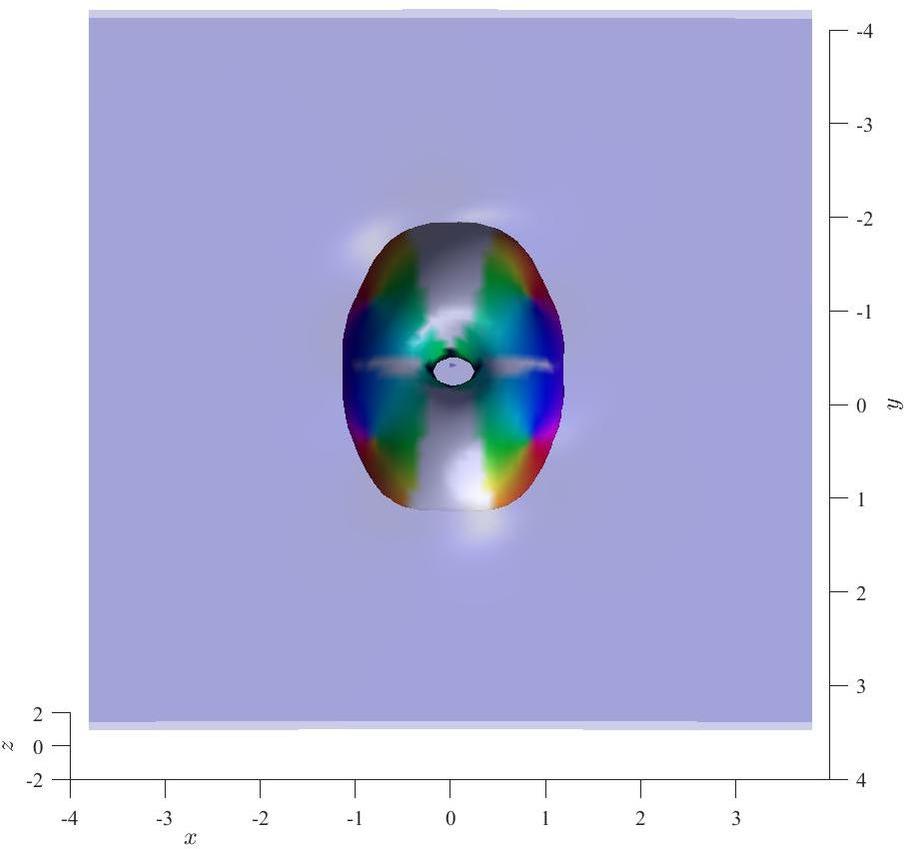}}
      \subfloat{\includegraphics[width=0.245\linewidth]{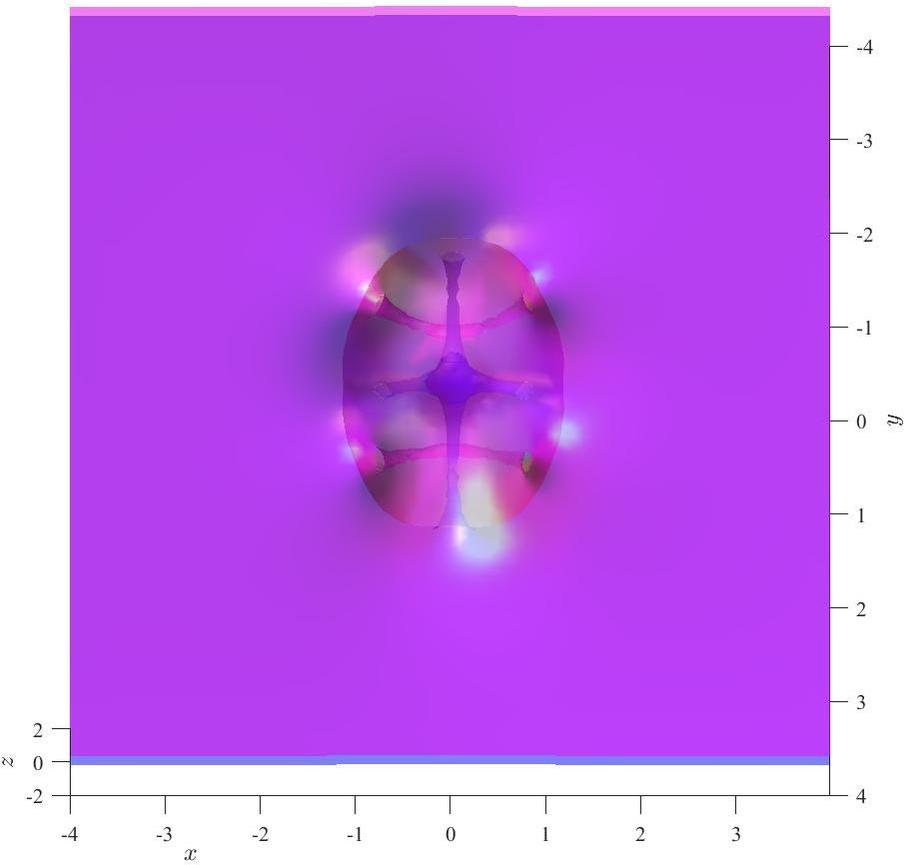}}
      \subfloat{\includegraphics[width=0.245\linewidth]{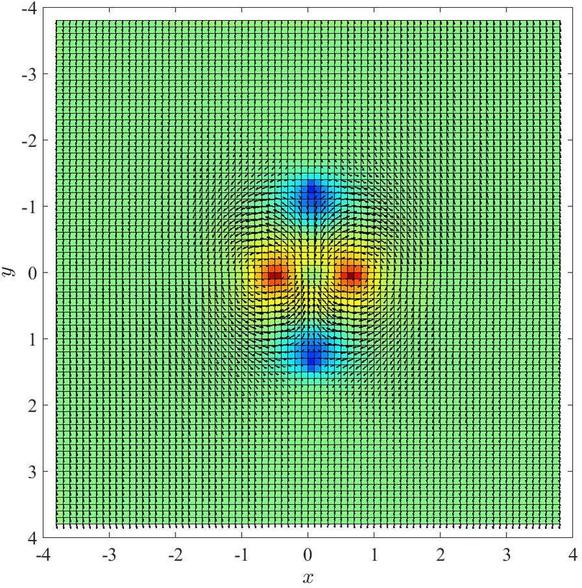}}
      \subfloat{\includegraphics[width=0.245\linewidth]{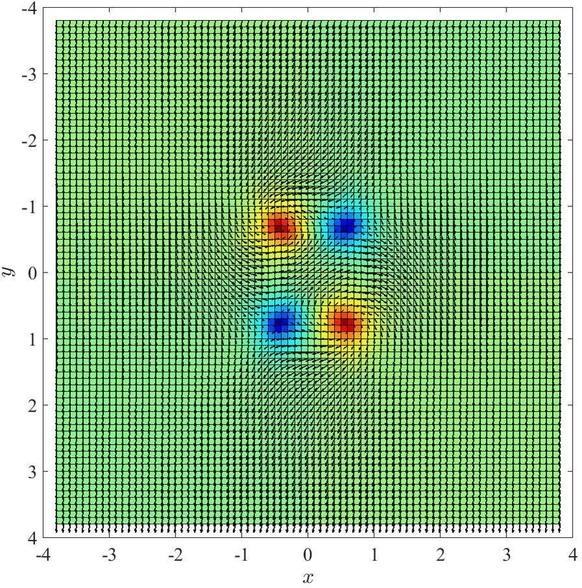}}}
    \mbox{\subfloat{\includegraphics[width=0.245\linewidth]{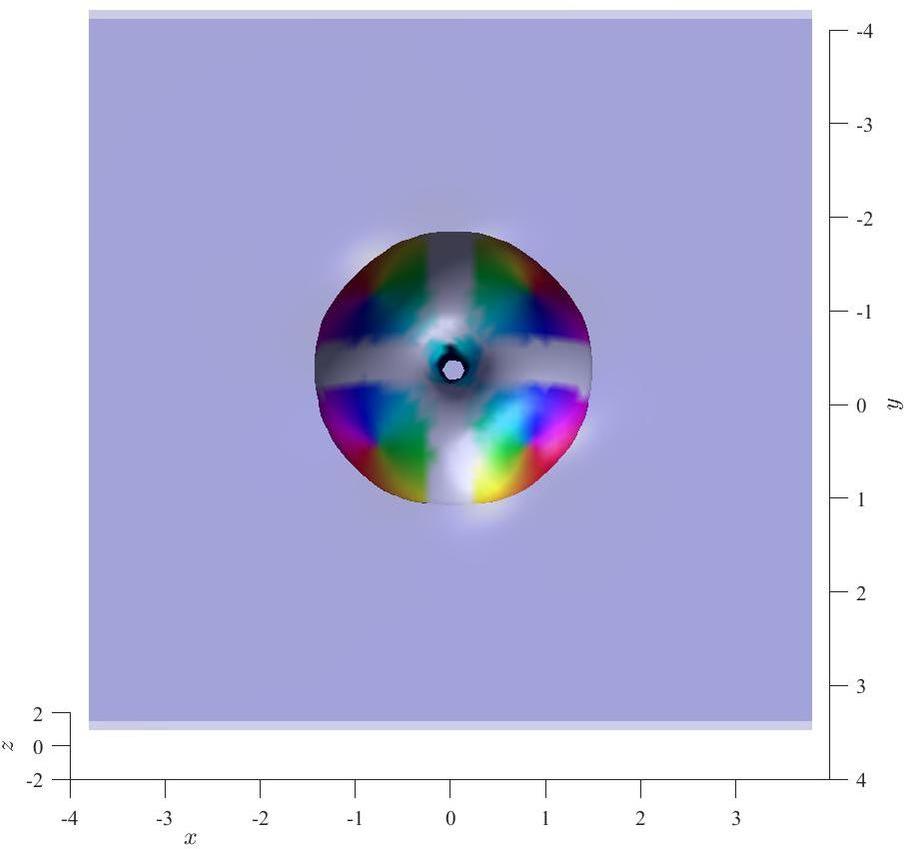}}
      \subfloat{\includegraphics[width=0.245\linewidth]{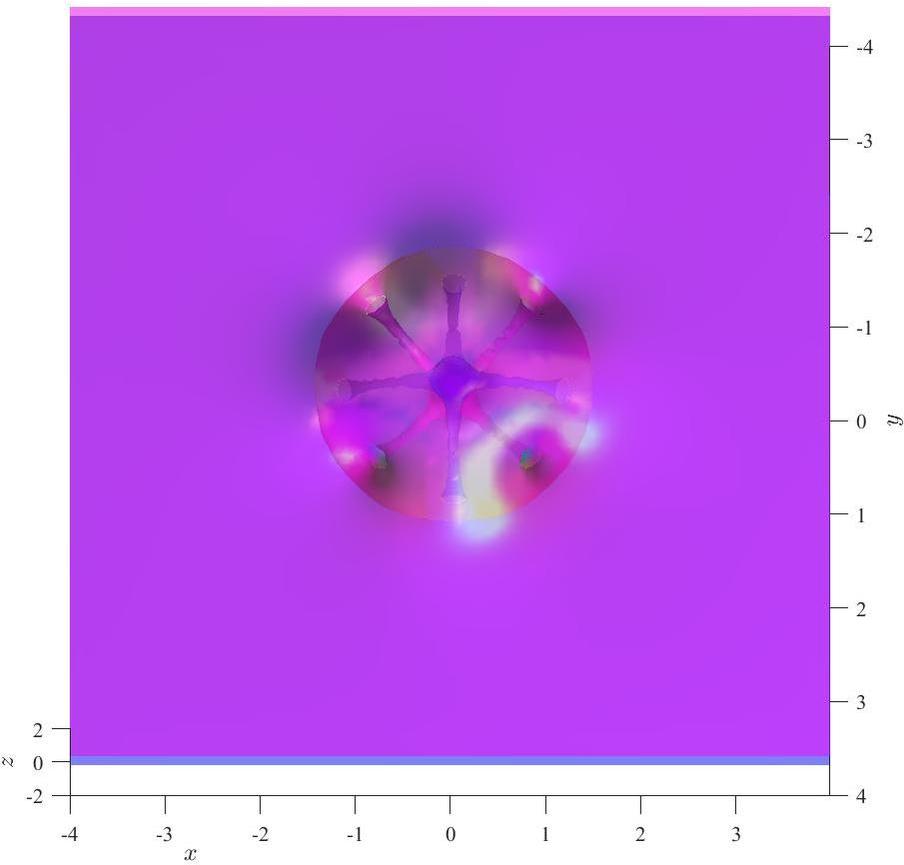}}
      \subfloat{\includegraphics[width=0.245\linewidth]{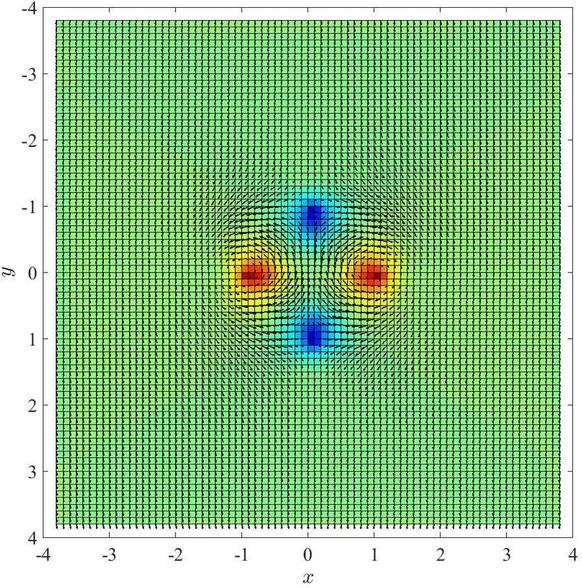}}
      \subfloat{\includegraphics[width=0.245\linewidth]{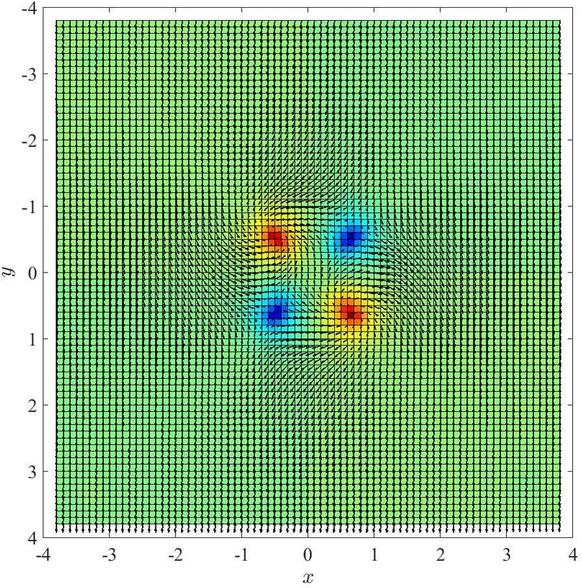}}}
    \caption{Skyrmion-Skyrmion interaction in the domain wall:
      a vortex handle interacts with another vortex handle in the
      attractive channel in the 2+4 model.
      The four columns display (1) the energy isosurfaces and baryon
      charge isosurface; (2) the zeros of $\phi_2$ (blue) and $\phi_1$
      (magenta); (3) the vortex (red) and antivortex (blue) pairs of
      $\phi_2$; (4) the vortex (red) and antivortex (blue) pairs of
      $\phi_1$. The rows correspond to (imaginary) relaxation time of
      the simulation. The rows are not equidistant in relaxation
      time. In this figure, we have taken $M=7$. }
    \label{fig:int7c4r}
  \end{center}
\end{figure}
    
\begin{figure}[!htp]
  \begin{center}
    \mbox{\subfloat{\includegraphics[width=0.49\linewidth]{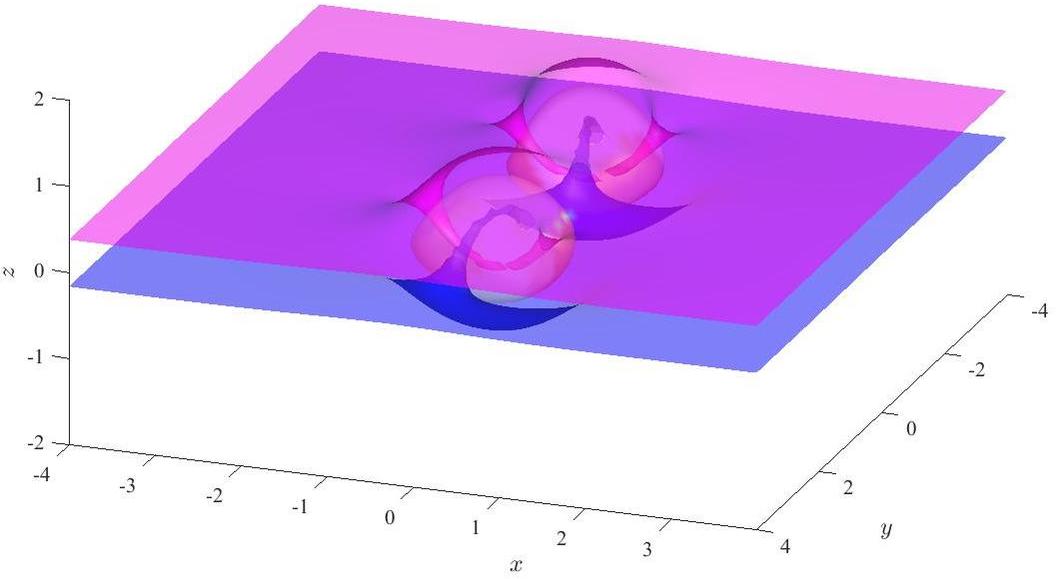}}
      \subfloat{\includegraphics[width=0.49\linewidth]{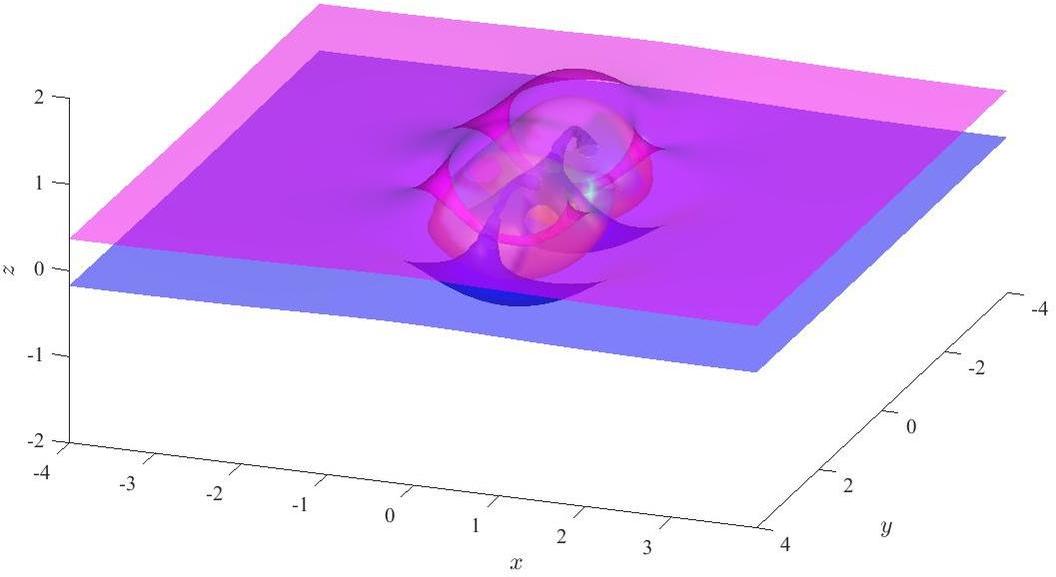}}}
    \mbox{\subfloat{\includegraphics[width=0.49\linewidth]{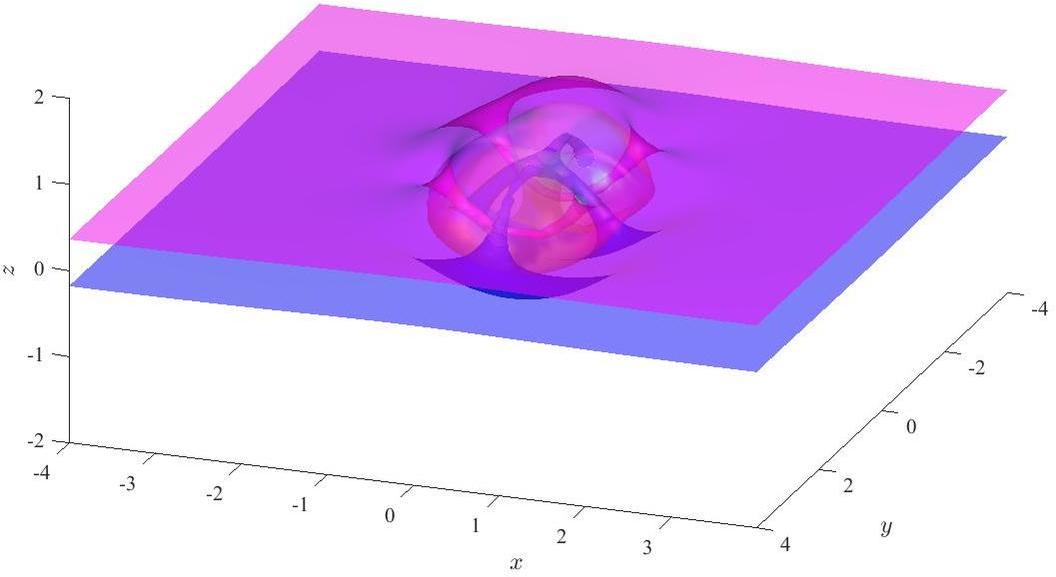}}
      \subfloat{\includegraphics[width=0.49\linewidth]{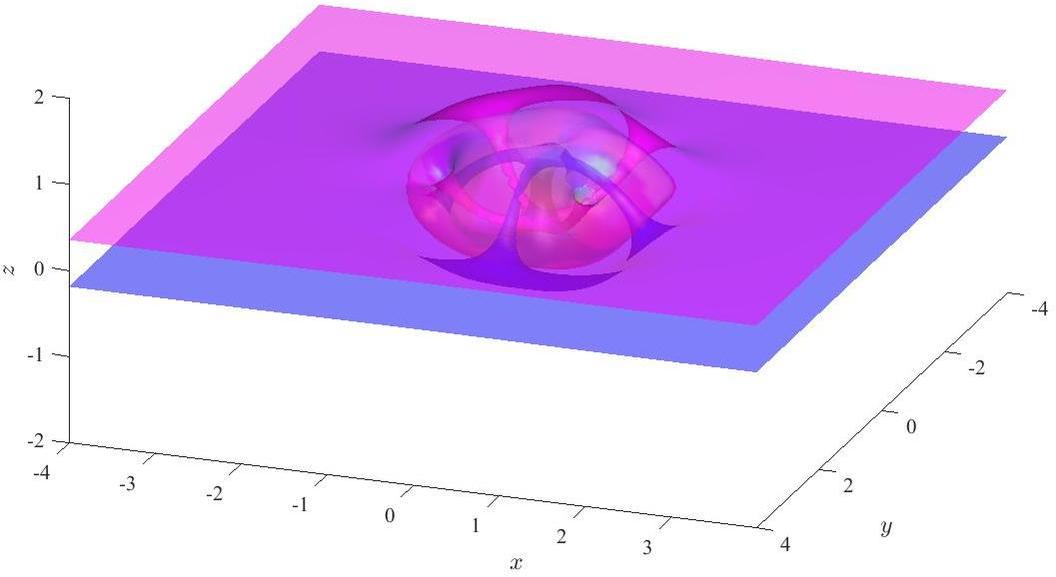}}}
    \caption{Same as the second column of Fig.~\ref{fig:int7c4r}, but
      with a titled view point. The time evolution of the simulation
      has the following order: upper-left, upper-right, lower-left,
      and lower-right.
    }
    \label{fig:int7c4rtilt}
  \end{center}
\end{figure}

\begin{figure}[!htp]
  \begin{center}
    \mbox{\subfloat{\includegraphics[width=0.245\linewidth]{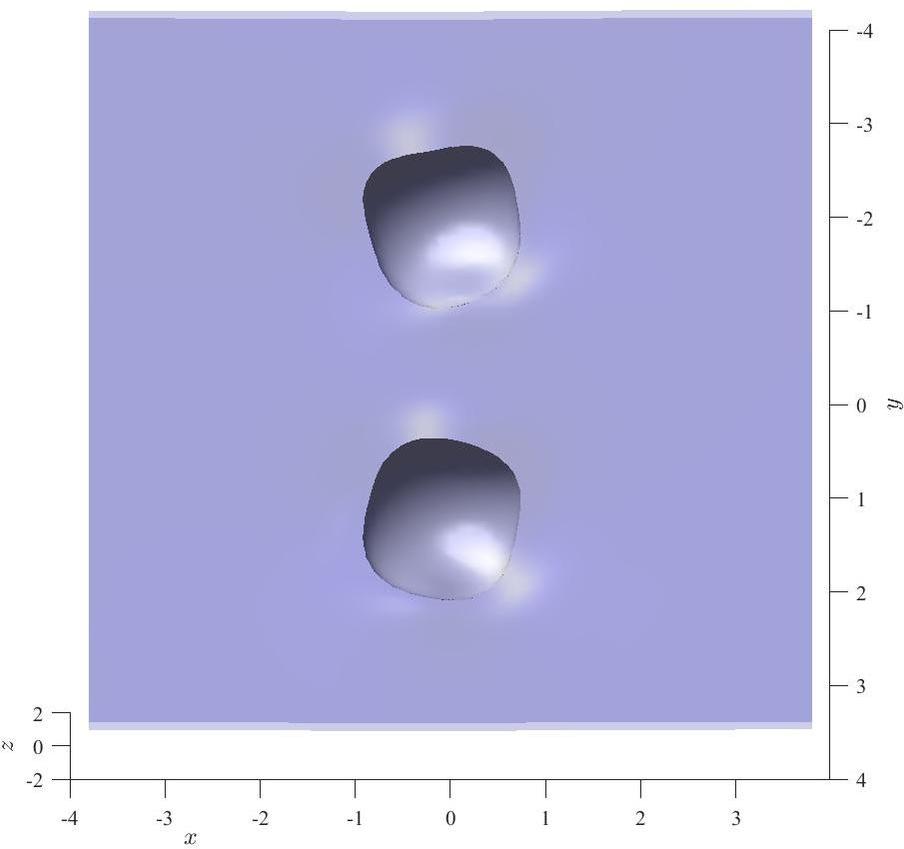}}
      \subfloat{\includegraphics[width=0.245\linewidth]{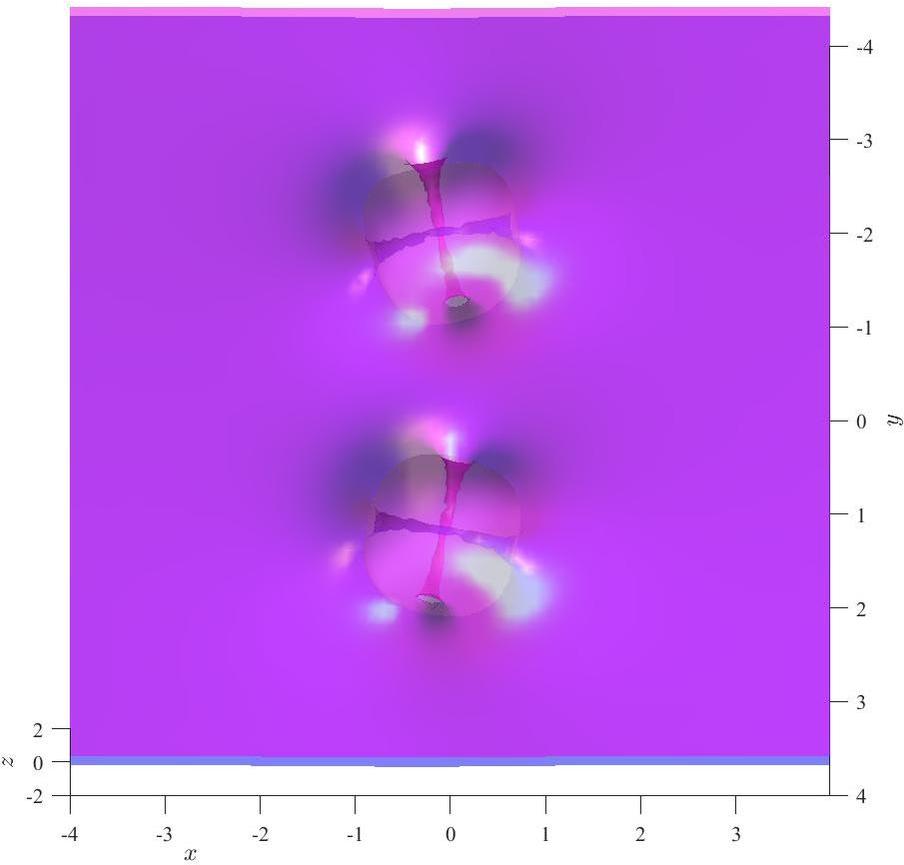}}
      \subfloat{\includegraphics[width=0.245\linewidth]{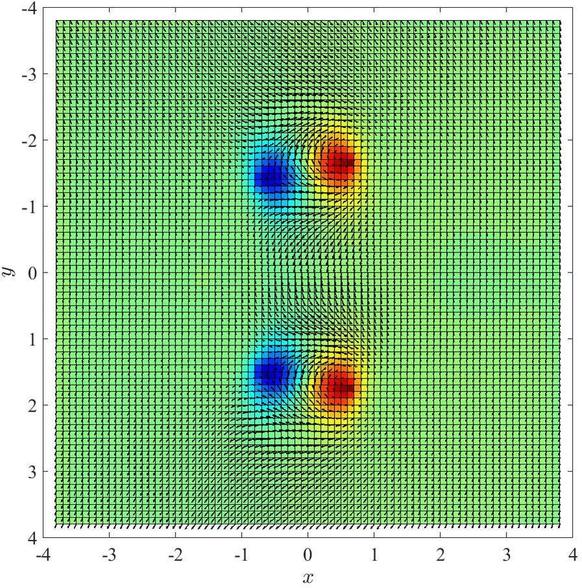}}
      \subfloat{\includegraphics[width=0.245\linewidth]{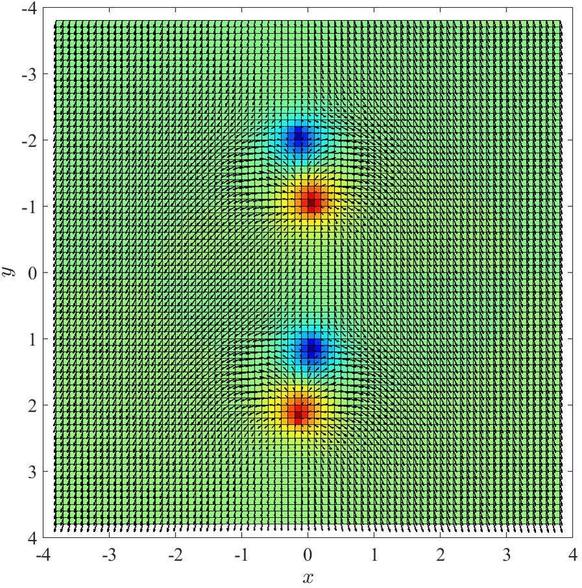}}}
    \caption{Skyrmion-Skyrmion interaction in the domain wall:
      a vortex handle interacts with another vortex handle in the
      repulsive channel in the 2+4 model.
      The four columns display (1) the energy isosurfaces and baryon
      charge isosurface; (2) the zeros of $\phi_2$ (blue) and $\phi_1$
      (magenta); (3) the vortex (red) and antivortex (blue) pairs of
      $\phi_2$; (4) the vortex (red) and antivortex (blue) pairs of
      $\phi_1$. In this figure, we have taken $M=7$. }
    \label{fig:int7c4}
  \end{center}
\end{figure}

As mentioned above, it is well known that there is an attractive
channel and a repulsive channel for Skyrmions, depending on their
mutual orientations in field space (target space).
In this case, where we consider the interaction in the world volume of
the domain wall, it will provide us with a new interpretation of what
is going on, at least in this flavor of the Skyrme-like model (note
that the potential is different and the Skyrmion has been transformed
due to the absorption by the domain wall).
Usual Skyrmions can be oriented in any direction in SO(3) or
equivalently SU(2), but the Skyrmion handle absorbed into the domain
wall can only be rotated in the plane, which reduces the possibilities
of spatial rotation to U(1).
Recall also that the vacuum manifold is
$\Og(2)\sim \U(1)\times \mathbb{Z}_2$ and so we really only
have the rotations in the plane to play with\footnote{It is possible
  to flip two of the spatial directions to get a different
  orientation, but that will not be essential here.}.
First we will orient the two 1-Skyrmions in the attractive channel and
see what happens next. 

The numerical result is shown in Fig.~\ref{fig:int7c4r}. The columns
display the energy and baryon charge density isosurfaces, the
vacua/vortex zeros, the (anti)vortices in the field $\phi_2$ on the
domain wall and the (anti)vortices in the field $\phi_1$ on the domain
wall, respectively.
The rows show the evolution in (imaginary) relaxation time.
As a simplified picture of the interactions that govern the physics on
the domain wall, we can look at the vortex-antivortex pairs of the two
fields $\phi_{2,1}$ in the third and fourth columns of
Fig.~\ref{fig:int7c4r}. 
What happens for the interaction of the two vortex handles in the
attractive channel is that they meet with one vortex-vortex pair (in
$\phi_2$) colliding in the domain wall (plane) and make a 90-degree
scattering into two new vortices, see the third column of
Fig.~\ref{fig:int7c4r}.
Since the vortices are global vortices, we would expect two vortices 
to mutually repel each other and the vortex to be attracted to the
antivortex.
A simplistic explanation for the attraction is that in the field
$\phi_2$, there are two facing vortices that want to repel each
other, but in the field $\phi_1$ there are 2 vortex-antivortex pairs
that both attract each other. The two attractive forces win over the
single repulsive force and the net force is attractive.

The resulting interaction attracts the two vortex handles and they
combine to form a torus in the plane of the domain wall. 
The torus configuration in the standard Skyrme model is well known of
course.
It is, nevertheless, interesting to look at the Skyrmion-Skyrmion
interactions in terms of the vortices in the two fields $\phi_{1,2}$.
In the second column of Fig.~\ref{fig:int7c4r}, we can see the vortex
lines ($\phi_1$ is magenta and $\phi_2$ is blue). In order to see the
direction of the vortex, i.e.~whether it is a vortex or an antivortex
in the $(x,y)$-plane, we have to refer to columns 3 and 4 of
Fig.~\ref{fig:int7c4r}.
The interaction in the full 3-dimensional picture is somewhat more
complicated than in the simplistic picture of the interactions seen
only in the plane of the domain wall.
The colliding vortex-vortex pair still occurs, of course, but what
happens more precisely is that after the vortex-vortex collision in
the plane has happened, a vortex string junction has been made that is
left in the $\otimes$ bulk (out of the plane of the domain wall).
After the creation of the string junction in the field $\phi_2$, there
are still just two independent strings in $\phi_1$ (magenta), but the
Skyrmion configuration is quite oval and once it relaxes into a more
symmetric (toroidal) shape, the two vortex strings in $\phi_1$ also
form the same string junction, but sitting in the $\odot$ bulk and
from a top view, the position of the (anti)vortices on the domain wall
are rotated by 45 degrees with respect to the $\phi_2$ ones, see
Fig.~\ref{fig:int7c4r}.
It is a bit difficult to see the 3-dimensional structure of the
configuration in the second column of Fig.~\ref{fig:int7c4r}, so we
have duplicated these images, but from a different view point in
Fig.~\ref{fig:int7c4rtilt}. 
Now we can better see that the configuration has two string junctions
with four vortices emanating (2 vortices and 2 antivortices) and the
two junctions are thus braiding their four fingers.

The above example illustrates well what happens in the attractive
channel.
We will now consider the case shown in Fig.~\ref{fig:int7c4}, where we
have rotated both the Skyrmion handles in such a way that there are
two repulsive interactions coming from a vortex-vortex pair and an 
antivortex-antivortex pair in the $\phi_2$ field, which dominates over
the vortex-antivortex attraction in $\phi_1$.
We have only shown a single snapshot of the configuration, because
what happens next is that they both run away from each other.

\subsection{Interactions between handle and a closed vortex string}\label{sec:ringint}

In this section, we will consider a different kind of interaction,
namely between the vortex handle on the domain wall and the vortex
ring in the ($\otimes$) bulk.

\begin{figure}[!htp]
  \begin{center}
    \mbox{\subfloat{\includegraphics[width=0.245\linewidth]{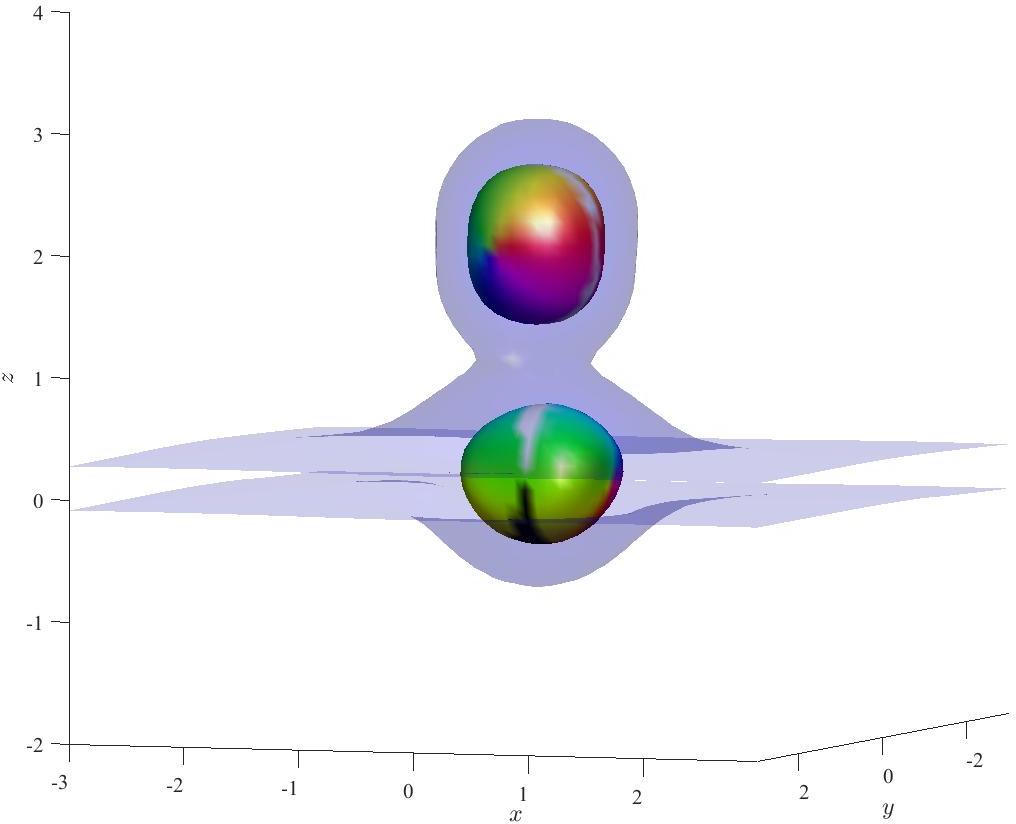}}
      \subfloat{\includegraphics[width=0.245\linewidth]{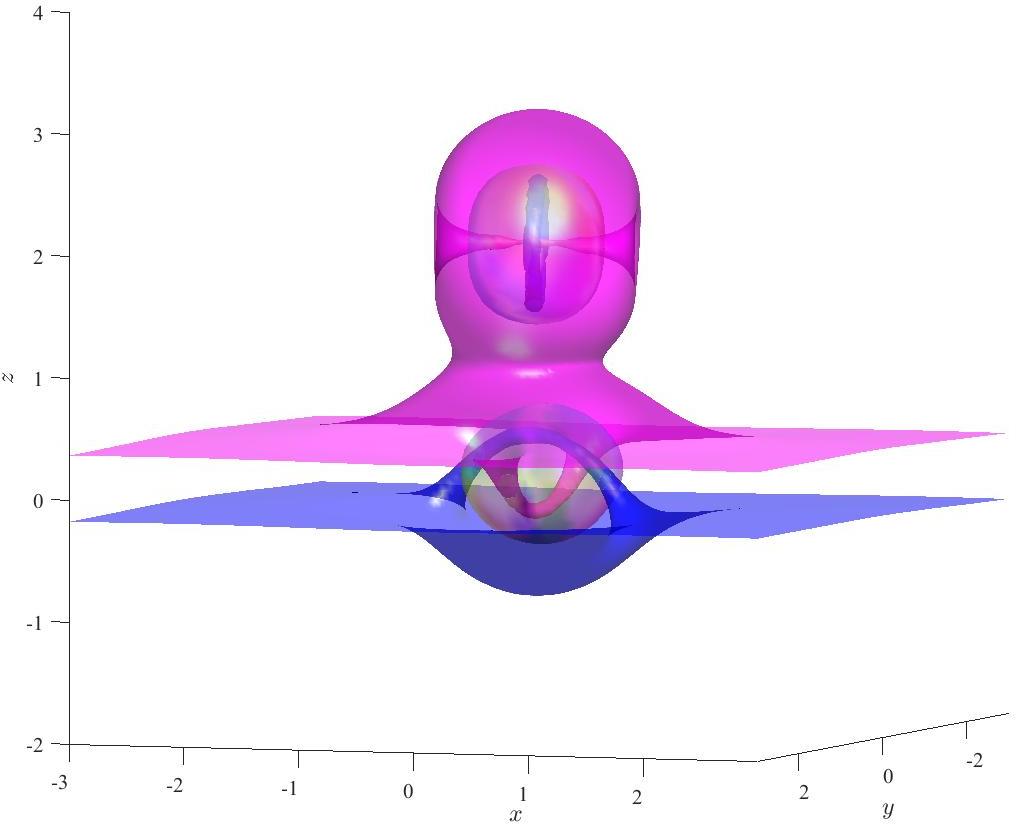}}
      \subfloat{\includegraphics[width=0.245\linewidth]{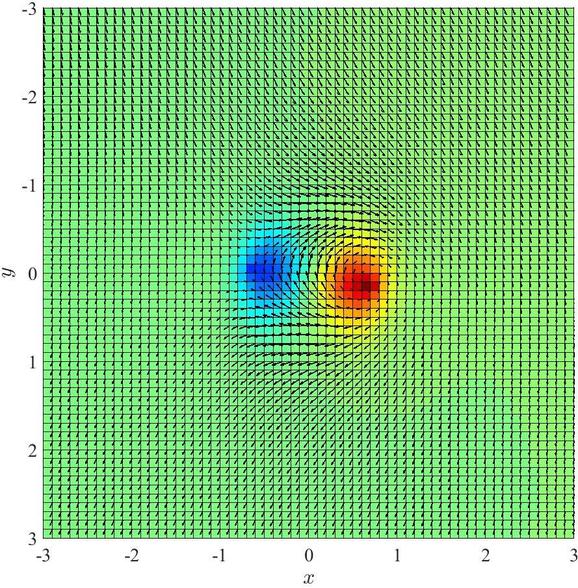}}
      \subfloat{\includegraphics[width=0.245\linewidth]{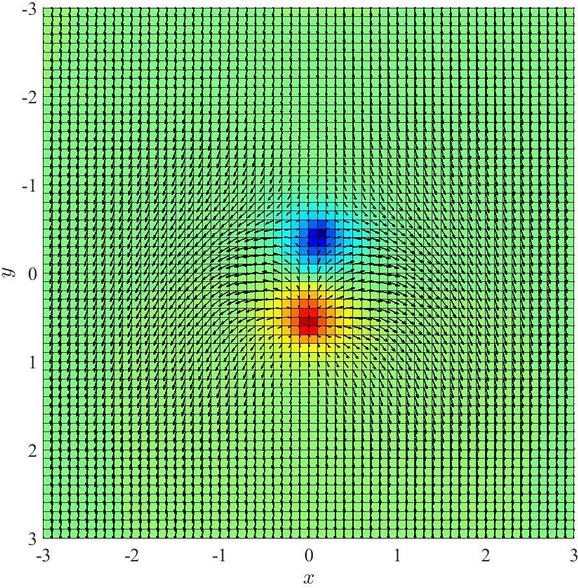}}}
    \mbox{\subfloat{\includegraphics[width=0.245\linewidth]{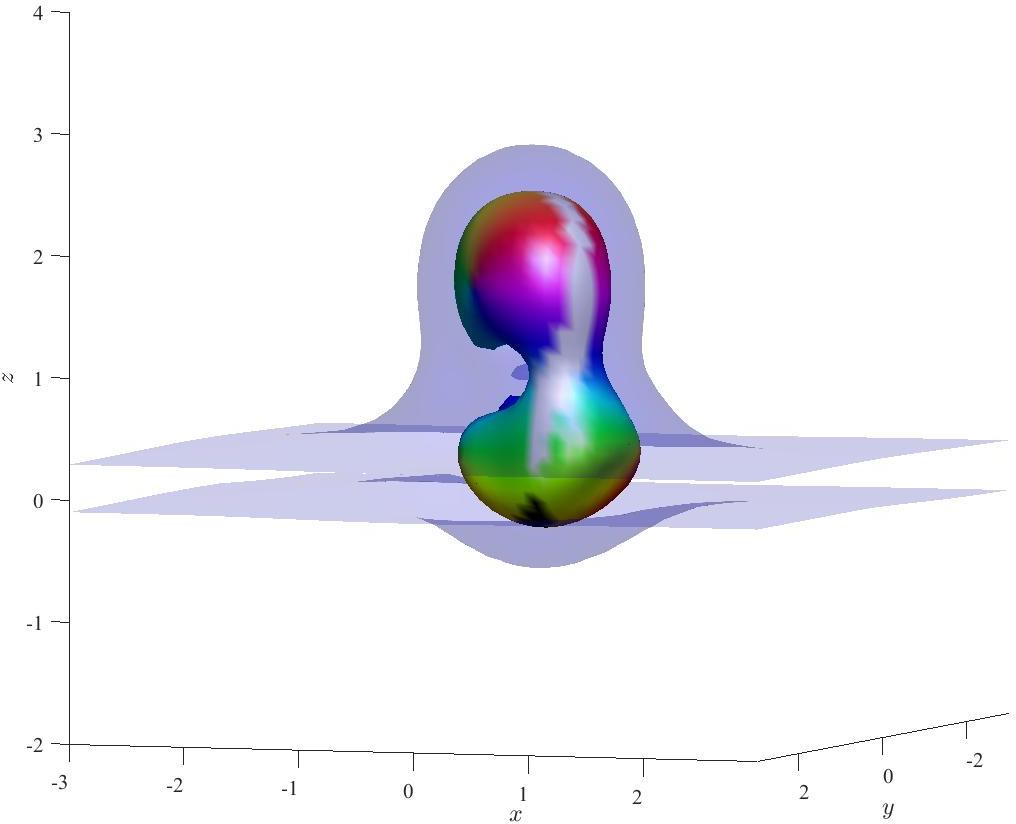}}
      \subfloat{\includegraphics[width=0.245\linewidth]{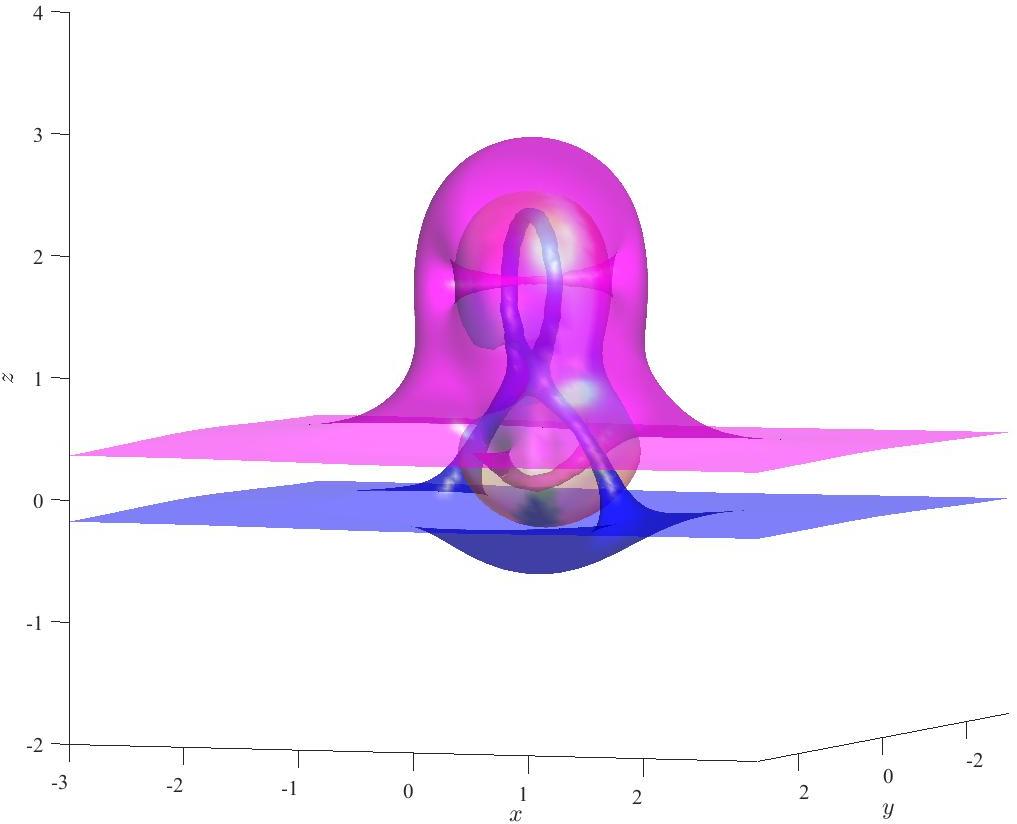}}
      \subfloat{\includegraphics[width=0.245\linewidth]{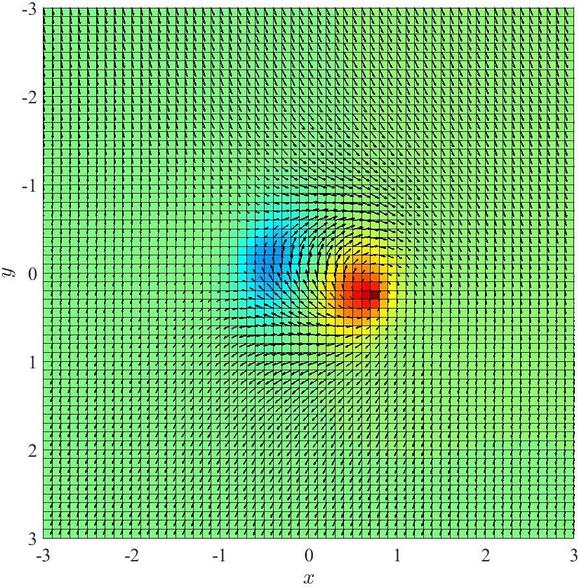}}
      \subfloat{\includegraphics[width=0.245\linewidth]{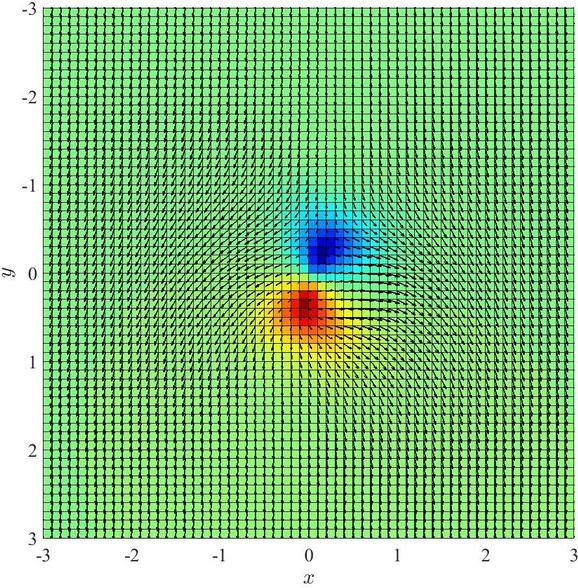}}}
    \mbox{\subfloat{\includegraphics[width=0.245\linewidth]{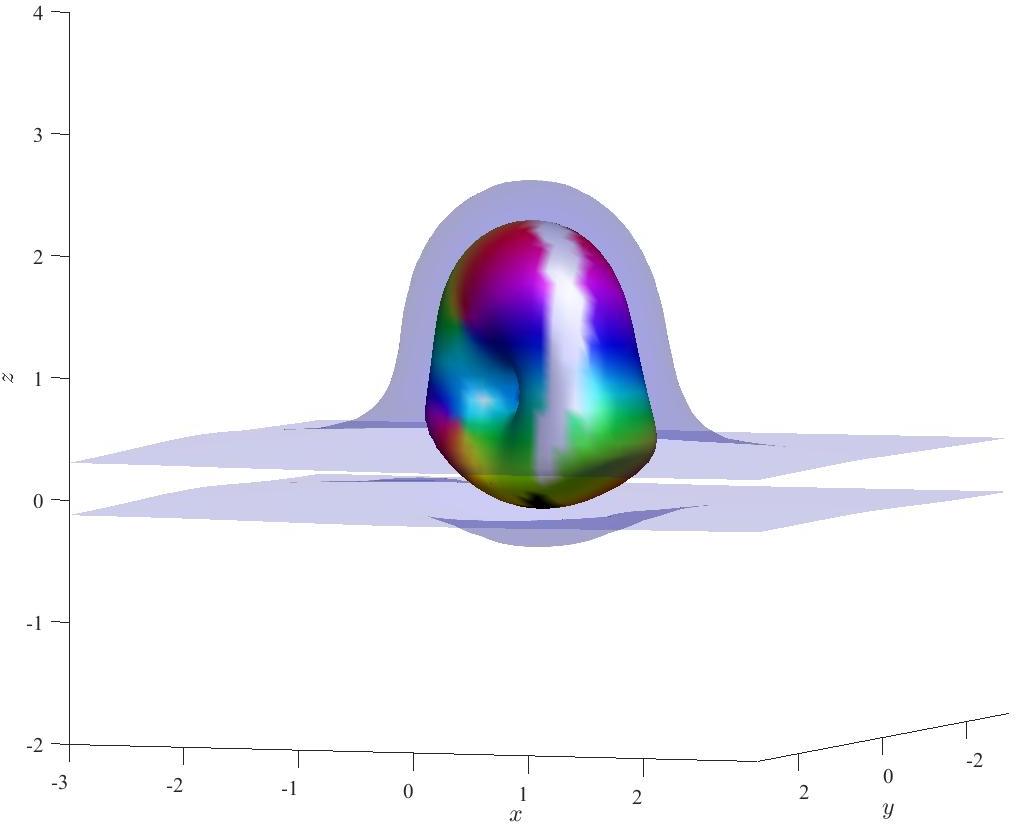}}
      \subfloat{\includegraphics[width=0.245\linewidth]{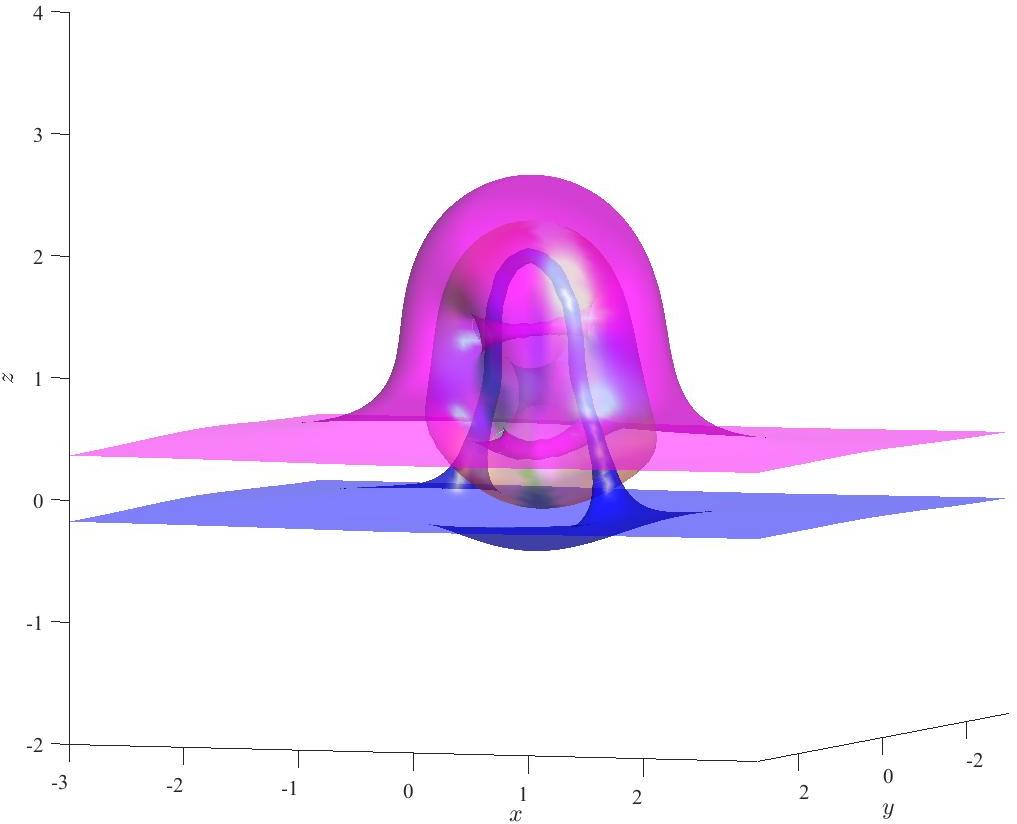}}
      \subfloat{\includegraphics[width=0.245\linewidth]{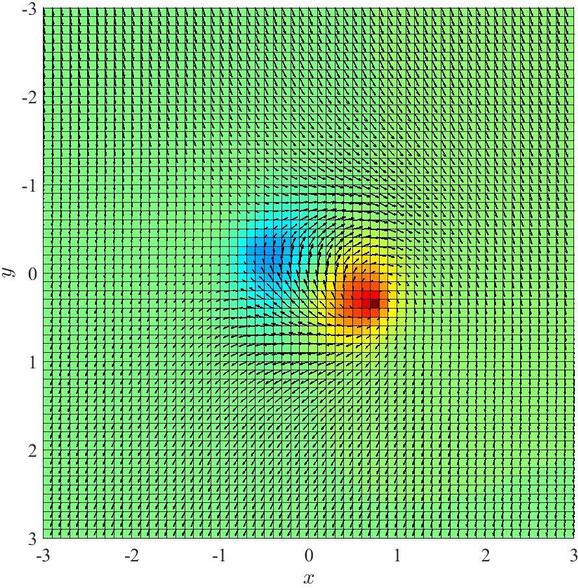}}
      \subfloat{\includegraphics[width=0.245\linewidth]{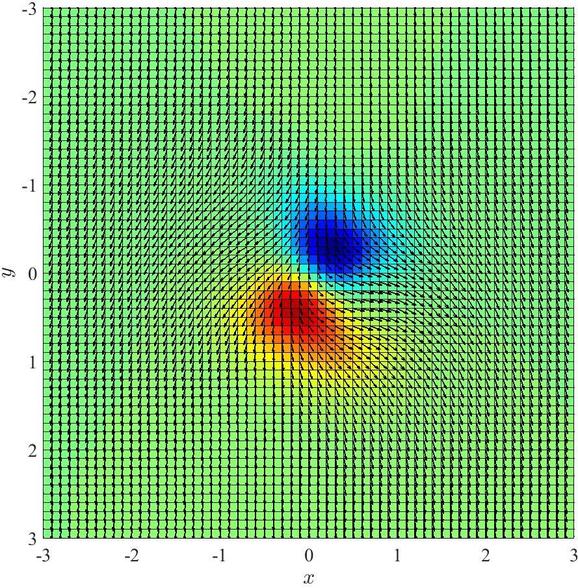}}}
    \mbox{\subfloat{\includegraphics[width=0.245\linewidth]{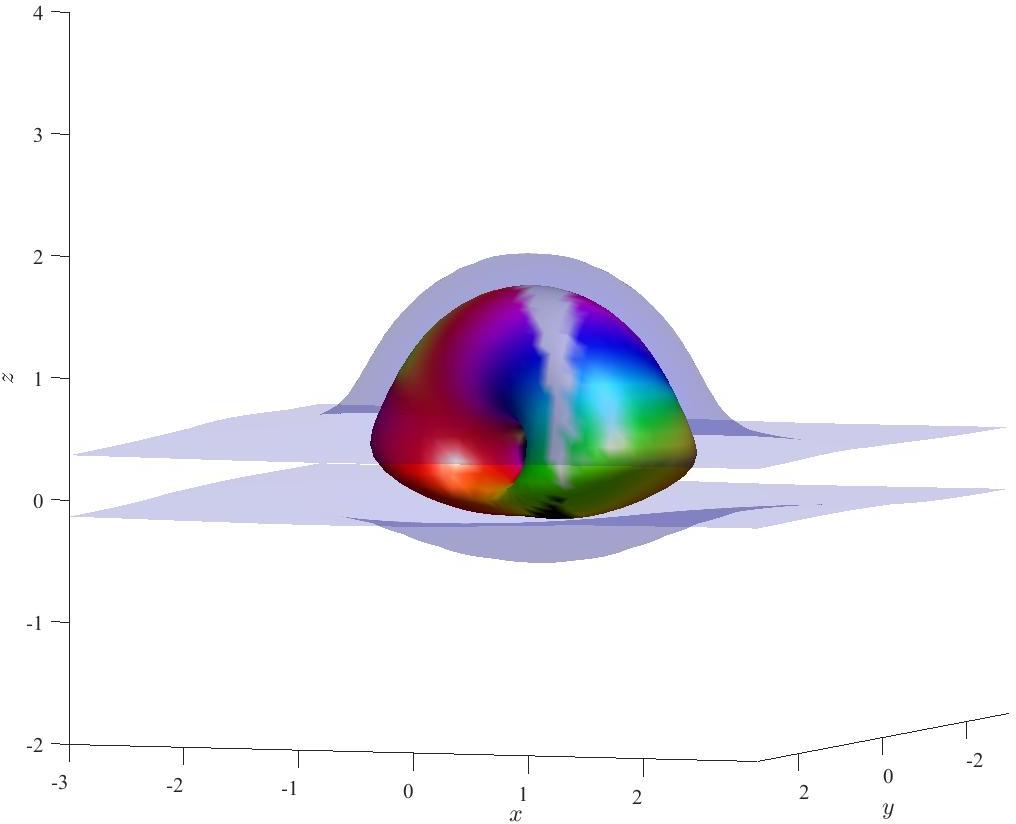}}
      \subfloat{\includegraphics[width=0.245\linewidth]{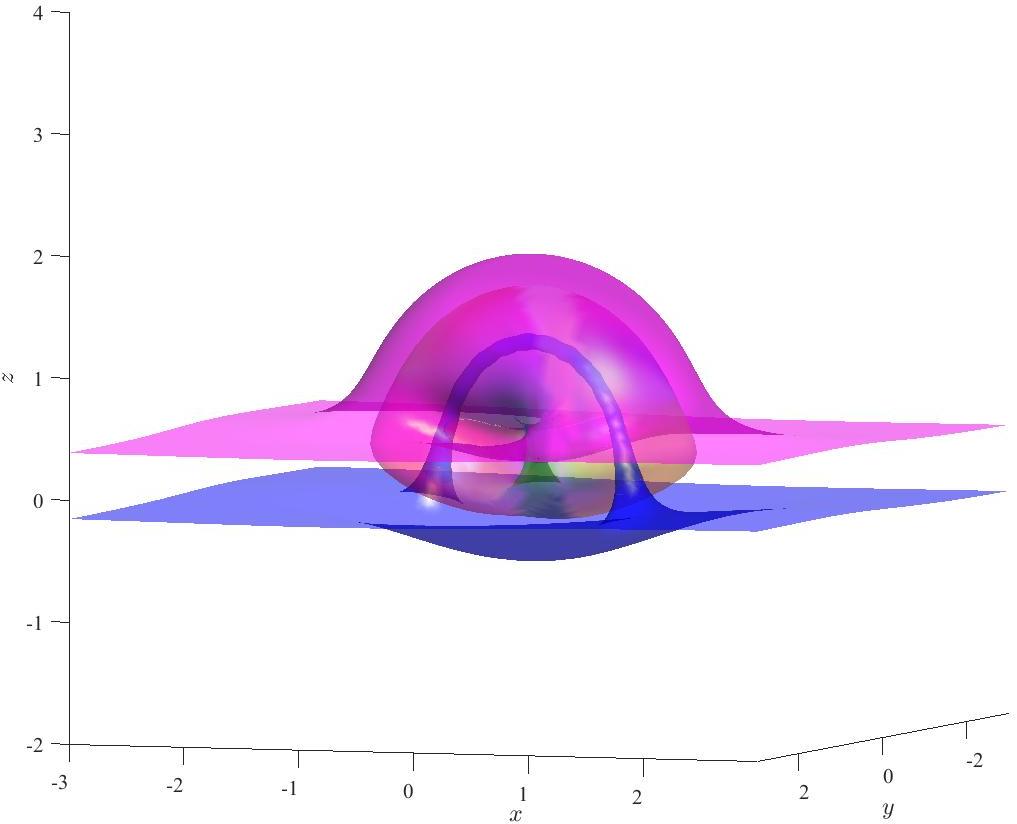}}
      \subfloat{\includegraphics[width=0.245\linewidth]{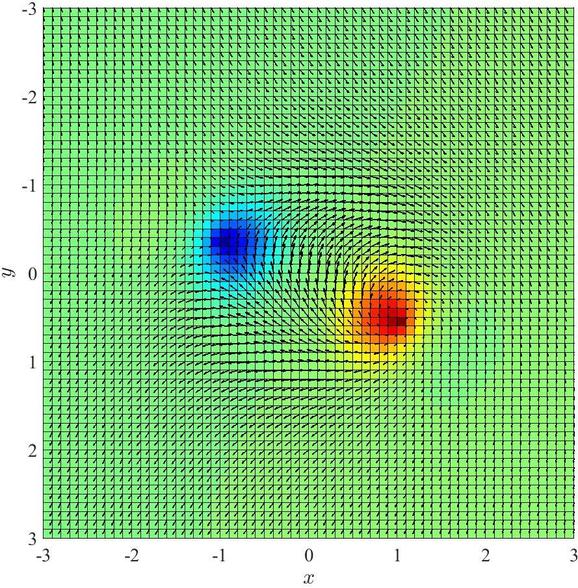}}
      \subfloat{\includegraphics[width=0.245\linewidth]{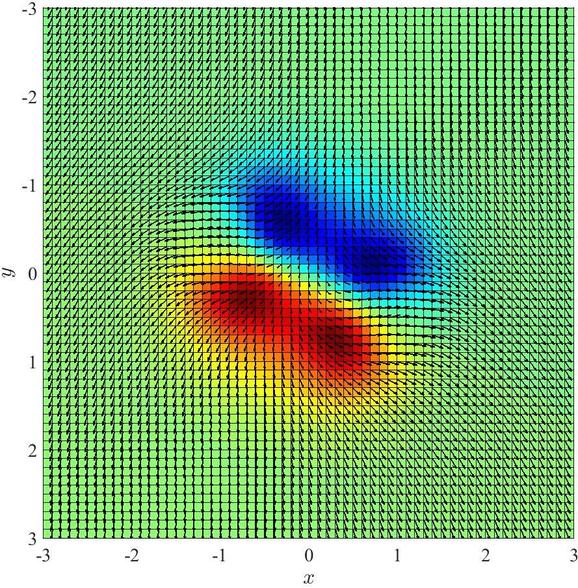}}}
    \caption{Bulk Skyrmion-Domain wall Skymion interaction: A Skyrmion
      vortex ring in the $\otimes$ bulk interacts with a vortex
      handle in the 2+4 model.
      The four columns display (1) the energy isosurfaces and baryon
      charge isosurface; (2) the zeros of $\phi_2$ (blue) and $\phi_1$
      (magenta); (3) the vortex (red) and antivortex (blue) pair of
      $\phi_2$; (4) the vortex (red) and antivortex (blue) pair of
      $\phi_1$. The rows correspond to (imaginary) relaxation time of
      the simulation. The rows are not equidistant in relaxation
      time. In this figure, we have taken $M=7$. }
    \label{fig:ringint7c4}
  \end{center}
\end{figure}

The numerical calculation is shown in Fig.~\ref{fig:ringint7c4}.
Similarly to the matrix in the figures in the previous section, we
show the configuration at four (imaginary) time steps as four rows in
the figure.
The initial orientation of either the vortex handle on the domain wall
or the vortex ring in the ($\otimes$) bulk is not too important,
because the vortex ring in the bulk will rotate into the attractive
channel as chosen as the initial configuration (first row) in
Fig.~\ref{fig:ringint7c4}.
The vortex ring (in $\phi_2$, blue) in the $\otimes$ bulk is initially 
oriented perpendicularly to the plane of the domain wall.
The first thing that happens in the interaction is that the vortex in
$\phi_2$ (blue) is pulled down towards the vortex handle on the domain
wall and it \emph{reconnects} such that the string becomes a longer
handle, see rows 1-3 of Fig.~\ref{fig:ringint7c4}.
The beauty of the reconnection is that it automatically produces a
vortex handle with double twist in the $\chi$ field (see the
parametrization \eqref{eq:phi_parametrization}). This can be seen from
the fact that the vortex handle in $\phi_2$ encloses (links with) two
dual strings in $\phi_1$, see the third row of
Fig.~\ref{fig:ringint7c4}.
The final phase in the relaxation just minimizes the energy by making
the the doubly twisted vortex handle shorter and more compact, see
the fourth row of Fig.~\ref{fig:ringint7c4}.

\subsection{Energy comparison of the two $B=2$ Skyrmions}\label{sec:energy}
It is interesting to see that the interaction between the two handles
in the plane of the domain wall created a different 2-Skyrmion (a
braided string junction torus) compared to the interaction of the
vortex handle and the vortex ring in the bulk, which created the
doubly twisted vortex handle that we constructed in
Sec.~\ref{sec:doublehandle}.
In order to know which configuration is the stable one, we will
compare their energies numerically.
As the domain wall has an infinite energy in the infinite space, we
will subtract off the domain wall energy and just calculate the energy
of the Skyrmion configurations, with zero being the empty domain
wall.

\begin{table}[!htp]
  \begin{center}
    \caption{Energies of various configurations in the 2+4 model.}
    \label{tab:En}
    \begin{tabular}{l||ccc}
      type & $B$ & $M$ & $E$\\
      \hline\hline
      handle                  & 1 & 3 & 69.70\\
      ring                    & 1 & 3 & 94.59\\
      \hline
      handle                  & 1 & 7 & 84.73\\
      ring                    & 1 & 7 & 112.4\\
      \hline
      braided string junction & 2 & 3 & 132.8\\
      doubly twisted handle   & 2 & 3 & 144.3\\
      \hline
      braided string junction & 2 & 7 & 163.3\\
      doubly twisted handle   & 2 & 7 & 173.6
    \end{tabular}
  \end{center}
\end{table}

The results of the numerical calculations for the energies are shown
in Tab.~\ref{tab:En}.
In particular, we first calculate the masses of the vortex handle for
$M=3,7$ compared to the vortex ring and see that there is a strong
binding energy for the Skyrmion to be gained by getting absorbed into
the domain wall.
Next we compare the braided string junction of Sec.~\ref{sec:int} with
the doubly twisted handles of Secs.~\ref{sec:doublehandle} and
\ref{sec:ringint}. The conclusion drawn from the energy measurements
is clear; the braided string junction of toroidal shape has the far
lowest energy of the two different 2-Skyrmions and thus is the stable
one.

\subsection{Higher-charged handles}\label{sec:higher}

As we have seen in the previous sections, the interaction dynamics is
quite intricate and the number of ways of combining handles and rings
with various numbers of twists is overwhelmingly large.
Therefore, a complete study of higher-charged Skyrmions in this theory
is beyond the scope of the paper.
However, let us mention that there are in principle 3 different
possibilities: multi-Skyrmions as multi-solitons in the domain wall
(as e.g.~the braided string junction as a 2-Skyrmion), multi-Skyrmions
as higher-twisted handles sticking into the bulk and finally, a
hybrid of the previous two options.

\begin{figure}[!htp]
  \begin{center}
    \mbox{\sidesubfloat[]{\includegraphics[width=0.45\linewidth]{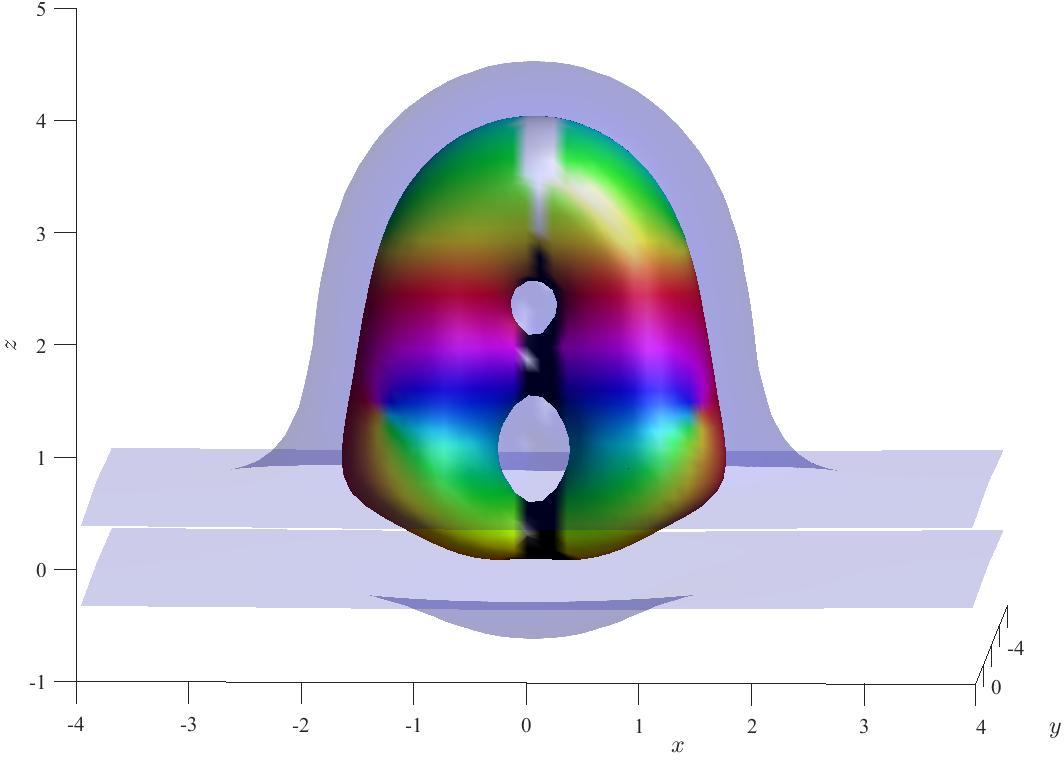}}
    \sidesubfloat[]{\includegraphics[width=0.45\linewidth]{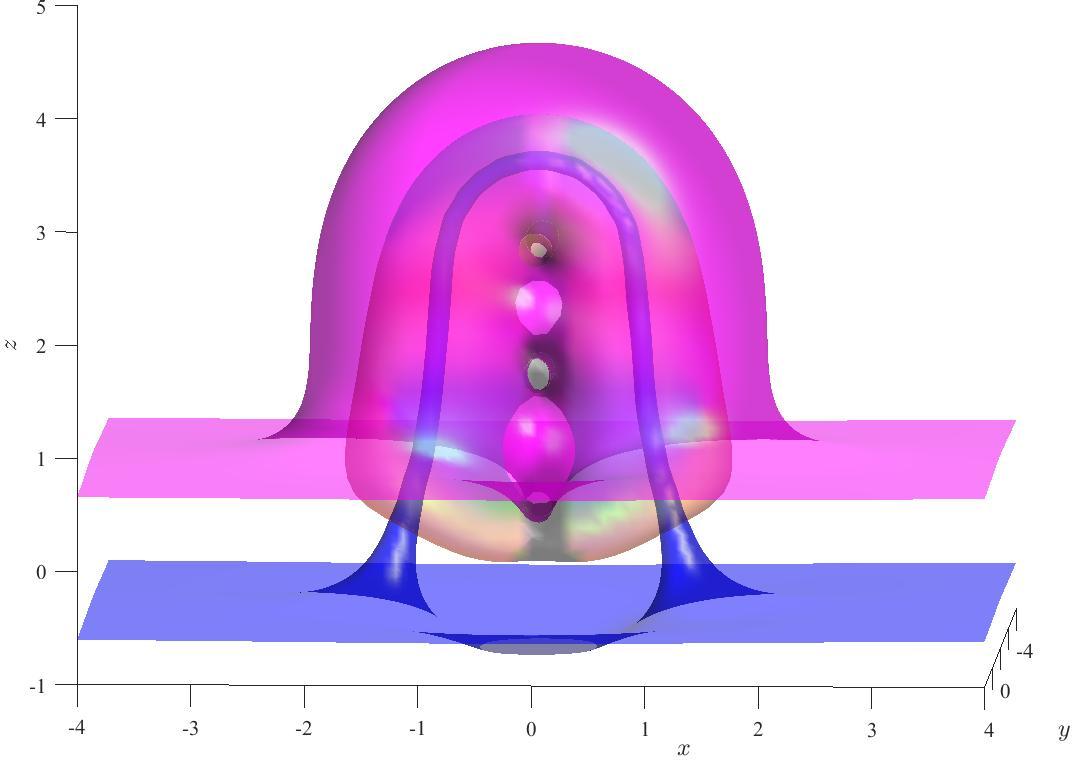}}}
    \mbox{\sidesubfloat[]{\includegraphics[width=0.4\linewidth]{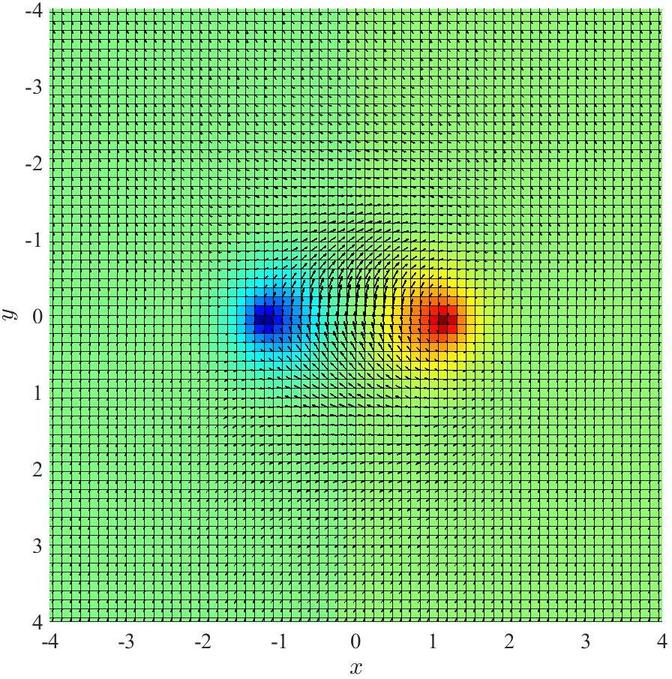}}\ \ 
      \sidesubfloat[]{\includegraphics[width=0.4\linewidth]{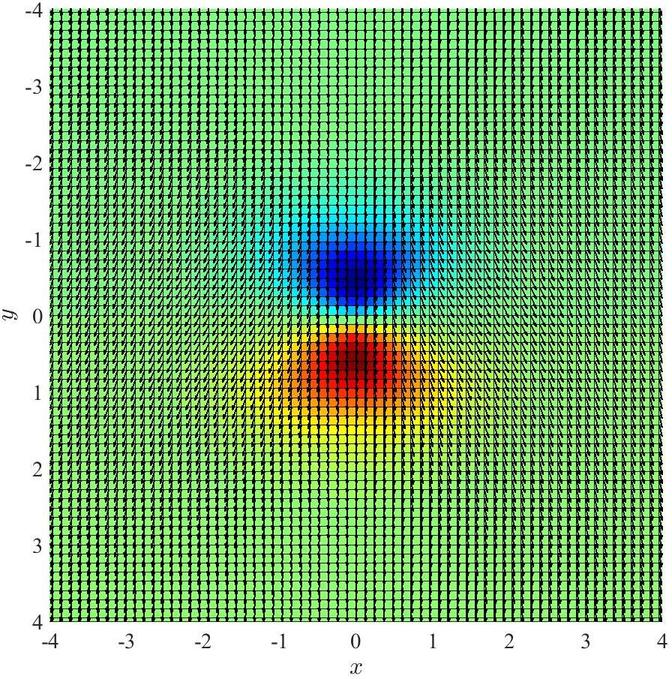}}}
    \caption{3-Skyrmion as a vortex handle with three twists on the
      domain wall in the 2+4 model. (a) the energy density is shown
      with blue transparent isosurfaces illustrating the domain wall
      and the baryon charge density is shown with an isosurface with
      the color scheme described in the text. (b) the blue isosurface
      (bottom) represents the zeros of $\phi_2$ and the magenta
      isosurface (top) is the zeros of $\phi_1$. The baryon charge is
      added transparently. (c) the vortex (red) and antivortex (blue)
      pair of $\phi_2$. (d) the vortex (red) and antivortex (blue)
      pair of $\phi_1$. In this figure, we have taken $M=3$.  }
    \label{fig:24B3}
  \end{center}
\end{figure}

\begin{figure}[!htp]
  \begin{center}
    \mbox{\sidesubfloat[]{\includegraphics[width=0.45\linewidth]{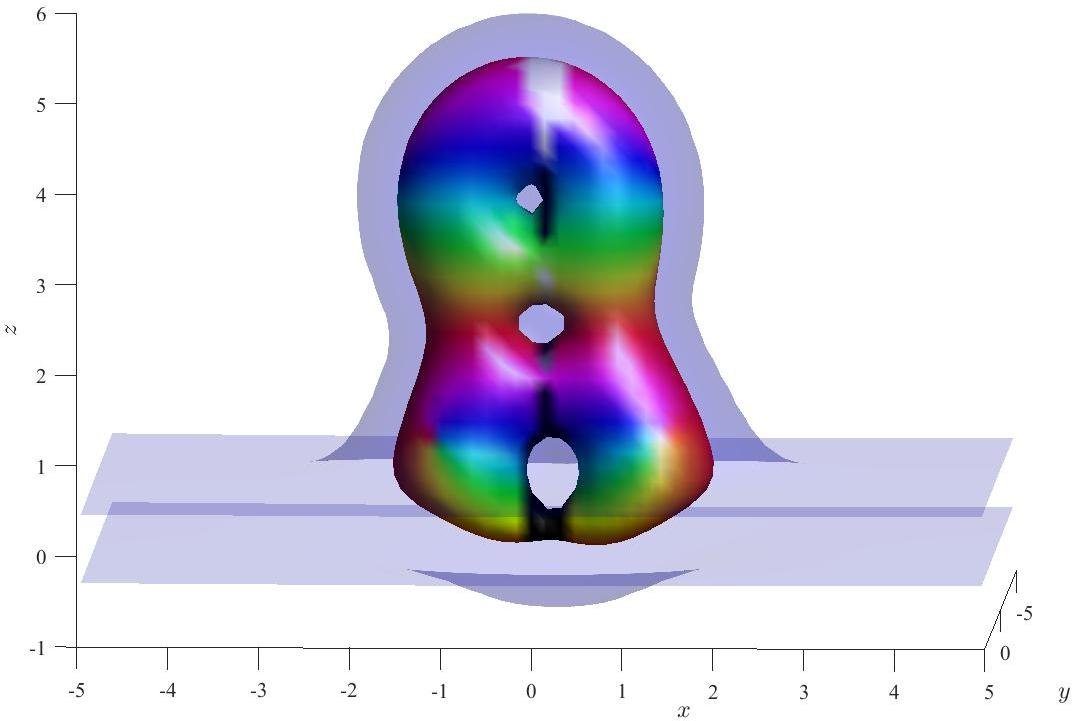}}
    \sidesubfloat[]{\includegraphics[width=0.45\linewidth]{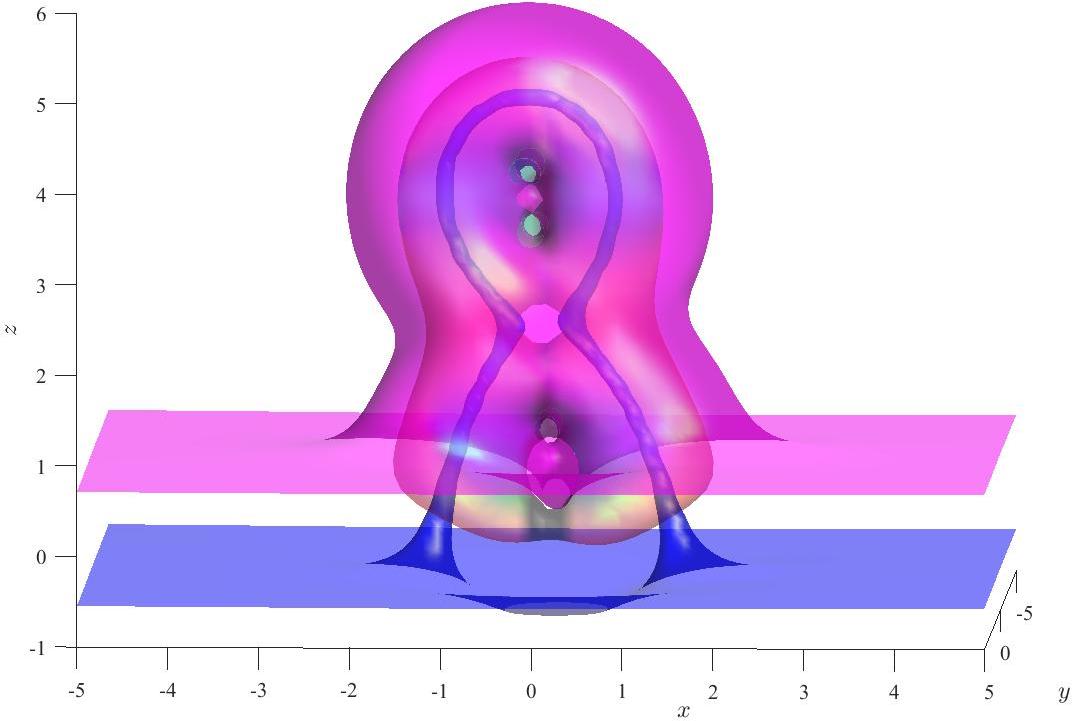}}}
    \mbox{\sidesubfloat[]{\includegraphics[width=0.4\linewidth]{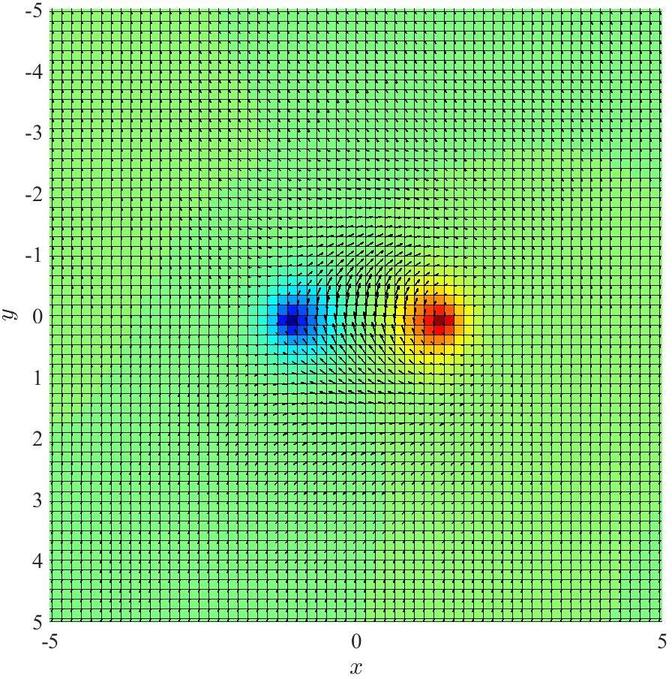}}\ \ 
      \sidesubfloat[]{\includegraphics[width=0.4\linewidth]{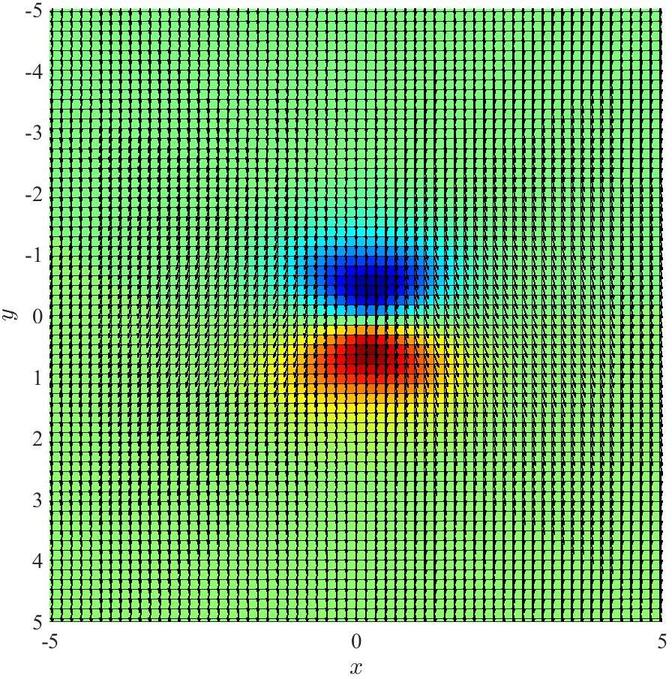}}}
    \caption{4-Skyrmion as a vortex handle with four twists on the
      domain wall in the 2+4 model. (a) the energy density is shown
      with blue transparent isosurfaces illustrating the domain wall
      and the baryon charge density is shown with an isosurface with
      the color scheme described in the text. (b) the blue isosurface
      (bottom) represents the zeros of $\phi_2$ and the magenta
      isosurface (top) is the zeros of $\phi_1$. The baryon charge is
      added transparently. (c) the vortex (red) and antivortex (blue)
      pair of $\phi_2$. (d) the vortex (red) and antivortex (blue)
      pair of $\phi_1$. In this figure, we have taken $M=3$.  }
    \label{fig:24B4}
  \end{center}
\end{figure}

\begin{figure}[!htp]
  \begin{center}
    \mbox{\sidesubfloat[]{\includegraphics[width=0.45\linewidth]{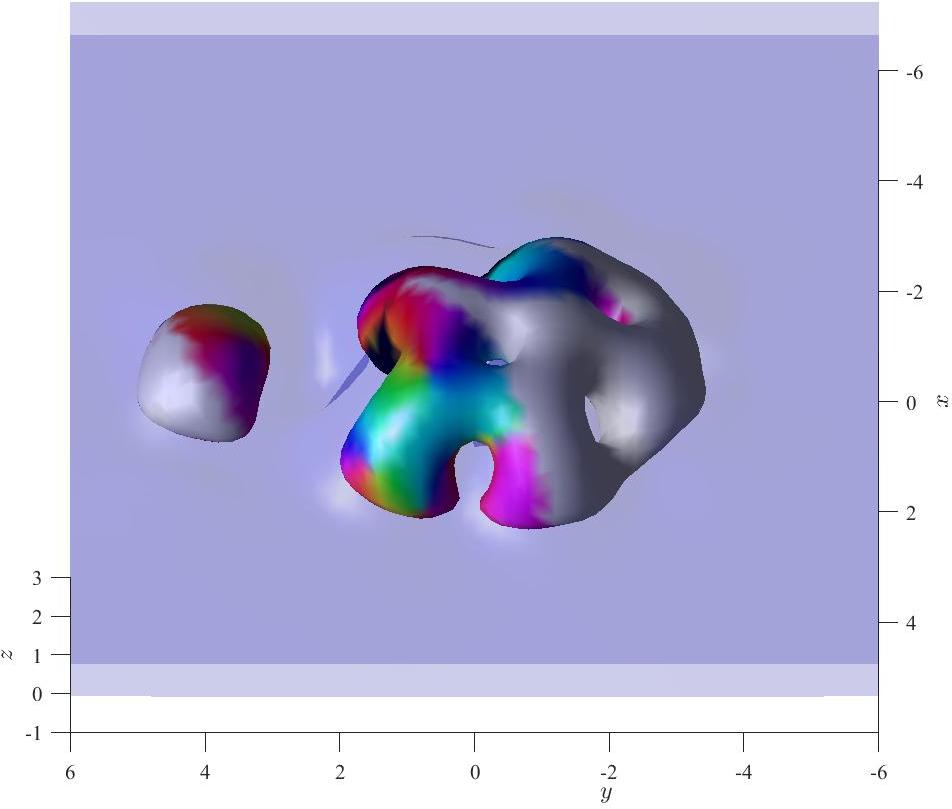}}
    \sidesubfloat[]{\includegraphics[width=0.45\linewidth]{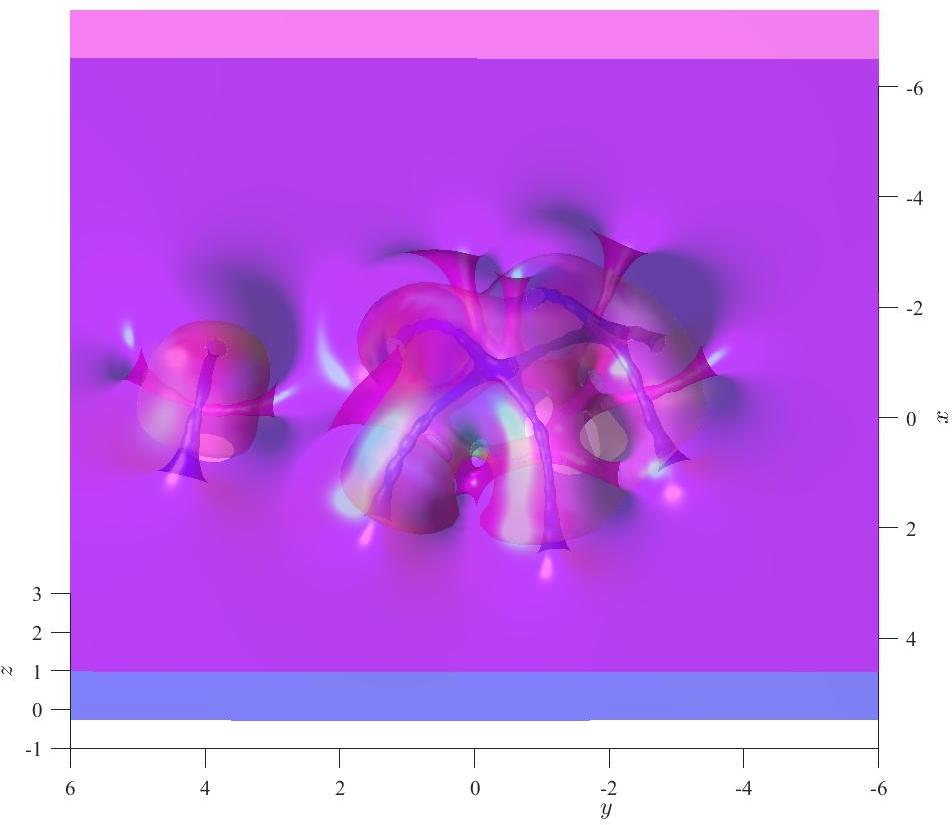}}}
    \mbox{\sidesubfloat[]{\includegraphics[width=0.4\linewidth]{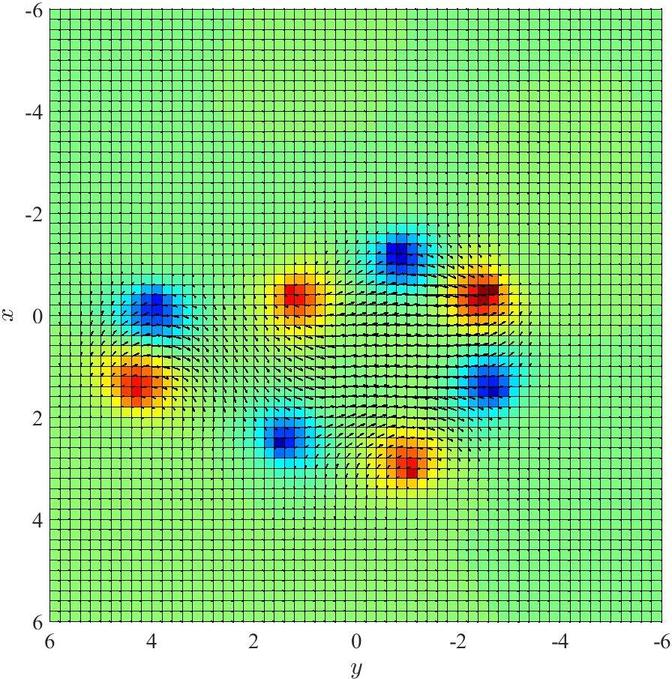}}\ \ 
      \sidesubfloat[]{\includegraphics[width=0.4\linewidth]{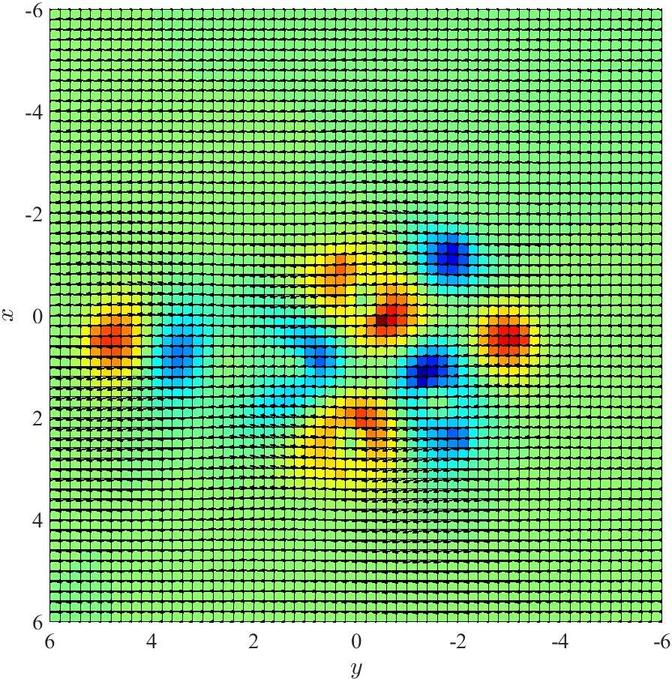}}}
    \caption{5-Skyrmion as a double vortex junction with twists on the
      domain wall next to a 1-Skyrmion as a handle, in the 2+4
      model. (a) the energy density is shown 
      with blue transparent isosurfaces illustrating the domain wall
      and the baryon charge density is shown with an isosurface with
      the color scheme described in the text. (b) the blue isosurface
      (bottom) represents the zeros of $\phi_2$ and the magenta
      isosurface (top) is the zeros of $\phi_1$. The baryon charge is
      added transparently. (c) the vortex (red) and antivortex (blue)
      pairs of $\phi_2$. (d) the vortex (red) and antivortex (blue)
      pairs of $\phi_1$. In this figure, we have taken $M=3$.  }
    \label{fig:24B5+1}
  \end{center}
\end{figure}

\begin{figure}[!htp]
  \begin{center}
    \mbox{\sidesubfloat[]{\includegraphics[width=0.45\linewidth]{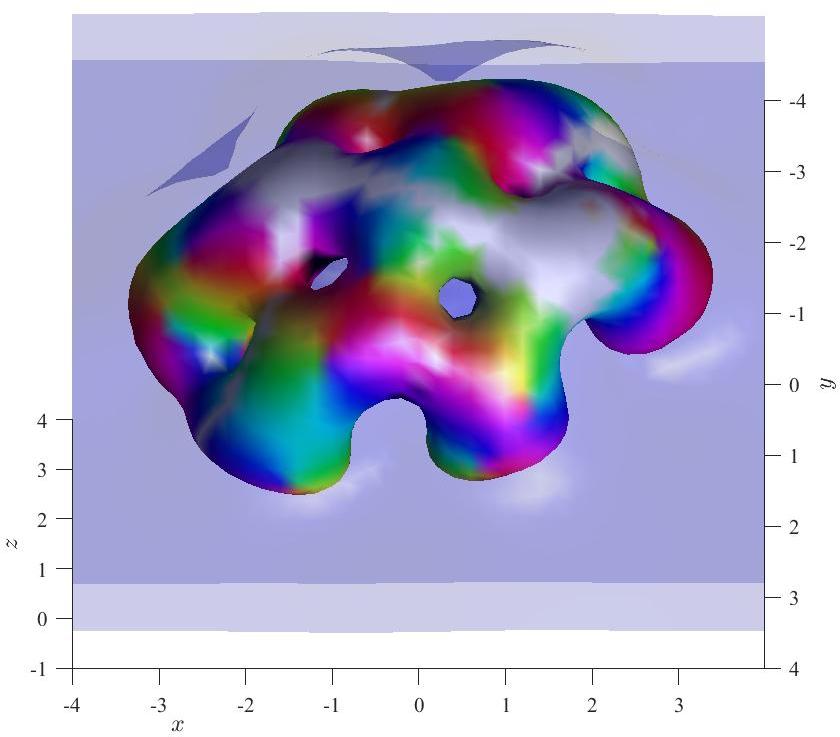}}
    \sidesubfloat[]{\includegraphics[width=0.45\linewidth]{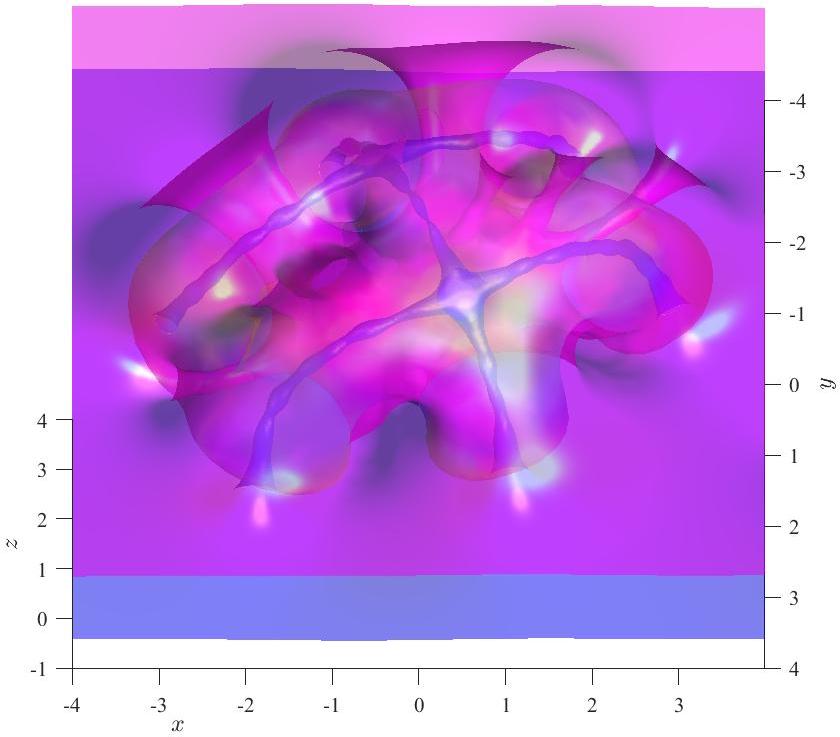}}}
    \mbox{\sidesubfloat[]{\includegraphics[width=0.4\linewidth]{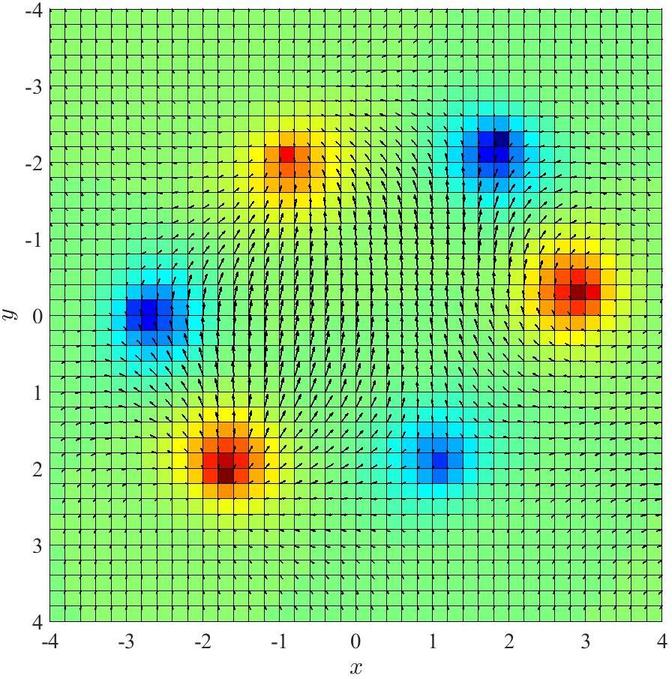}}\ \ 
      \sidesubfloat[]{\includegraphics[width=0.4\linewidth]{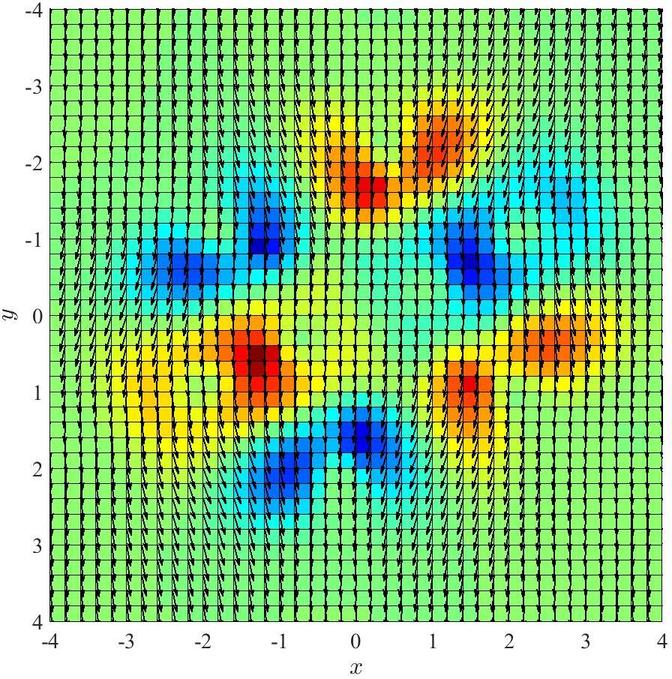}}}
    \caption{7-Skyrmion as a double vortex junction with twists on the
      domain wall in the 2+4 model. (a) the energy density is shown 
      with blue transparent isosurfaces illustrating the domain wall
      and the baryon charge density is shown with an isosurface with
      the color scheme described in the text. (b) the blue isosurface
      (bottom) represents the zeros of $\phi_2$ and the magenta
      isosurface (top) is the zeros of $\phi_1$. The baryon charge is
      added transparently. (c) the vortex (red) and antivortex (blue)
      pairs of $\phi_2$. (d) the vortex (red) and antivortex (blue)
      pairs of $\phi_1$. In this figure, we have taken $M=3$.  }
    \label{fig:24B7}
  \end{center}
\end{figure}

In this section, we only explore one possibility as the initial
condition, but then use the numerical calculations and relax them to
the nearest metastable configuration; that is, we consider a single
open vortex handle with various twists as the initial guess.

The first numerical result is for $B=3$, i.e.~a single vortex in
$\phi_2$ that is twisted 3 times yielding a 3-Skyrmion, see
Fig.~\ref{fig:24B3}.
It turns out to be metastable and it has exactly one vortex handle in
the field $\phi_2$ and it is linked with 3 individual dual strings of
$\phi_1$.

Next, we try to add a twist to the previous initial guess in order to
create a 4-Skyrmion as a vortex handle on the domain wall.
The result is shown in Fig.~\ref{fig:24B4} and it also turns out to be
metastable.
A curiosity is that the 4 dual string are located near the top
``hole'' of the Skyrmion and the bottom ``hole'', but not near the
middle ``hole,'' which seems to appear due to a stereo cusp in the
$\phi_2$ vortex handle.

We now add two more twists to the initial guess and try to search for
a 6-Skyrmion, but this time the metastability of the configuration was
not found (or the guess was too far away from such a solution).
The numerical result is shown in Fig.~\ref{fig:24B5+1}, where the
vortex handle has collapsed into a single handle and a 5-Skyrmion that
consists of a double string junction in the field $\phi_1$ with
complicated twists around it (dual strings). 

The last attempt at looking for a higher-charged handle is to add
another twist to the former guess, yielding a total of 7 twists.
The numerical result is shown in Fig.~\ref{fig:24B7} and it also
collapsed from a vortex handle into a double string junction,
intricately braided with dual strings, yielding a 7-Skyrmion. 

The true minimizers of the energy for $B\geq 3$ may not be the
solutions found here.

\section{Discussion and outlook}\label{sec:discussion}

In this paper we have considered a BEC-inspired potential in the
Skyrme model and a sextic version of the Skyrme model, which due to
the potential possesses vortex strings, and in particular we have
studied the setting in the presence of a domain wall -- which is also
possessed by the theory with the BEC-inspired potential.
The vortex strings contain a U(1) modulus, that once twisted by $2\pi$
yields a unit baryon charge.
The first intuitive picture is that we wind the vortex string once to
get a 1-Skyrmion in the bulk and when it is placed in the vicinity of
the domain wall, we have found that attractive forces absorb the
Skyrmion and it becomes a vortex handle sitting on the domain wall.
It turns out that it inevitably also has a dual string, i.e.~a vortex
in the complementary component of the 2-component complex scalar field
$\boldsymbol{\phi}$, and the configuration is made of two linked
open strings ending on the domain wall. 
We have further studied the interactions between two handles and found
that when oriented in the attractive channel, they attract and combine
as a torus-shaped braided string junction.
Another way to create a 2-Skyrmion is to place a vortex ring in the
vicinity of a vortex handle sitting on the domain wall, which also
attract each other and create a doubly twisted vortex handle.
The former has the lowest energy, nevertheless.

An obvious direction for further studies is to consider higher-charged
Skyrmions, which may yield a large number of metastable
configurations; in particular we have not necessarily found the true
energy minimizers for $B\geq 3$. This would require a large search for
configurations based on many different initial guesses. We will leave
this for future work.

An interesting observation is that all the Skyrmions of baryon number
$B$ seem to be composed of vortex zeros of $\phi_1$ and of $\phi_2$
that link each other $B$ times.
It could be interesting to study this fact further and investigate
whether this is always the case.

An important development would be to see how many of the results in 
this model can be carried over to BECs and under what circumstances.

\subsection*{Acknowledgments}

This work is supported by the Ministry of Education, Culture, Sports,
Science (MEXT)-Supported Program for the Strategic Research Foundation
at Private Universities ``Topological Science'' (Grant No.~S1511006) and
by a Grant-in-Aid for Scientific Research on Innovative Areas
``Topological Materials Science'' (KAKENHI Grant No.~15H05855) from
MEXT, Japan.
The work of M.~N.~is also supported in part by the Japan Society for
the Promotion of Science (JSPS) Grant-in-Aid for Scientific Research
(KAKENHI Grant No.~16H03984 and No.~18H01217).
The TSC-computer of the ``Topological Science'' project at Keio
University was used for the numerical calculations.

\end{document}